\documentclass{article}

\usepackage{arxiv}
\usepackage[utf8]{inputenc}
\usepackage[T1]{fontenc}
\usepackage{hyperref}
\usepackage{url}
\usepackage{booktabs}
\usepackage{amsfonts}
\usepackage{nicefrac}
\usepackage{microtype}
\usepackage{graphicx}
\usepackage{amsmath}
\usepackage[round]{natbib}
\usepackage{doi}
\usepackage{tcolorbox}
\usepackage{subcaption}
\graphicspath{ {./figures/} }
\usepackage{color}

\newcommand{\D}{\mathbf{D}}
\newcommand{\F}{\mathbf{F}}

\newcommand{\bgamma}{\boldsymbol{\gamma}}

\title{Trust-Aware Multimodal Data Fusion for Yield Estimation: A Case Study of the 2020 Beirut Explosion}

\author{
 Lekha Patel \\
  Scientific Machine Learning \\
  Sandia National Laboratories\\
  Albuquerque, NM 87123 \\
\texttt{lpatel@sandia.gov} \\
   \And
   Craig Ulmer \\
  Scalable Modeling \& Analysis \\
  Sandia National Laboratories\\
  Livermore, CA 94550 \\
\texttt{cdulmer@sandia.gov} \\
\And
Stephen J. Verzi \\
Computational Decision Science \\
Sandia National Laboratories\\
Albuquerque, NM 87123 \\
\texttt{sjverzi@sandia.gov} \\
\And
Daniel Krofcheck \\
Complex Systems Risk \& Resilience \\
Sandia National Laboratories\\
Albuquerque, NM 87123 \\
\texttt{djkrofc@sandia.gov} \\
\And
Indu Manickam \\
Complex Systems Risk \& Resilience \\
Sandia National Laboratories\\
Albuquerque, NM 87123 \\
\texttt{imanick@sandia.gov} \\
\And
Asmeret Naugle \\
Computational Decision Science \\
Sandia National Laboratories\\
Albuquerque, NM 87123 \\
\texttt{abier@sandia.gov} \\
\And 
 Jaideep Ray \\ 
 Computational Data Science \\
   Sandia National Laboratories\\ 
   Livermore, CA 94550 \\ 
\texttt{jairay@sandia.gov} \\
}

\begin{document}
\maketitle

% !TEX root = main.tex
\begin{abstract}
The estimation of explosive yield from heterogeneous observational data presents fundamental challenges in inverse problems, particularly when combining traditional physical measurements with modern artificial intelligence-interpreted modalities. We present a novel Bayesian fractional posterior framework that fuses seismic waves, crater dimensions, synthetic aperture radar imagery, and vision-language model interpreted ground-level images to estimate the yield of the 2020 Beirut explosion. Unlike conventional approaches that may treat data sources equally, our method learns trust weights for each modality through a Dirichlet prior, automatically calibrating the relative information content of disparate observations. Applied to the Beirut explosion, the framework yields an estimate of 0.34--0.48 kt TNT equivalent, representing 12 to 17 percent detonation efficiency relative to the 2.75 kt theoretical maximum from the blast's stored ammonium nitrate. The fractional posterior approach demonstrates superior uncertainty quantification compared to single-modality estimates while providing robustness against systematic biases. This work establishes a principled framework for integrating qualitative assessments with quantitative physical measurements, with applications to explosion monitoring, disaster response, and forensic analysis.
\end{abstract}

% !TEX root = main.tex 
\section{Introduction}  
Explosive yield estimation serves critical functions in international security monitoring, disaster response planning, and forensic accident investigation. For treaty verification under frameworks such as the Comprehensive Nuclear-Test-Ban Treaty, accurate yield determination enables verification of compliance with weapons testing restrictions \citep{pasyanos2018coupled,pasyanos2022full}. Following industrial accidents involving energetic materials, yield estimates inform emergency response decisions, guide evacuation planning, and support forensic analyses to prevent future incidents \citep{pasman2020beirut,sadek2022impacts}. In both, timely and accurate yield assessment can mean the difference between effective response and catastrophic outcomes.

Modern disasters generate unprecedented volumes of heterogeneous observational data across multiple sensing modalities. Satellite synthetic aperture radar (SAR) imagery captures spatial damage patterns across entire cities, social media platforms collect thousands of geotagged photographs documenting ground-level structural damage, and seismic networks record body waves and surface waves at regional and teleseismic distances. Recent advances in vision-language models have further enabled automated interpretation of imagery for damage assessment without requiring extensive training datasets \citep{ahn2024generalizable,beltre2025performance,gupta2019creating}. These models can classify structural damage severity from photographs, estimate overpressure from visual cues, and extract quantitative information from unstructured image collections. However, this data proliferation presents both opportunities and challenges. While diverse observations potentially contain complementary information that could reduce yield estimation uncertainty when properly combined, integrating, for example, qualitative visual assessments with quantitative physical measurements requires principled statistical frameworks that account for varying uncertainties and information content across modalities.

Traditionally, yield estimation methods rely on well-calibrated empirical relationships between physical observables and explosive energy. For instance, seismic magnitude scales have been developed through decades of underground nuclear testing \citep{pasyanos2015determining,pasyanos2019seismoacoustic}, crater dimensions provide direct evidence of energy deposition \citep{ambrosini2004crater,holsapple1993scaling}, and atmospheric infrasound measurements capture acoustic energy propagation \citep{kim2022yield}. However, each single-modality approach carries substantial systematic uncertainties. Seismic coupling efficiency varies by orders of magnitude depending on geological conditions and emplacement configuration \citep{khalturin1998review}, crater scaling relationships depend critically on substrate properties, moisture content, and burst height \citep{ambrosini2002size}, and atmospheric propagation models require accurate knowledge of temperature and wind profiles that are rarely available in operational settings \citep{kim2022yield}.

Recent studies have attempted to leverage the complementary information from multiple modalities through joint Bayesian inference. 
\cite{ford2021joint} successfully estimated yield using seismic, acoustic, and optical observations through standard Bayesian inference, demonstrating computational feasibility of multimodal fusion. \cite{kim2022yield} achieved comparable results for the 2020 Beirut explosion using seismoacoustic data with physics-based propagation models, obtaining yield estimates of 0.4--1.4 kt. However, both approaches employ conventional posterior forms $p(\theta|y) \propto p(\theta)\prod_i L(y_i|\theta)$, thus implicitly assuming equal reliability across fundamentally different measurement types. This assumption proves problematic when integrating physics-based forward models with data-driven model outputs. For instance, a seismic measurement in poorly-coupled marine sediments receives the same inferential weight as a direct crater diameter measurement, despite the order-of-magnitude difference in model uncertainty. Similarly, vision-language model (VLM) damage classifications \citep{krofcheck2025} that inherently embed perceptual and semantic ambiguities contribute equally to AI-derived overpressure estimates as traditional physical measurements with well-characterized error models. When one modality exhibits systematic bias or model misspecification, this conventional approach lacks mechanisms to down-weight its contribution, allowing corrupted observations to degrade the overall inference. The challenge intensifies when incorporating machine learning interpretations into physics-based inference frameworks, raising fundamental questions regarding how to quantify the reliability of qualitative assessments relative to traditional measurements, how to appropriately weight contributions from disparate data sources, and how to prevent systematic biases in one modality from corrupting overall inference. This motivates our development of learned, data-driven trust weights that automatically calibrate relative modality contributions.

To address this heterogeneity in reliability across modalities, we develop a generalized fractional Bayesian framework in which each data stream contributes a tempered likelihood factor with a learned exponent. Generalized Bayes provides robustness to misspecification through a principled loss-based updating rule \citep{bissiri2016general}, with established schemes for choosing temperature parameters to stabilize inference \citep{grunwald2017inconsistency,holmes2017assigning}. We extend this framework to the multimodal setting by placing a Dirichlet prior on modality-specific exponents constrained to the simplex, enabling the data itself to learn relative trust weights while preserving full uncertainty propagation. Rather than assuming equal reliability or manually specifying weights, our approach allows the observations  to determine their relative contributions through data-driven calibration that naturally down-weights modalities exhibiting model-data discrepancies whilst amplifying those providing consistent, informative constraints.

\subsection{The 2020 Beirut Explosion}
The August 4, 2020 explosion at the Port of Beirut provides an ideal test case for multimodal yield estimation. Approximately 2750 tonnes of ammonium nitrate detonated following a warehouse fire, producing 220 fatalities, over 7,000 injuries, and an estimated 15 billion dollars in economic damage \citep{pasman2020beirut,sadek2022impacts}. The event generated diverse observational data across multiple sensing modalities: regional seismic networks recorded moment magnitude between 3.3 and 3.9, high-resolution satellite imagery documented a 140-meter diameter crater and extensive structural damage extending several kilometers from the detonation point \citep{pilger2021beirut}, synthetic aperture radar (SAR) interferometry captured coherence loss patterns indicating damage severity, and thousands of geotagged photographs provide ground-level documentation of structural damage.

Previous studies have estimated yields ranging from 0.2 to 1.4 kilotons TNT equivalent using various single-modality approaches \citep{aouad2021beirut,diaz2021explosion,goldstein2020beirut,goldstein2021reconciling,kim2022yield,pilger2021beirut,rigby2020preliminary,stennett2020estimate}. Table \ref{tab:previous_estimates} summarizes these estimates and demonstrates substantial variation across methodologies, highlighting the need for integrated multimodal approaches that utilize principled statistical data fusion.

\begin{table}[h]
\centering
\caption{Summary of previous Beirut explosion yield estimates from literature}
\label{tab:previous_estimates}
\begin{tabular}{lcc}
\toprule
Study & Method & Yield (kt TNT) \\
\midrule
\cite{rigby2020preliminary} & Video footage & 0.5--1.12 \\
\cite{aouad2021beirut} & Fireball evolution & 0.5--1.1 \\
\cite{diaz2021explosion} & Image analysis & 0.3--0.5 \\
\cite{stennett2020estimate} & Multiple methods & 0.5--1.1 \\
\cite{goldstein2020beirut,goldstein2021reconciling} & Mushroom cloud & 0.2--0.4 \\
\cite{pilger2021beirut} & Open source data & 0.5--1.1 \\
\cite{kim2022yield} & Infrasound & 0.4--1.4 \\
\bottomrule
\end{tabular}
\end{table}

% The Beirut blast offers several advantages for methodological development. The known quantity of ammonium nitrate provides an upper bound on possible yield, the surface burst configuration may simplify modeling relative to buried or airburst scenarios, the urban setting provides extensive damage documentation unavailable for remote test sites, and the open-source nature of much observational data enables reproducible research and methodology validation.

In this paper, we make three novel primary contributions to explosive yield estimation methodology. First, we develop a Bayesian fractional posterior approach with Dirichlet-distributed trust weights that enables principled fusion of heterogeneous data while quantifying the relative information content of each modality. Second, we demonstrate how vision-language model damage assessments can be incorporated into physics-based inference through appropriate forward models and uncertainty propagation, converting qualitative damage classifications into probability distributions over overpressure ranges whilst maintaining full uncertainty quantification. Third, we provide the first unified yield estimate for the Beirut explosion that combines seismic observations, crater measurements, SAR damage patterns, and ground-level image damage assessments.

% The methodology developed here extends beyond explosion analysis to any inverse problem requiring integration of traditional measurements with unstructured data. Applications include climate model validation using satellite imagery, structural health monitoring combining sensor networks with visual inspection, and materials characterization fusing multiple diagnostic modalities with varying information content.

This paper is organized as follows. In Section~\ref{sec:data-sources}, we detail the data sources and collection processes. Section~\ref{sec:theoretical_framework} presents the hierarchical Bayesian fusion framework. In Section~\ref{sec:forward_models}, we describe the individual forward models and data utilized for each modality. Section~\ref{sec:results} reports empirical yield estimates for each modality separately and the results of the Bayesian data fusion. Finally, Section~\ref{sec:discussion} discusses implications and extensions of our methodology and describes future work.
% !TEX root = main.tex
\section{Data Sources}
\label{sec:data-sources}
The Beirut explosion generated diverse observational data across multiple sensing modalities. This section describes the data sources employed in the yield estimation framework, with detailed preprocessing procedures provided in the Supplement. The availability of heterogeneous observations spanning traditional physical measurements and modern imagery enables development of robust multimodal fusion methodologies.

\subsection{Seismic Waveform Data}
\label{sec:data-seismic}

\begin{figure}[ht!]
    \centering
    \includegraphics[trim={5cm 0 0 0},clip,width=0.5\linewidth]{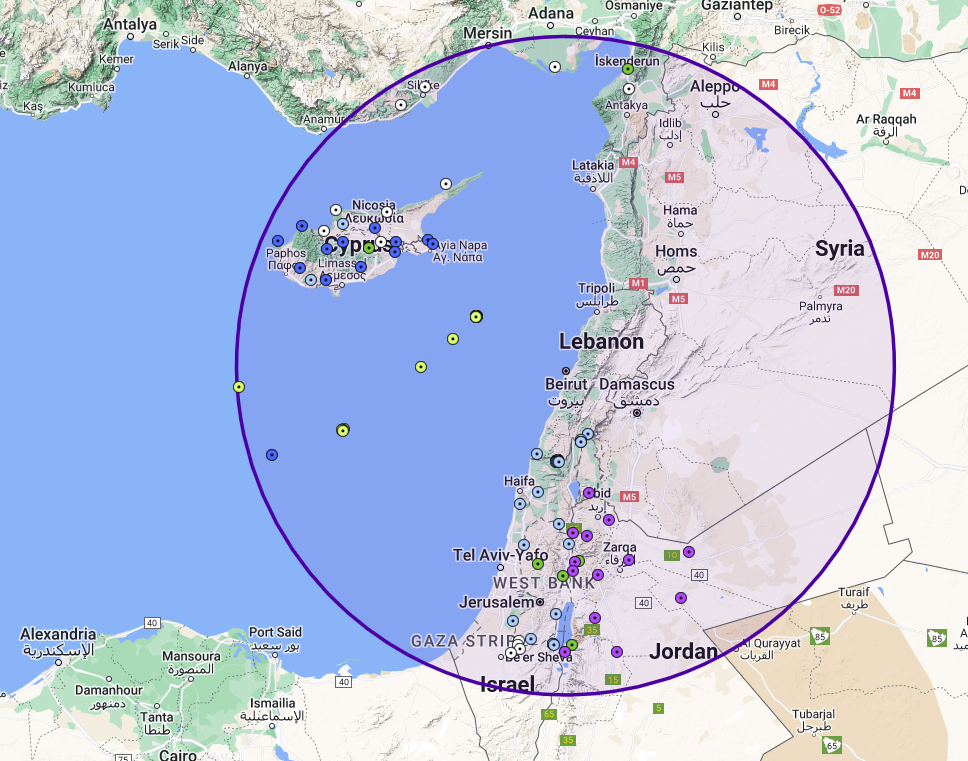}
    \vspace{2mm}
    \includegraphics[width=\linewidth]{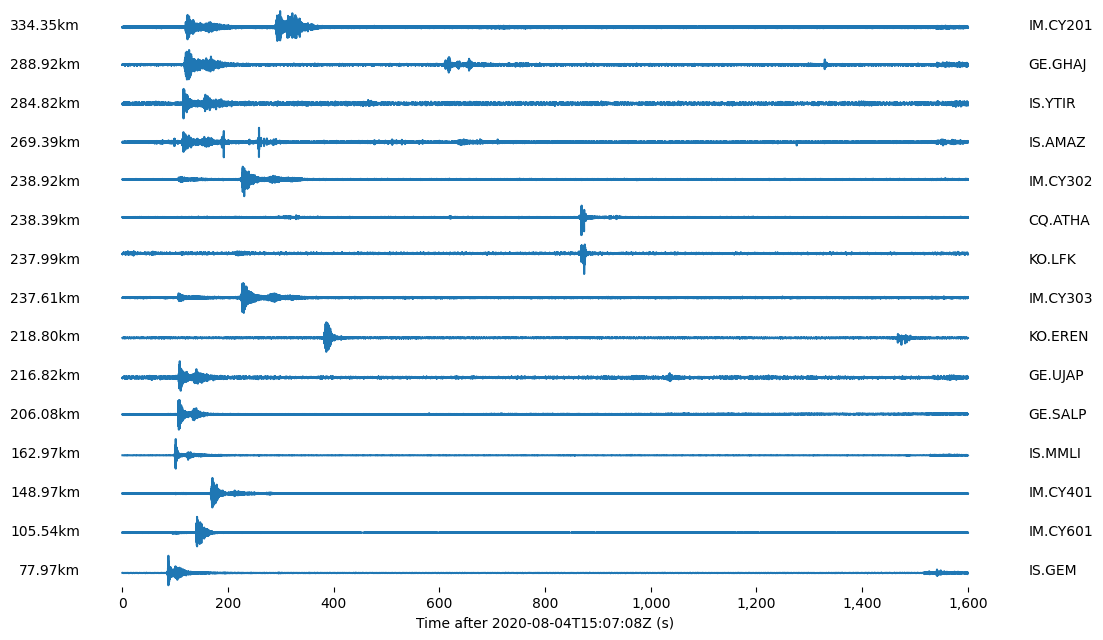}
    \caption{Seismic data from regional stations. Top: Station locations color-coded by distance from epicenter, providing good azimuthal coverage. Bottom: Waveform data sorted by distance, showing clear P-wave and S-wave arrivals with amplitude decay consistent with geometric spreading.}
    \label{fig:sage}
\end{figure}

We obtained regional seismic waveform data from NSF's Seismological Facility for the Advancement of Geoscience (SAGE). Of 93 stations within a three-degree radius of Beirut, we successfully retrieved one hour of data (15:00--16:00 UTC, August 4, 2020) from 43 stations at epicentral distances of 100--500 km. These stations include all 24 used by the analysis in \cite{pilger2021beirut}, providing comprehensive azimuthal coverage (Figure~\ref{fig:sage} a). The Python ObsPy library~\citep{obspy} was used to extract P-wave and S-wave arrival times, with clear body wave arrivals visible across the regional network (Figure~\ref{fig:sage} b). The primary challenge for seismic yield estimation is coupling efficiency in Beirut's near-shore environment, where reclaimed land and marine sediments exhibit substantially different mechanical properties than competent bedrock typical of underground nuclear test sites.

% \begin{figure}[ht!]
%     \centering
%     \includegraphics[trim={5cm 0 0 0},clip,width=0.38\linewidth]{seismic-station-map.png}
%     \hfill
%     \includegraphics[width=0.60\linewidth]{seismic_plot.png}
%     \caption{Seismic data from regional stations. (a) Station locations color-coded by distance from epicenter, providing good azimuthal coverage. (b) Waveform data sorted by distance, showing clear P-wave and S-wave arrivals with amplitude decay consistent with geometric spreading.}
%     \label{fig:sage}
% \end{figure}

\subsection{Crater Dimensions from Satellite Imagery}
\label{sec:data-crater}
\begin{figure}[ht!]
    \centering
    \includegraphics[trim={0 0 38cm 0},clip,width=0.8\linewidth]{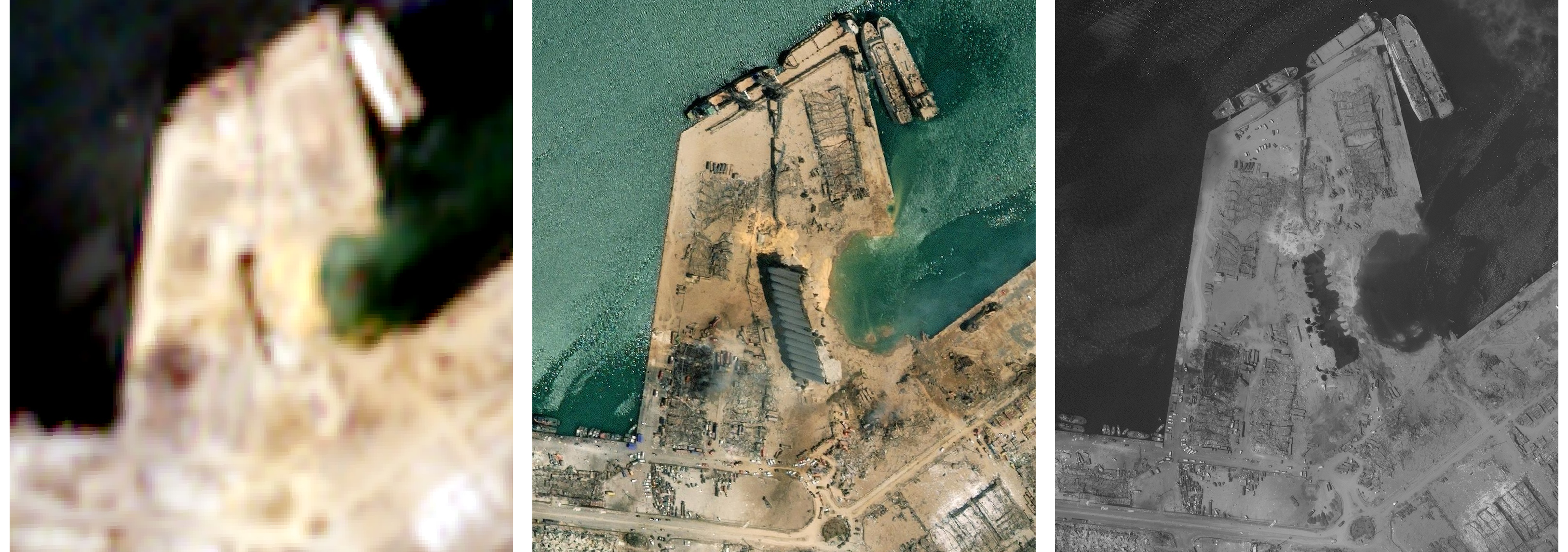}
    \caption{Resolution comparison for satellite imagery. Individual vehicles are clearly resolved in 1-meter commercial imagery (right), while Sentinel-2 at 4-meter resolution (left) provides adequate large-scale coverage.}
    \label{fig:satellite_visible}
\end{figure}

High-resolution satellite imagery captured an elliptical crater at the detonation site, providing direct evidence of explosive energy deposition. We obtained visible-spectrum imagery from two sources: Maxar Technologies (1 m resolution)~\cite{maxar-technologies-opendata} and ESA Sentinel-2 (4 m multispectral). Table~\ref{tab:satellite_sources} summarizes acquisition dates and characteristics. The commercial 1-meter imagery (Figure~\ref{fig:satellite_visible}) clearly resolves individual vehicles and structural features, enabling precise crater boundary delineation. 

\subsection{SAR Coherence Change Detection}
\label{sec:data-sar}
SAR \citep{enwiki:1319276477} coherence analysis quantifies structural damage by measuring decorrelation in repeat-pass acquisitions. We processed four Sentinel-1 acquisition pairs (ascending and descending orbits, July 30--August 5, 2020) to compute interferometric coherence. Coherence ranges from 1 (no change) to 0 (complete decorrelation), with loss exceeding 0.2 reliably indicating damage~\citep{sudhaus2021damage}. 
% Binary damage maps were created through coherence thresholding, then despeckled using SpikeAD adaptive filtering~\cite{verzi2017optimization}. Damage percentages were aggregated into 100 m $\times$ 100 m boxes, with zonal percentile thresholding (95th percentile within 15 concentric annuli) selecting ~150--200 high-damage locations experiencing near-direct blast exposure. This procedure addresses spatial heterogeneity from shielding, reflection, and structural variability not fully captured by simplified blast models. The primary challenge is the indirect relationship between coherence loss and structural damage, requiring learned vulnerability curves to connect radar observations to overpressure.

\subsection{Ground-Level Imagery}
\label{sec:data-vlm}

\begin{figure}[ht!]
    \centering
    \includegraphics[width=0.48\linewidth]{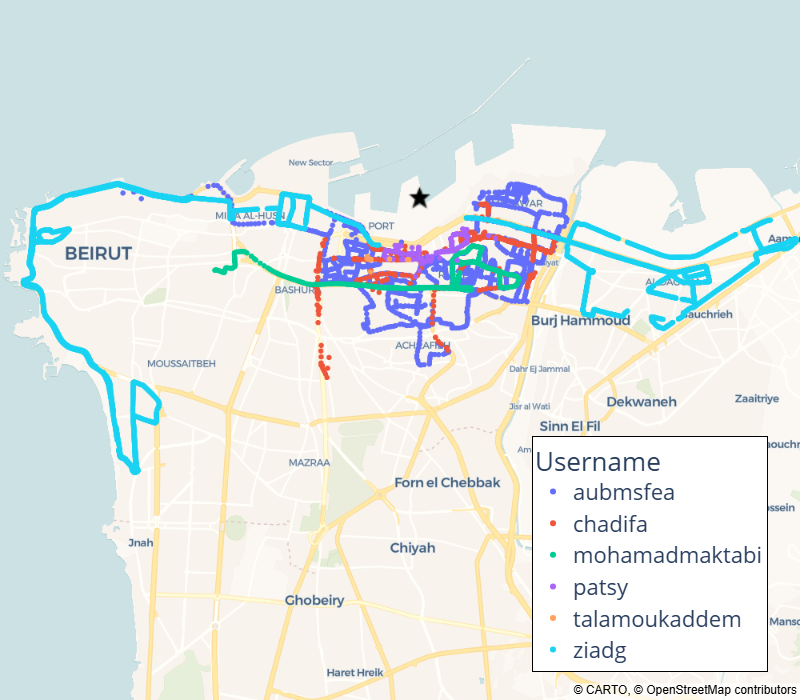}
    \hfill
    \includegraphics[width=0.48\linewidth]{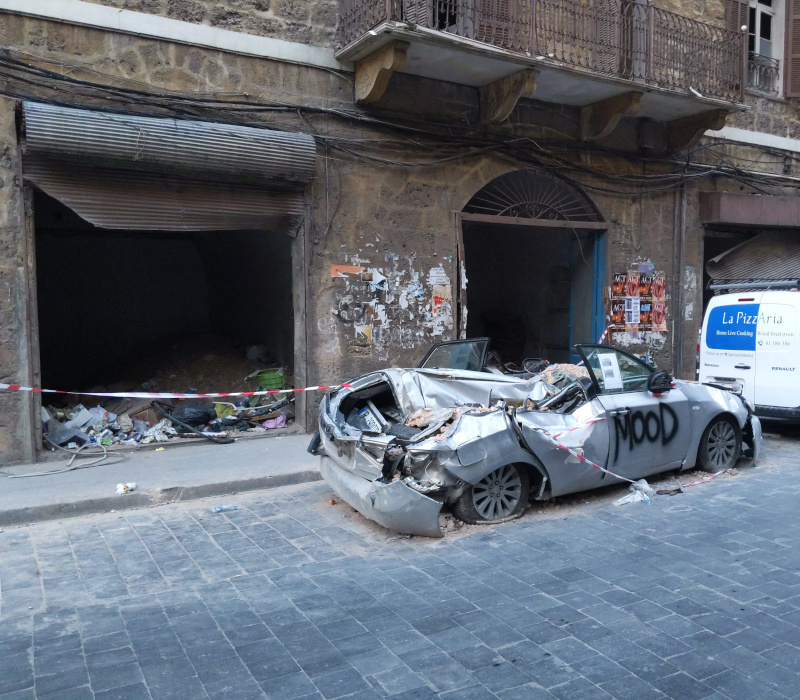}
    \caption{Ground-level imagery from Mapillary. (a) Geographic distribution showing photosphere concentration in damaged areas near the port. (b) An example of a ground-level photo with multiple damage indicators (damaged vehicle, broken roll-up garage door, piles of debris, and safety barrier markers). Image from Mapillary.com.}
    \label{fig:mapillary_locations}
\end{figure}

Researchers at the American University of Beirut collected 360° photosphere\footnote{A photosphere image is a picture that captures views in all directions; see ~\citep{vidanapathirana2019cognitiveanalysis360degree}.} imagery from Beirut's streets a few months after the explosion~\cite{prj3030}. Anonymized imagery and metadata was subsequently uploaded to the street-image hosting website Mapillary~\citep{mapillary}, a community website that allows users to store and view planar, 360° panoramic, and 360° photosphere images that are tagged with geotemporal information. We used Mapillary's free API to retrieve 685 photosphere images (238 from August 8--23, 447 from October 1--15) and 4,259 planar images with coordinates (Figure~\ref{fig:mapillary_locations}a). The photosphere images concentrate in heavily damaged areas and provide omnidirectional coverage including building facades and upper stories. We converted photosphere images to four planar views using oblique viewing angles (10° upward pitch, yaws at 45°, 135°, 225°, 315°) and a resolution of 1600$\times$1200 pixels, and then used vision-language models to identify signs of damage in the images. Visual damage indicators include broken windows with visible glass fragments, debris piles containing masonry and building materials, damaged vehicles, construction barriers marking hazardous structures, and crushed roll-up garage doors (Figure~\ref{fig:mapillary_locations}b). The primary challenges are temporal delays allowing cleanup efforts (before imaging could be done), spatial sampling gaps, and inferring quantitative overpressure from qualitative visual damage classifications. The Supplement details more of the preprocessing workflow and test suite creation.
    
The four modalities provide complementary information: seismic waveforms offer established yield-magnitude relationships but suffer from uncertain coupling efficiency (with the ground); crater dimensions provide direct physical evidence with low systematic uncertainty; SAR enables city-wide damage mapping through an indirect overpressure relationship; and ground-level imagery captures detailed damage states but requires converting qualitative assessments to quantitative constraints. The multimodal fusion framework we develop integrates these observations while appropriately accounting for their varying reliability and information content.
% !TEX root = main.tex
\section{Theoretical Framework} \label{sec:theoretical_framework}

The estimation of explosive yield from the Beirut explosion presents a classic inverse problem: inferring causal parameters from observed effects. Unlike traditional single-modality inversions, this analysis leverages multiple heterogeneous data sources that observe different physical manifestations of the same event. This section develops the mathematical framework for rigorously combining these disparate observations while accounting for their varying reliability and information content.

\subsection{The Multimodal Inverse Problem}
The fundamental task is to estimate the explosive yield $Y \in \mathbb{R}_{+}$ (in kilotons TNT equivalent) given observations $\mathcal{D} = \{\D_1, \D_2, \ldots, \D_M\}$ from $M$ different modalities, where $\D_i \in \mathbb{R}^{n_i}$ and the dimensions $n_i \in \mathbb{N}$ may differ across datasets. 
% Each modality observes the explosion through distinct physical processes: seismic waves propagate through the earth's crust and are recorded by regional sensor networks, the blast excavates a crater whose dimensions reflect the deposited energy, pressure waves damage structures in predictable patterns captured by satellite radar, and ground-level photographs document building damage states that vary with distance from the epicenter.

Each dataset $\D_i$ relates to the yield through a forward model that encodes the relevant physics:
$
\D_i = \mathcal{F}_i(Y; \boldsymbol{\theta}_i) + \boldsymbol{\epsilon}_i,
$
where $\mathcal{F}_i: \mathbb{R} \to \mathbb{R}^{n_i}$ represents the forward operator mapping yield to expected observations, $\boldsymbol{\theta}_i \in \Theta_i \subseteq \mathbb{R}^{p_i}$ are modality-specific model parameters (such as seismic coupling coefficients, crater scaling exponents, or building fragility curves), and $\boldsymbol{\epsilon}_i \in \mathbb{R}^{n_i}$ represents observation noise.
Here, $p_i$ denotes the dimensionality of the parameter space, where in general, $n_i \neq p_i$.

% The forward models can be categorized into two classes based on their physical directness. Seismic waves and crater dimensions provide direct observations of energy release, connected to yield through well-established empirical scaling laws derived from decades of weapons testing. In contrast, structural damage observed in synthetic aperture radar imagery and ground-level photographs requires intermediate modeling steps that first relate yield to overpressure through blast propagation models, then connect overpressure to observable damage states through structural vulnerability functions.
% The challenge of multimodal fusion arises because these forward models encode different systematic uncertainties and varying degrees of physical understanding. For instance, seismic coupling depends strongly on local geology and can vary by an order of magnitude for identical yields, particularly in complex near-shore environments where the Beirut explosion occurred. 
% Crater dimensions are influenced by substrate properties, moisture content, and pre-existing structures that are rarely characterized with sufficient detail. 
% On the other hand, damage assessments from imagery face complications from building heterogeneity, progressive collapse, cleanup efforts between the explosion and data acquisition, and the subjective nature of damage classification schemes. Any principled fusion framework must therefore account for these modality-specific uncertainties without allowing systematic biases in one observation type to corrupt the overall inference.

The challenge of multimodal fusion arises because these forward models encode different systematic uncertainties and varying degrees of physical understanding. For instance, seismic coupling can vary by an order of magnitude depending on local geology, crater scaling depends on substrate properties and geometry, and damage assessments require mapping qualitative visual evidence to quantitative overpressure through building vulnerability models. Any principled fusion framework must therefore account for these modality-specific uncertainties without allowing systematic biases in one observation type to corrupt the overall inference.

% In this realm, a modeling framework that leverages conditional independence is appropriate for this application given the fundamentally different observation geometries of our modalities. In particular, while SAR coherence detects changes in radar backscatter from roofs and upper structures from above, ground-level imagery studied by the VLM captures facade damage, window breakage, and street-level debris invisible to satellite radar. These complementary perspectives observe distinct physical manifestations of blast damage rather than the same underlying damage state, supporting a conditional independence assumption that may prove essential for tractable multimodal fusion.

\subsection{Limitations of Conventional Bayesian Fusion}
The standard Bayesian approach to multimodal inference assumes conditional independence and treats all observations equally: $
p(Y|\mathcal{D}) \propto \pi(Y) \prod_{i=1}^M L(\D_i|Y),$
where $\pi (Y)$ represents prior knowledge about plausible yield values and each likelihood function $L(\D_i|Y)$ quantifies how probable the observations from modality $i$ are given a hypothesized yield. This formulation provides a coherent framework for combining evidence, but it implicitly assumes that all modalities merit equal trust regardless of their information content or model fidelity.

Recent applications have demonstrated both the promise and limitations of this conventional approach. \cite{ford2021joint} successfully estimated yield using seismic, acoustic, and optical observations through standard Bayesian inference implemented in Stan, demonstrating computational feasibility and showing that multimodal fusion can reduce uncertainty relative to single-modality estimates. \cite{kim2022yield} achieved comparable results for the Beirut explosion specifically, combining seismoacoustic data with physics-based propagation models to constrain the yield. However, both studies employ the conventional posterior form shown above, which lacks mechanisms to adapt the relative contribution of each modality based on model-data consistency.

% This equal-weighting assumption proves problematic when integrating observations with vastly different reliability. Consider a seismic moment magnitude measurement recorded in poorly-coupled marine sediments that systematically underestimates yield by a factor of three, combined with a crater diameter measurement that directly reflects energy deposition with well-understood scaling relationships. The conventional posterior weights both observations equally, allowing the biased seismic estimate to pull the inferred yield toward an incorrect value despite the crater providing stronger physical evidence. Similarly, when incorporating vision-language model damage classifications that inherently embed perceptual ambiguities and semantic uncertainties, the conventional approach grants these qualitative assessments the same inferential weight as precise physical measurements with well-characterized error models. When one modality exhibits undetected systematic bias or model misspecification, the framework has no principled mechanism to reduce its influence, resulting in degraded estimates and overconfident uncertainty quantification.

Existing general data fusion methodologies offer alternative approaches to combining heterogeneous observations, yet each carries limitations when applied to problems requiring integration of physics-based forward models with data-driven outputs. Bayesian model averaging provides principled uncertainty quantification by computing posterior probabilities over a discrete set of models and averaging predictions weighted by these probabilities \citep{hoeting1999bayesian,fragoso2018bayesian}. However, this approach assumes models are exchangeable and typically requires manual specification of prior model probabilities, which may not be reliable for novel data modalities like vision-language model interpretations where validation datasets are limited. The framework also struggles when the set of candidate models does not include the true data-generating mechanism, as is the case when simplified physics-based forward models are employed.

% Weighted likelihood methods generalize classical fusion through adaptive exponents, forming combined likelihoods as $L(\theta; y_1,\ldots,y_M) = \prod_i L_i(y_i;\theta)^{\alpha_i}$ where the weights $\alpha_i$ reflect data source relevance \citep{wang2012weighted}. While this formulation explicitly acknowledges varying information content across sources, weight selection remains non-trivial. 
% Cross-validation strategies require substantial validation data that are often unavailable for rare events like industrial explosions, while information-theoretic approaches based on Kullback-Leibler divergence assume well-specified models that may not hold when integrating simplified physics with AI-driven outputs. 

Covariance intersection provides optimal conservative fusion when cross-correlations between estimates are unknown, guaranteeing that fused covariance bounds encompass all possible correlation structures \citep{julier2007using}. However, this conservatism significantly degrades performance when estimates are highly correlated, as occurs when multiple modalities observe the same underlying physical quantity through different measurement processes. The method also requires Gaussian assumptions that may be inappropriate for damage observations or other ordinal data.
To deal with this, Dempster-Shafer theory explicitly utilizes belief assignments over hypothesis power sets, enabling fusion of uncertain evidence without requiring precise probability distributions \citep{shafer1976mathematical}, with recent applications to disaster management demonstrating capabilities for handling conflicting information \citep{fan2024dempster}. Nonetheless, the exponential computational complexity in the number of hypotheses limits scalability, while the framework lacks natural mechanisms for incorporating continuous-valued physics-based forward models.

Additionally, meta-learning approaches enable rapid adaptation to varying sensor configurations by learning optimal fusion strategies from related tasks \citep{finn2017model}, showing promise when sufficient task diversity exists for generalization, which may be hindered for yield estimation due to data sparsity that limits the availability of similar training examples.
Nevertheless, all of these methodological limitations share the difficulty in appropriately weighting heterogeneous data sources with fundamentally different noise characteristics, model uncertainties, and information content. To the best of our knowledge, no existing framework systematically addresses this heterogeneity through data-driven reliability quantification while maintaining rigorous uncertainty propagation.

\subsection{Fractional Posteriors for Robust Inference}
Bayesian fractional posterior methods offer a principled solution to data source heterogeneity through the tempering of likelihood contributions \citep{wang2012weighted}. Rather than treating all observations equally, the fractional posterior down-weights likelihoods that exhibit poor model-data agreement while amplifying those that provide consistent constraints. The fractional posterior takes the form
$
p(\theta|x) \propto p(\theta) L(x|\theta)^{\gamma},
$
where $x$ denotes the datum, $\theta$ the parameter of interest, and $0 < \gamma \leq 1$ controls the relative influence of the likelihood versus the prior \citep{bissiri2016general}. When $\gamma = 1$, this reduces to the standard Bayesian posterior, while $\gamma < 1$ reduces the likelihood's influence, making inference more robust to model misspecification.
This modification provides robust inference when the assumed model deviates from the true data-generating process. Moreover, \cite{bhattacharya2019bayesian} rigorously established that fractional posteriors achieve posterior contraction rates even under misspecification, proving that the posterior concentrates around the parameter value that minimizes a specific loss function rather than diverging or producing nonsensical inferences. 

An important consideration with this framework is temperature parameter selection, which has evolved from ad-hoc choices to principled data-driven approaches through SafeBayes \citep{grunwald2017inconsistency} and information-matching \citep{holmes2017assigning}, with recent work  demonstrating that appropriate tempering does not impact posterior predictions in moderate-to-large samples \citep{mclatchie2023robust}. However, these developments address single data sources with one global temperature parameter.

While weighted likelihood methods \citep{wang2012weighted} employ modality-specific exponents, these weights are typically fixed a priori or selected through cross-validation requiring substantial validation data. The extension of fractional posterior theory to learn modality-specific exponents through hierarchical Bayesian priors has yet to be explored for multimodal data fusion. 
This distinction proves critical for explosive yield estimation: when combining seismic measurements with crater dimensions and damage assessments, different modalities warrant different degrees of tempering based on their model fidelity and systematic uncertainties. Our approach enables the data itself to determine these weights while maintaining the theoretical robustness guarantees of fractional posteriors under misspecification.

\subsection{Learning Trust Weights Through Hierarchical Modeling}
\sloppy
To this end, we formulate multimodal yield estimation via modality-specific exponents that leverages conditional independence between modalities: 
$
p(Y|\mathcal{D}, \boldsymbol{\gamma}) \propto \pi(Y) \prod_{i=1}^M L_i(\D_i|Y)^{\gamma_i},$
where $L_i(\D_i|Y)$ denotes the likelihood for modality $i$ and trust weights $\boldsymbol{\gamma} = (\gamma_1, \ldots, \gamma_M) \in (0,1]^M$ quantify the relative information content of each modality.
The product form assumes conditional independence between modalities given yield, an appropriate assumption since the modalities that we consider observe distinct physical manifestations of damage from different geometries rather than the same underlying damage state.
% Rather than fixing these weights manually, we place a Dirichlet prior on the simplex:
% \begin{equation}
% \boldsymbol{\gamma} \sim \text{Dirichlet}(\boldsymbol{\alpha}),
% \end{equation}
% yielding the joint posterior:
% \begin{equation}
% p(Y, \boldsymbol{\gamma}|\mathcal{D}) \propto p(Y) p(\boldsymbol{\gamma}) \prod_{i=1}^M L_i(\D_i|Y)^{\gamma_i}.
% \end{equation}
This hierarchical specification enables data-driven calibration: modalities providing consistent constraints on yield receive higher posterior weights, while those exhibiting model-data discrepancies are naturally down-weighted. The modeling of $\bgamma$ using Dirichlet priors will be discussed in Section~\ref{sec:inference-procedure}.
% The concentration parameters $\boldsymbol{\gamma}$ control prior assumptions about weight distributions, with $\gamma_i < 1$ favoring sparse weightings, $\alpha_i > 1$ encouraging uniformity, and $\alpha_i = 1$ allowing the data to determine the structure. 
% This flexibility distinguishes the approach from methods requiring external weight specification, as the trust weights emerge as natural parameters of the generalized posterior distribution.
% !TEX root = main.tex
\section{Forward Models} 
\label{sec:forward_models}
The success of our multimodal fusion framework depends critically on appropriate construction of forward models $\mathcal{F}_i$ and likelihood functions $\mathcal{L}_i$ for each modality. These components encode both our physical understanding of how yield manifests in observable quantities and the uncertainties inherent in this mapping. This section details the forward model construction, prior specifications, and likelihood formulations for four data modalities: seismic waves, crater dimensions, SAR damage assessment, and vision-language model interpretation of ground-level imagery.

\subsection{Overview and Common Elements}
The forward models can be categorized into two classes based on their physical directness. Direct measurements such as seismic moment magnitude $M_w$ and crater diameter $D$, relate to yield through empirical scaling laws established from decades of physical testing. Damage proxies such as the structural damage changes observed in SAR imagery and ground-level photographs, require intermediate modeling that first relates yield to blast overpressure $P$ through blast--propagation models, then connects overpressure to damage states through vulnerability functions.

All modalities share a common prior on yield $
\log_{10} Y \sim \text{TruncatedNormal}(\mu=0.0, \sigma=0.1, \text{upper}=\log_{10} 2.75),$
where $Y$ is expressed in kilotons TNT equivalent. This prior centers on 1 kt with approximately one order of magnitude uncertainty, truncated at 2.75 kt to reflect the theoretical maximum from complete detonation of the stored ammonium nitrate with TNT-like efficiency and energetic yield. 
% The logarithmic formulation naturally enforces positivity while enabling efficient sampling across multiple orders of magnitude.

\subsubsection{Blast Overpressure Model: Kingery-Bulmash}
\label{sec:kb_model}
Both damage-based modalities (SAR and VLM) require conversion from yield to blast overpressure. Here, we employ the Kingery-Bulmash (KB) correlations \citep{kingery1984airblast} as refined by \cite{swisdak1994simplified} for hemispherical surface bursts. The incident (side-on) peak overpressure $P_{\text{inc}}$ in psi depends on scaled distance $Z = r/W^{1/3}$, where $r$ is range in meters and $W$ is TNT-equivalent mass in kilograms ($W = Y \times 10^6$ kg for $Y$ in kt).

Converting units, we have $Z_{\text{en}} = Z_{\text{SI}} \times 3.28084 / (2.20462)^{1/3} \approx 2.521 Z_{\text{SI}}$. Here, the subscripts ``en'' and ``SI'' stand for English and SI units. The overpressure thereby follows piecewise polynomial expressions in log-space:
$
\log_{10} P_{\text{inc}} = A + B \log Z_{\text{en}} + C (\log Z_{\text{en}})^2 + D (\log Z_{\text{en}})^3 + E (\log Z_{\text{en}})^4,$
with coefficients specified for three scaled distance regimes (Table~\ref{tab:kb_coefficients}.
This piecewise structure captures distinct physical regimes: strong shock near the detonation point, Mach stem formation at intermediate ranges, and weak shock propagation at large distances. The model provides overpressure in psi, which is, in the following, converted to kiloPascals via $P_{\text{kPa}} = P_{\text{psi}} \times 6.89476$.

\subsection{Seismic Waves}
\label{sec:seismic}
Explosive yield can be inferred from radiated seismic waves through two related source-strength measures. The seismic moment $M_0$ (N$\cdot$m or dyne$\cdot$cm) is a physically grounded scalar quantifying the strength of seismic radiation. For tectonic earthquakes, $M_0 = \mu A D$ where $\mu$ is shear modulus, $A$ is fault rupture area, and $D$ is average slip. Here, the moment directly measures the mechanical work performed during shear failure. For explosions, the source mechanism is approximately isotropic rather than shear, and the effective $M_0$ scales with the pressure-volume work performed on the surrounding medium during cavity expansion. 
On the other hand, the moment magnitude $M_w$ provides a logarithmic, energy-consistent measure designed to avoid the saturation problems that plague body-wave magnitudes at large sizes:
$M_w = \frac{2}{3}\log_{10}(M_0[\text{dyne}-\text{cm}]) - 10.7. $
This relationship establishes $M_w$ as an approximately linear function of log-energy across all source sizes. In practice, seismic catalogs report $M_w$ estimated through spectral inversion of regional waveforms, which we employ as the observable in our inference framework.

For surface and near-surface chemical explosions, only a fraction of the explosive energy couples into seismic radiation since the coupling efficiency depends on local geology, water saturation, and confinement geometry. The same explosive yield can therefore produce substantially different moment magnitudes depending on these factors \citep{khalturin1998review}. The Beirut surface burst in a near-shore environment presents particular challenges, as unconsolidated marine sediments exhibit significantly poorer coupling relative to material typical of underground nuclear test sites, which most existing data cover \citep{pasyanos2022full}. This geological variability thus motivates a likelihood formulation with learned variance parameter rather than assuming deterministic yield-magnitude relationships.

In order to define a suitable forward model, we employ a two-stage approach: first, we extract raw seismic waveform features and predict $M_w$ using a 5-member neural network ensemble trained on the dataset of \cite{pasyanos2022full} of mixed surface-level chemical and U.S. underground nuclear tests, then we subsequently relate $M_w$ to yield through empirical regression.

The neural network ensemble processes regional seismic waveforms (measured at 100--500 km epicentral distance) to predict $\log_{10} M_0$ directly through five independently trained models with identical architectures but different random initializations. The ensemble prediction is computed as the equally-weighted average:
$
\log_{10} M_0^{\text{ensemble}} = \frac{1}{5} \sum_{k=1}^{5} \log_{10} M_0^{(k)},$
where $\log_{10} M_0^{(k)}$ is the prediction from the $k$-th ensemble member. This ensemble approach reduces prediction variance by approximately 40\% compared to single-model predictions.
Full architectural and training details are provided in the Supplement.

Cross-validation using group 10-fold (ensuring events do not leak between train/test sets) demonstrates that individual ensemble members achieve a mean absolute error of approximately 0.3 orders of magnitude in $M_0$ prediction, with the ensemble mean further reducing this uncertainty. Figure~\ref{fig:seismic-nn-performance} shows the training history and predictions from a representative ensemble member on a log scale, demonstrating good convergence and alignment between predicted and true values across multiple orders of magnitude spanning the full range of yields in the training dataset. The variance visible in lower-$M_0$ predictions reflects increased uncertainty for surface chemical explosions compared to buried nuclear tests that dominate the training data.
The yield-magnitude relationship subsequently follows an affine transformation on the log scale:
$\log Y = \alpha + \beta M_w,$
with regression coefficients $\alpha = -14.587$ and $\beta = 3.004$ calibrated from surface-level blasts in 
%\citet{khalturin1998calibration}.
\cite{khalturin1998review}.

For the observed moment magnitude $M_w^{\text{obs}}$, the expected value given yield $Y$ follows from inversion: $
M_w^{\text{pred}}(Y) = \frac{\log Y - \alpha}{\beta}.$
We posit a likelihood that assumes Gaussian-distributed residuals:$
\mathcal{L}_m(M_w^{\text{obs}} | Y, \sigma_m) = \mathcal{N}(M_w^{\text{obs}} | M_w^{\text{pred}}(Y), \sigma_m^2)$
where the uncertainty parameter $\sigma_m$ is learned rather than fixed, with prior:
$\sigma_m \sim \text{TruncatedNormal}(\mu=0.13, \sigma=0.01, \text{lower}=0.05, \text{upper}=0.30).$
This prior centers on the empirical variance in $M_w$ from the ensemble models, while allowing data-driven refinement to account for differences between the Beirut surface explosion and the underground nuclear tests used for training.

Applied to Beirut, the neural network processes waveforms from regional stations spanning 100--500 km epicentral distance, providing good azimuthal coverage  (Figure~\ref{fig:beirut-seismic-stations}), yielding $M_w = 4.50 \pm 0.13$. 

\subsection{Crater Geometry}
\label{sec:crater}
Explosive excavation creates craters whose dimensions scale with yield through dimensional analysis. For gravity-dominated cratering in competent materials, the fundamental relationship relates the crater diameter $D$ to yield, $D \propto Y^{1/3}$, reflecting the volumetric nature of energy deposition \citep{nordyke1961nuclear,schmidt1987gravity}. 
% This scaling holds across six orders of magnitude in yield, from conventional munitions to megaton nuclear devices.

In logarithmic space, the scaling relationship becomes linear: 
$\log_{10} D = (\log_{10} Y + 6.0)/3.0,$
where $D$ is crater diameter in meters and $Y$ is yield in kilotons \citep{cooper1996explosives}. The constant 6.0 reflects unit conversions and average geological conditions for surface bursts.

The Beirut crater exhibits elliptical geometry, reflecting either directional blast effects or substrate anisotropy. Rather than modeling these directional effects explicitly, we treat width and length as independent observations of the idealized circular diameter, acknowledging that both dimensions carry yield information. 

The likelihood treats log-transformed dimensions of crater width $w$ and length $l$, as independent Gaussian observations:
$\mathcal{L}_c(w, \ell | Y, \sigma_c) = \prod_{d \in \{w,\ell\}} \mathcal{N}(\log_{10} d | \mu_c(Y), \sigma_c^2),$
where $\mu_c(Y) = (\log_{10} Y + 6.0)/3.0$. The uncertainty parameter accounts for both measurement error (satellite resolution, boundary identification) and geological variability:
$
\sigma_c \sim \text{TruncatedNormal}(\mu=0.08, \sigma=0.02, \text{lower}=0.02, \text{upper}=0.15),$
centering around 0.08 dex (approximately 20\% in linear space), consistent with crater scaling studies \citep{sadek2022impacts}. 

Crater dimensions were extracted from Sentinel-2 imagery acquired August 8, 2020, using image segmentation and ellipse fitting techniques; more details are available in the Supplement. An ellipse fitted to the contour yields major axis 108.1 m and minor axis 46.7 m, giving equivalent circular diameter $D_{\text{equiv}} = \sqrt{108.1 \times 46.7} \approx 71$ m. Rather than performing separate analyses for the major and minor axes as sensitivity bounds, our Bayesian framework treats both dimensions as independent observations in the likelihood, thereby naturally incorporating the elliptical geometry and directional effects into the yield uncertainty quantification. This measurement closely aligns with independent estimates of 120 m for the major axis \citep{prj3030}. Further details are in Section ~\ref{sec:app-crater-geom}.

\subsection{Synthetic Aperture Radar Damage Assessment}
\label{sec:sar}
Interferometric SAR coherence analysis quantifies structural damage by comparing radar backscatter before and after an event. The coherence between acquisitions from identical viewing geometries ranges from 1 (no change) to 0 (complete decorrelation), with loss exceeding 0.2 reliably indicating damage \citep{sudhaus2021damage}. We aggregate coherence measurements into $10 \times 10$ pixel boxes (approximately $100$ m $\times 100$ m ground footprint) and define the observable $D_{\%,i} \in [0,100]$ as the percentage of pixels flagged as damaged within box $i$ after despeckling and quality filtering.

The forward model $\mathcal{F}_{\text{SAR}}(Y; \boldsymbol{\theta}_{\text{SAR}})$ relates yield to damage percentage through blast overpressure, employing a learned damage-pressure vulnerability curve to accommodate scene-specific structural heterogeneity.

For each box at distance $r_i$ from the epicenter, we compute incident overpressure using the Kingery-Bulmash correlations (Section~\ref{sec:kb_model}):
$
P_{r,i}(Y) = \mathcal{P}_{\text{KB}}^{\text{inc}}(r_i, Y), P_{\text{kPa},i}(Y) = 6.89476 \, P_{r,i}(Y).$
Instead of fixing the damage-pressure relationship through engineering models \citep{pilger2021beirut} that may not capture site-specific construction practices, we learn a two-parameter logistic curve in log-pressure space:
$
\mu_i(Y) = 100 \, \sigma\big(K \, [\log_{10} P_{\text{kPa},i}(Y) - \log_{10} P_{50}]\big), \sigma(u) = (1+e^{-u})^{-1}, $where $P_{50}$ (kPa) represents the midpoint overpressure producing 50\% expected damage, and $K > 0$ controls the transition gradient. This formulation thus naturally captures the transition from low to high damage while absorbing scene heterogeneity including variations in building age, construction quality, and structural systems into the learned parameters.
% To emphasize boxes experiencing near-direct blast exposure while maintaining radial coverage, we partition the damage map into 15 concentric annuli and retain only high-percentile boxes (95th by default) within each annulus. This procedure addresses a critical challenge: tall structures in direct line-of-sight experience maximum damage, surrounded by areas of lower damage from shielding, reflection, and refraction effects not captured by the idealized KB model. By selecting the 95th percentile, we focus on locations where the simplified physics model is most applicable.

SAR observations exhibit heavy-tailed variability from multiple sources: mixed construction within aggregation boxes (the $10 \times 10$ pixel boxes with a ground footprint of about 100 meter square), shielding and occlusion effects, registration errors between acquisition pairs, and despeckling artifacts. A Gaussian error model may therefore prove overly sensitive to inevitably large residuals, allowing a few outlier boxes to dominate the likelihood. To deal with this, we instead adopt a Student-$t$ observation model on the logit scale, which naturally accommodates outliers through heavier tails whilst leaving the core fit unchanged when data follow near-Gaussian behavior.

To do so, we define the logit-transformed observations and predictions $
y_i \equiv \frac{D_{\%,i}}{100} \in (0,1), z_i \equiv \text{logit}(y_i) = \log (y_i/1-y_i), z_{\mu,i}(Y) \equiv \text{logit}\mu_i(Y)/100.$    
Further, conditional on yield and curve parameters, we assume:
$z_i \mid Y, P_{50}, K, \sigma_{\text{SAR}}, \nu \sim \text{Student-}t\left(\nu, \mu=z_{\mu,i}(Y), \sigma=\sigma_{\text{SAR}}/100\right), \quad \nu > 2,$
where $\nu$ controls the tail heaviness (smaller $\nu$ produces heavier tails; $\nu \to \infty$ recovers Gaussian). 
% Here, the probability density for the Student-$t$ distribution is written in its location-scale form
% $f(t) = \frac{\Gamma \left( \frac{\nu+1}{2}\right)}{\sqrt{\pi \nu} \Gamma \left( \frac{\nu}{2}\right)} \left( 1 + \frac{t^2}{2}\right)^{- (\nu+1)/2}, \hspace {2mm} \mbox{where} \hspace{2mm} 
%     t = \frac{z_i - \mu}{\sigma}.$

The degrees of freedom parameter is learned from data rather than being fixed, allowing the model to adapt to the actual outlier frequency.
The average log-likelihood across retained boxes prevents sample-size dominance in multimodal fusion:
$\ell_{\text{SAR}}(Y, P_{50}, K, \sigma_{\text{SAR}}, \nu) = N_{\text{SAR}}^{-1} \sum_{i=1}^{N_{\text{SAR}}} \log \text{Student-}t\left(z_i \mid \nu, z_{\mu,i}(Y),\sigma_{\text{SAR}}/100\right).$

This formulation incorporates the learned vulnerability curve parameters $\boldsymbol{\theta}_{\text{SAR}} = \{P_{50}, K, \sigma_{\text{SAR}}, \nu\}$ into the inference, with broad prior specifications:
$
P_{50} \sim \text{LogNormal}(\log 60, 0.8), K \sim \text{HalfNormal}(3), \sigma_{\text{SAR}} \sim \text{TruncatedNormal}(20, 10; [5, 60]), \nu \sim 2 + \text{Exp}(1/5),$
where the $P_{50}$ prior centers on 60 kPa with broad uncertainty spanning 15--240 kPa (interquartile range), $K$ favors moderate steepness, $\sigma_{\text{SAR}}$ accommodates scene heterogeneity, and $\nu$ allows data-driven tail behavior with mild preference for heavier-than-Gaussian tails. 

% \SVcomment{Sorry, but I will need to edit this a bit to address Jaideep's comments (below). I have revisited the CSV files used for our analysis, and they were directly from Pilger processed our way, without using SpikeAD. We also used SpikeAD, but the results were more uncertain.}
The Beirut SAR analysis processes four Sentinel-1 acquisition pairs from ascending and descending orbits spanning July 23 to August 6, 2020. Multiple image pairs provide redundant observations that reduce sensitivity to individual acquisition artifacts, though they introduce registration challenges from slightly different viewing geometries. Since part of this data (4 images) was obtained from ~\cite{pilger2021beirut}, we addressed the registration challenges by using the 0-point GPS and x-y size of the combined damage image (5th image) in \cite{pilger2021beirut} to down-sample the 4 separate images. We also collected our own (GRD) data from the Copernicus Browser for all July and August of 2020~\citep{sentinelhub2020}, where each image collected has the exact same 0-point GPS for all four combinations of orientation and relative orbit~\ref{app:sar-processing}. Since the data from \cite{pilger2021beirut} was already despeckled, we used it as is. We also conducted our own experiments for denoising using a spiking neural network model, SpikeAD~\citep{verzi2017optimization,verzi2018neural,verzi2018computing}, see the Supplementary Materials.  
% \JRcomment{Can we make sure that Appendix~\ref{sec:app-forward} contains how coherence maps are computed? The bullent list contains that.}  

The four individual coherence damage maps were aggregated by summation and normalized by the maximum to keep the composite damage map between 0 and 1, inclusive. Then these pixels were further aggregated into $10 \times 10$ windows to produce a finalized damage map (of percentage values between 0 and 100). The resultant composite damage map can be seen in Figure~\ref{fig:SAR-composite-damage-map} in the Supplement.

\subsection{Vision-Language Model Damage Assessment}
\label{sec:vlm}
Ground-level photosphere imagery from Mapillary (see Section~\ref{sec:data-vlm}) provides omnidirectional views of building facades and upper stories at hundreds of geotagged locations throughout Beirut. We employ vision-language models (VLMs), specifically Google's \texttt{gemma-3-27b-it} model \citep{gemma_2025}, to classify structural damage severity at each location, then relate these qualitative assessments to quantitative overpressure estimates through physically-grounded forward models. 
The vision-language model methodology employed is described in a companion study \citep{krofcheck2025},  demonstrating the feasibility of VLM-based blast damage assessment through systematic evaluation of three Google Gemma-3 model variants (4B, 12B, 27B parameters) on the Mapillary images from the Beirut explosion. This study establishes the methodological foundation, showing that VLMs can produce probabilistically-calibrated damage classifications with Shannon entropies ranging from 1.43 to 1.82 bits, with the 27B model providing the most confident assessments and superior spatial coherence.

For each image location $i$, a VLM  processes the photosphere (converted to four planar views as described in Section~\ref{sec:data-vlm}) and produces a probability mass function $\mathbf{q}_i = (q_{i0}, q_{i1}, \ldots, q_{i8})$ over nine damage categories (and their descriptions) ranging from ``None'' (category 0) to ``Complete Destruction'' (category 8), that relate to specific overpressure intervals, following established post-disaster damage assessment scales used in structural engineering (see Table \ref{tab:vlm-damage-rubric} in the Supplement and \cite{lees2012lees}). 
% with intermediate levels capturing progressive failure modes: superficial damage (broken windows, facade cracks), moderate damage (partial wall collapse, structural distortion), and severe damage (major structural failure, near-total collapse).
The VLM output represents epistemic uncertainty in damage classification since the model may assign probability mass to multiple adjacent categories when visual evidence is ambiguous. 
% For example, an image showing extensive broken glass and minor facade cracks might yield $\mathbf{q} = (0.0, 0.1, 0.6, 0.3, 0.0, \ldots, 0.0)$, indicating high confidence in "Light" (category 2) with uncertainty extending to "Moderate" (category 3). 
This probabilistic representation proves essential for proper uncertainty propagation, as deterministic classifications would artificially narrow the posterior and fail to account for subjective assessment boundaries.

The forward model $\mathcal{F}_{\text{VLM}}(Y; \theta_{\text{VLM}})$ relates yield to expected damage classifications through two sequential mappings: yield $\to$ overpressure via Kingery-Bulmash, then overpressure $\to$ damage category probabilities via logistic binning, as a result of the VLM's softmax outputs over the 9 damage categories; more details are presented in the Supplement. 

For each image at distance $r_i$ from the epicenter, we compute incident peak overpressure (side-on pressure, appropriate for damage to building surfaces) $
P_i(Y) = \mathcal{P}_{\text{KB}}(r_i, Y), P_{\text{psi},i}(Y) \text{ in psi}.$
We then map overpressure to a probability distribution over the nine damage categories through logistic binning. We define bin edges in overpressure space: $
\mathbf{E}_{\text{psi}} = [0.00, 0.04, 0.16, 0.40, 1.10, 2.10, 3.10, 5.10, 10.0, \infty] \text{ psi},$
corresponding to engineering thresholds for progressive structural failure (see Table~\ref{tab:vlm-damage-rubric}). Here, category $k$ nominally corresponds to overpressures in the range $[E_k, E_{k+1})$.

Rather than imposing sharp boundaries that would create artificial discontinuities in the likelihood, we employ soft logistic binning controlled by a spread parameter $\sigma_{\text{dex}}$ (in log-space):
$
\pi_{ik}(P_i) = \Phi\left(\frac{\log_{10} E_{k+1} - \log_{10} P_i}{\sigma_{\text{dex}}}\right) - \Phi\left(\frac{\log_{10} E_k - \log_{10} P_i}{\sigma_{\text{dex}}}\right),$
where $\Phi(\cdot)$ denotes the logistic cumulative distribution function: $\Phi(x) = 1/(1+e^{-x})$. The vector $\boldsymbol{\pi}_i = (\pi_{i0}, \ldots, \pi_{i8})$ represents the expected probability distribution over damage categories for a structure at distance $r_i$ experiencing overpressure $P_i(Y)$, with spread $\sigma_{\text{dex}}$ controlling transition sharpness. Larger $\sigma_{\text{dex}}$ produces broader transitions reflecting greater uncertainty in the damage-overpressure relationship from structural variability, whereas smaller values approach deterministic assignment.

This formulation appropriately captures two sources of uncertainty. First, the VLM classification uncertainty (represented by $\mathbf{q}_i$) reflects perceptual ambiguity in assigning damage categories from visual evidence. Second, the spread parameter $\sigma_{\text{dex}}$ quantifies physical variability in structural response since identical overpressures produce different damage levels depending on building orientation, construction quality, and local shielding effects. Both uncertainties propagate through the forward model into the yield posterior.

The likelihood treats the observed VLM damage PMFs as draws from a multinomial distribution with probabilities determined by the forward model. For image location $i$, the log-likelihood contribution follows:
$
\ell_{i}(Y, \sigma_{\text{dex}}) = \sum_{k=0}^{8} q_{ik} \log \pi_{ik}(P_i(Y), \sigma_{\text{dex}}),$
representing the cross-entropy between the observed classification distribution $\mathbf{q}_i$ and the model-predicted distribution $\boldsymbol{\pi}_i$. 
% This formulation naturally handles probabilistic classifications: when the VLM is uncertain and spreads mass across multiple categories, the likelihood accommodates this through the cross-entropy structure.

% \textbf{Entropy-Based Weighting:} 
Not all VLM classifications provide equal information. Images with high classification entropy (uniform distributions over many categories) indicate ambiguous damage states that provide weak constraints on overpressure. Conversely, low-entropy classifications (peaked distributions) reflect confident assessments that should receive greater weight. To incorporate this informativeness structure, we apply entropy-based reweighting. This empirical Bayes approach naturally down-weights ambiguous classifications while preserving the full probabilistic structure of the VLM output: 
$
w_i = (1 + H(\mathbf{q}_i))^{-1}, H(\mathbf{q}_i) = -\sum_{k=0}^{8} q_{ik} \log_2 q_{ik},$
where $H(\mathbf{q}_i)$ denotes the Shannon entropy in bits. We normalize weights by their median and clip to $[0.25, 4.0]$ to prevent extreme down-weighting of moderately uncertain classifications while still emphasizing confident assessments.
The average log-likelihood across all images subsequently prevents sample-size effects in multimodal fusion:
$
\ell_{\text{VLM}}(Y, \sigma_{\text{dex}}) = \sum_{i=1}^{N_{\text{VLM}}} w_i \ell_i(Y, \sigma_{\text{dex}})/\sum_{i=1}^{N_{\text{VLM}}} w_i.$

The spread parameter receives a prior reflecting typical uncertainty in damage-pressure relationships:
$
\sigma_{\text{dex}} \sim \text{TruncatedNormal}(\mu=0.15, \sigma=0.05, \text{lower}=0.05, \text{upper}=0.60),$
centered at 0.15 dex (approximately 40\% variability in pressure) with broad support allowing data-driven calibration. Smaller values indicate tight damage-overpressure correlations appropriate for uniform construction, while larger values accommodate heterogeneous building stocks.

The Beirut VLM dataset comprises $N_{\text{VLM}}= $ 685 photosphere images distributed across the city (Figure~\ref{fig:mapillary_locations}). The VLM identifies damage indicators including broken windows with visible glass fragments, debris piles containing building materials, structural deformation, and facade collapse. 
As demonstrated in our companion article \citep{krofcheck2025}, the Google \texttt{gemma-3-27b-it} model achieves consistent damage classification performance with reasonable processing throughput on a single A100. This study confirms that entropy-based weighting successfully identifies ambiguous classifications, with reasonable model agreement across both the 12B and 27B models potentially indicating convergent behavior across model scales. 

\section{Results}
\label{sec:results}
This section presents the yield estimation results obtained through Bayesian inference over the fractional posterior framework. 
% We first describe the joint likelihood formulation and computational inference procedure, then present the learned hyperparameters for each modality, followed by individual modality posteriors, the fused multimodal estimate and posterior predictive validation. 

\subsection{Joint Likelihood and Inference Procedure}
\label{sec:inference-procedure}
The joint posterior distribution over yield $Y$, trust weights $\boldsymbol{\gamma} = (\gamma_{\text{seismic}}, \gamma_{\text{crater}}, \gamma_{\text{SAR}}, \gamma_{\text{VLM}})$, and modality-specific hyperparameters $\boldsymbol{\theta} = \{\boldsymbol{\theta}_{\text{seismic}}, \boldsymbol{\theta}_{\text{crater}}, \boldsymbol{\theta}_{\text{SAR}}, \boldsymbol{\theta}_{\text{VLM}}\}$ is: 

\begin{equation}
p(Y, \boldsymbol{\gamma}, \boldsymbol{\theta} | \D) \propto \pi(Y) \, \pi (\boldsymbol{\gamma}) \prod_{i} \pi(\boldsymbol{\theta}_i) \prod_{i} \mathcal{L}_i(D_i | Y, \boldsymbol{\theta}_i)^{\gamma_i}, \label{eq:posterior}
\end{equation}

where $\D = \{\D_{\text{seismic}}, \D_{\text{crater}}, \D_{\text{SAR}}, \D_{\text{VLM}}\}$ denotes the complete observational dataset, $\pi(\cdot)$ represents prior distributions, and $\mathcal{L}_i$ denotes the likelihood function for modality $i$ as specified in Section 4. The trust weights are constrained to the simplex through a Dirichlet prior with concentration parameter $\boldsymbol{\alpha} = (
\alpha_{\text{seismic}}, \alpha_{\text{crater}}, \alpha_{\text{SAR}}, \alpha_{\text{VLM}}) $:
$\boldsymbol{\gamma} \sim \text{Dirichlet}(\boldsymbol{\alpha}), $
with concentration parameters $\boldsymbol{\alpha} = \mathbf{1}_4$, where  $\mathbf{1}_4$ corresponds to the 4-vector of ones, corresponding to a uniform prior over the simplex that expresses no preference for any particular weighting scheme.

The modality-specific hyperparameters comprise: $\boldsymbol{\theta}_{\text{seismic}} = \{\sigma_m\}$ representing seismic magnitude uncertainty; $\boldsymbol{\theta}_{\text{crater}} = \{\sigma_c\}$ capturing crater scaling variability; $\boldsymbol{\theta}_{\text{SAR}} = \{P_{50}, K, \sigma_{\text{SAR}}, \nu\}$ parameterizing the damage-overpressure vulnerability curve, scene heterogeneity, and tail behavior; and $\boldsymbol{\theta}_{\text{VLM}} = \{\sigma_{\text{dex}}\}$ quantifying damage classification uncertainty. Accordingly, the joint inverse problem targets $(Y,\boldsymbol{\gamma},\boldsymbol{\theta})$ with effective dimension $1+3+\sum_{i=1}^4 p_i$ (one yield, $3-$dimensional simplex-constrained trust weights, and $\sum_i p_i$ modality hyperparameters); in our four-modality case with $p=(1,1,4,1)$ this totals $11$ unknowns.

% The individual likelihood functions and their forward models are detailed in Section 4, where each $\mathcal{L}_i$ connects observations to yield through the appropriate physical relationships.

\subsubsection{Markov Chain Monte Carlo Sampling}   
We perform Bayesian inference, sampling from \eqref{eq:posterior} using the No-U-Turn Sampler (NUTS) \citep{hoffman2014nuts}, an adaptive Hamiltonian Monte Carlo algorithm implemented in Stan \citep{carpenter2017stan}. The sampling configuration consists of four independent chains initialized from dispersed starting points, with 8,000 total iterations per chain. The first 2,000 iterations serve as burn-in (warm-up), yielding 6,000 posterior samples per chain for a total of 24,000 samples. We set the target acceptance probability to 0.95 and maximum tree depth to 12 to ensure thorough exploration of the posterior geometry while avoiding divergent transitions.

% Convergence diagnostics confirm successful exploration of the posterior distribution. All parameters achieve Gelman-Rubin statistics $\hat{R} < 1.01$, indicating convergence across chains. Effective sample sizes exceed 1,000 for all parameters of interest, ensuring adequate sampling precision for posterior inference. Trace plots (Figure \ref{fig:mcmc-traces}) demonstrate good mixing without systematic trends, and autocorrelation functions decay rapidly within 20 lags, confirming that successive samples provide nearly independent draws from the posterior.

% \begin{figure}[htbp]
%     \centering
%     \includegraphics[width=\textwidth]{figs/chp05/mcmc_traces_yield_gamma.png}
%     \caption{MCMC trace plots for explosive yield $Y$ and trust weights $\boldsymbol{\gamma}$ across four independent chains. The chains exhibit excellent mixing and convergence to a stationary distribution, with no systematic trends or divergent transitions. Different colors represent different chains.}
%     \label{fig:mcmc-traces}
% \end{figure}

\subsection{Hyperparameter Posteriors}
\label{sec:hyperparameter-posteriors}
The learned hyperparameters reveal important information about measurement uncertainties and model-data consistency for each modality. Figure \ref{fig:hyperparameter-posteriors} in the Supplement displays the posterior distributions for all modality-specific hyperparameters, using the fused data, including the estimated damage to overpressure relationship learned using SAR imagery. 
The seismic magnitude uncertainty parameter exhibits posterior mean $\sigma_m = 0.159$ with 95\% highest density interval (HDI) [0.067, 0.245] magnitude units, while the crater scaling uncertainty converges to $\sigma_c = 0.080$ with 95\% HDI [0.046, 0.121] in log$_{10}$-space, corresponding to approximately 20\% uncertainty in linear dimensions.
The SAR modality learns a damage-overpressure vulnerability curve through four parameters. The 50\% posterior mean damage threshold $P_{50} = 83.6$ kPa (95\% HDI [3.6, 223.3] kPa) suggests that port-area construction exhibited greater blast resistance than typical unreinforced masonry, potentially reflecting reinforced concrete warehouses designed for industrial loads. The transition steepness $K = 2.16$ (95\% HDI [0.003, 5.03]) indicates gradual rather than sharp damage thresholds, reflecting heterogeneous structural vulnerability. The scene heterogeneity parameter $\sigma_{\text{SAR}} = 23.5\%$ (95\% HDI [6.3, 39.1]\%) quantifies substantial variability not explained by distance alone, potentially encompassing mixed construction, shielding effects, and directional blast patterns. The degrees of freedom parameter $\nu = 3.68$ (95\% HDI [0.001, 11.4]) indicates notably heavy tails, demonstrating that SAR observations contain significant outliers requiring robust statistical treatment to prevent distortion of yield inference.
The VLM damage classification spread parameter converges to $\sigma_{\text{dex}} = 0.158$ with 95\% HDI [0.066, 0.246] in log-space. The posterior shift from the prior mean of 0.15 reflects both the subjective nature of damage classification and inherent variability in structural response to blast loading, appropriately capturing the challenges of inferring quantitative overpressure from qualitative visual inspection.

The learned hyperparameters demonstrate that the Bayesian framework successfully calibrates modality-specific uncertainties from data while incorporating prior knowledge from physical principles. The tight constraints on crater and seismic uncertainties reflect the maturity of these measurement techniques, while broader SAR and VLM uncertainties acknowledge inherent challenges in damage-based inference.

\subsection{Individual Modality Yield Estimates}
\label{sec:individual-estimates}

\begin{figure}[htbp]
    \centering
    \includegraphics[width=\linewidth]{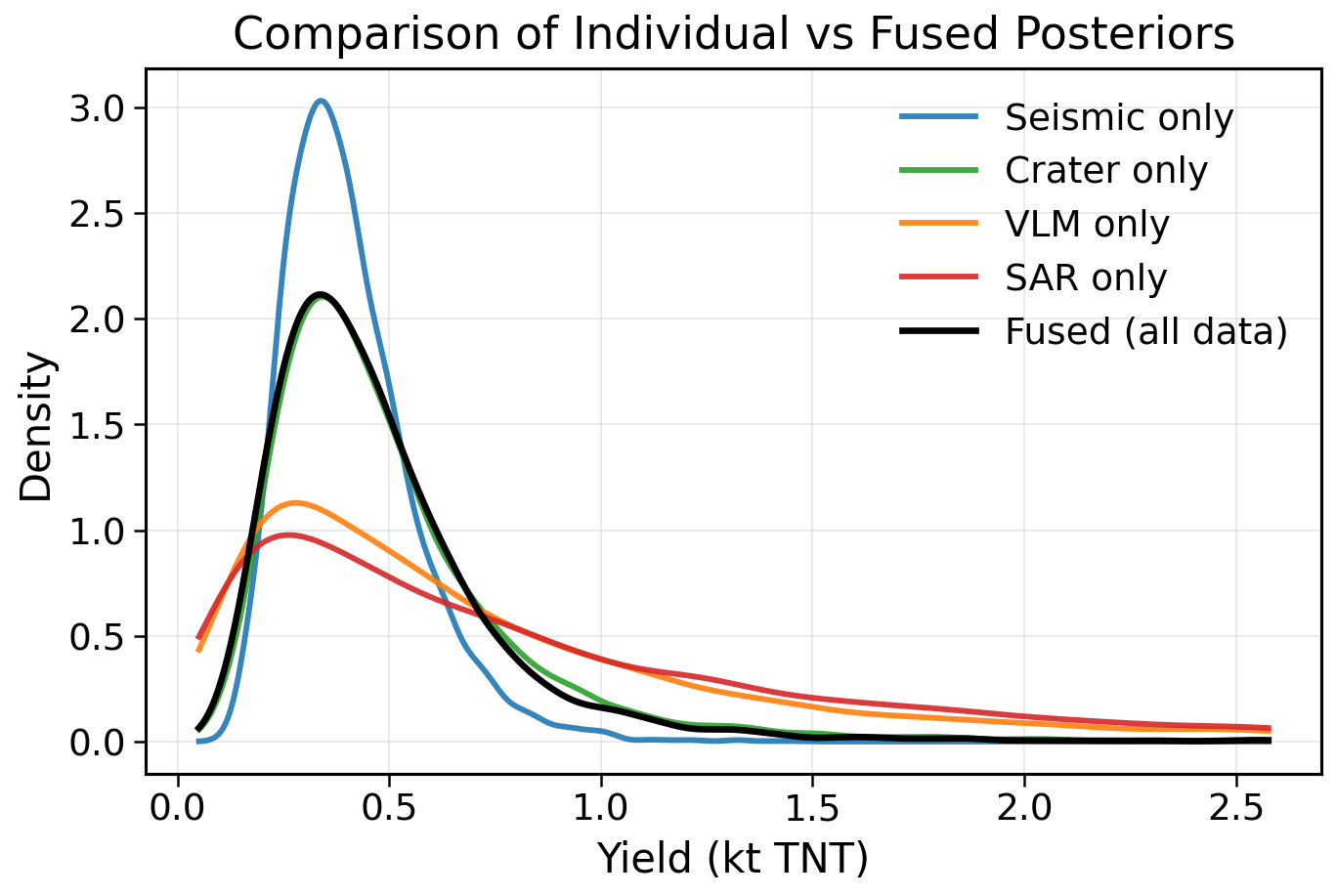}
    \caption{Individual modality posterior distributions for explosive yield against fused posterior.}
    \label{fig:individual-posteriors}
\end{figure}

Before presenting the multimodal fusion results, we examine yield estimates obtained from each modality independently to establish baseline performance and identify systematic trends. Table \ref{tab:individual-yields} summarizes the single-modality posteriors, which is also visualized in Figure \ref{fig:individual-posteriors} against the fused posterior.

\begin{table}[htbp]
\centering
\caption{Individual modality yield estimates with posterior statistics}
\label{tab:individual-yields}
\begin{tabular}{lcccc}
\hline
Modality & Mode (kt) & Median (kt) & Mean (kt) & 95\% HDI (kt) \\
\hline
Seismic & 0.34 & 0.38 & 0.41 & [0.16, 0.71] \\
Crater & 0.34 & 0.43 & 0.50 & [0.11, 1.06] \\
VLM & 0.28 & 0.54 & 0.71 & [0.03, 1.97] \\
SAR & 0.26 & 0.59 & 0.77 & [0.02, 2.08] \\
\hline
\end{tabular}
\end{table}

The seismic and crater modalities demonstrate remarkable convergence with identical modal estimates of 0.34 kt, providing strong independent validation of the explosive yield. The seismic posterior (mode 0.34 kt, 95\% HDI [0.16, 0.71] kt) exhibits relatively tight constraints, with the neural network ensemble successfully generalizing from underground nuclear tests to the surface chemical explosion despite the substantial differences in emplacement conditions. The crater-based posterior (mode 0.34 kt, 95\% HDI [0.11, 1.06] kt) provides independent physical confirmation through direct excavation evidence, with broader uncertainty that appropriately accounts for substrate heterogeneity and the observed elliptical crater geometry.
The damage-based modalities exhibit substantially greater uncertainty, as expected when vulnerability parameters must be learned jointly with yield from limited observational data. The VLM posterior (mode 0.28 kt, median 0.54 kt, 95\% HDI [0.03, 1.97] kt) shows considerable spread, appropriately reflecting the compound uncertainties in mapping qualitative visual damage assessments through overpressure to yield. Similarly, the SAR damage-based posterior (mode 0.26 kt, median 0.59 kt, 95\% HDI [0.02, 2.08] kt) exhibits the broadest uncertainty, capturing both the indirect relationship between yield and coherence loss and the epistemic uncertainty from simultaneously learning the vulnerability curve parameters ($P_{50}$, $K$, $\sigma_{\text{SAR}}$, $\nu$).

Despite the heterogeneous uncertainty levels, all four modalities show general consistency with modal estimates clustering in the 0.26--0.34 kt range. The exact agreement between seismic and crater modes is particularly compelling, as these modalities rely on fundamentally different physical processes, yet converge to identical point estimates. The damage-based modalities (VLM and SAR) contribute valuable spatial information covering the entire affected urban area, complementing the point measurements with distributed observations despite their wider posteriors that honestly reflect the combined uncertainties in both yield-to-pressure and pressure-to-damage relationships.

Figure \ref{fig:fused-posterior}  shows the fused posterior for explosive yield obtained by combining all four modalities. The distribution is unimodal with a mild right tail; its center aligns with the 0.34-kt modes of the seismic and crater posteriors in Figure  \ref{fig:individual-posteriors}, while SAR and VLM contributions broaden the upper tail. We use this fused distribution for all subsequent summaries and validation.

\subsection{Learned Trust Weights}
\label{sec:trust-weights}
The posterior distributions for trust weights reveal the relative information content contributed by each modality. Figure \ref{fig:trust-weights} displays the trust weight posteriors.
The learned trust weights exhibit posterior mean 0.344 for crater, 0.281 for seismic, 0.221 for VLM, and 0.154 for SAR. Crater observations receive the greatest contribution due to the direct physical relationship between explosive energy and excavation, relatively low systematic uncertainty, and precise satellite-derived measurements. Seismic observations earn substantial weight despite systematic underestimation because the posterior appropriately accounts for geological uncertainty through $\sigma_m$, with the tight magnitude measurement ($M_w = 4.50 \pm 0.13$) providing strong constraints. VLM damage assessments contribute moderate weight, demonstrating that qualitative AI-interpreted observations provide useful constraints when uncertainties are properly characterized. SAR receives the lowest contribution due to elevated scene heterogeneity ($\sigma_{\text{SAR}} = 23.5\%$), significant outliers requiring heavy-tailed error models ($\nu = 3.68$), and the indirect connection to yield through learned vulnerability curves. The learned damage curve parameters suggest that SAR coherence loss may underestimate structural damage compared to visual assessment, possibly due to radar penetration effects or differing damage mechanisms.

The trust weight posteriors exhibit substantial overlap due to the simplex constraint $\sum_{i=1}^4 \gamma_i = 1$, with wide credible intervals indicating that multiple weight combinations produce similar posterior yields. This degeneracy reflects that the fused estimate is robust to the precise weighting scheme, with the yield remaining consistent across different weight allocations since all modalities provide complementary constraints.

\subsection{Model Validation}
To verify that our multimodal fusion framework produces well-calibrated estimates, we performed comprehensive validation through posterior predictive checks and leave-one-out analysis in the Supplement. Posterior predictive checks confirm that data generated from the fitted model closely resemble observed data across all modalities, with $p$-values near 0.5 indicating absence of systematic bias (seismic: $p=0.78$, crater: $p=0.51$). The learned vulnerability curves then successfully reproduce the spatial decay of damage for SAR and VLM observations, with most data points falling within 90\% credible intervals.
We additionally perform Leave-one-modality-out (LOO) validation which reveals the nuanced relationship between modality agreement and posterior influence. Despite receiving the lowest trust weight, removing SAR produces the largest KL divergence, demonstrating that even down-weighted modalities contribute meaningfully by expanding uncertainty bounds to reflect model disagreement.
Crucially, learned trust weights decrease for corrupted modalities, with $\boldsymbol{\gamma}$ being inversely ranked to LOO KL divergence metrics, confirming that discordant evidence is down-weighted while preserving consensus information.

We additionally perform simulation stress tests that evaluate heavy tails, systematic bias, label noise, cross-modality dependence and other fusion methodologies. Compared with a plain product posterior, a single-temperature model,  fixed-$\gamma$ via cross-validation, Bayesian Model Avergaing (BMA) and covariance intersection, the Dirichlet-tempered fusion maintains near-nominal 95\% coverage with competitive or smaller interval width and lower RMSE under misspecification. 
% !TEX root = main.tex
\section{Discussion} \label{sec:discussion}
The fractional Bayesian framework estimates the Beirut explosion yield at 0.34--0.48 kt TNT equivalent (mode-mean, 95\% HDI [0.11, 1.00] kt), corresponding to 12--17\% detonation efficiency relative to the theoretical maximum of 2.75 kt from complete ammonium nitrate combustion. 
% This efficiency aligns with historical precedents: the 1947 Texas City disaster achieved 10--15\% efficiency under similar unconfined conditions, while the 2015 Tianjin explosion exhibited 5--10\% efficiency with containerized storage. 
The incomplete detonation reflects three primary factors characteristic of industrial accidents: unconfined surface burst geometry allowing lateral energy dispersal, heterogeneous storage conditions with variable moisture content and material degradation over years of improper storage, and the large charge diameter exceeding critical dimensions for sustained detonation.

Our central estimate overlaps substantially with the 0.4--1.4 kt range reported by \cite{kim2022yield} and others, validating their seismoacoustic approach while providing tighter constraints through multimodal fusion. The narrower range of our mode-median estimate demonstrates the value of incorporating complementary observations, where crater measurements provide a direct physical anchor and damage-based modalities contribute spatial information across the affected area. The framework successfully reconciles the wide range of literature estimates (Table~\ref{tab:previous_estimates}, 0.2--1.4 kt) by revealing that the apparent discrepancy stems from different uncertainty quantification approaches rather than fundamental disagreement.

The success of the fractional fusion approach rests on three critical design decisions that address fundamental challenges in multimodal inference. 
First, modality-specific tempering through learned trust weights provides an automatic mechanism for handling systematic biases without requiring manual calibration or extensive validation data. The learned weights reveal an informative hierarchy reflecting both physical directness and model-data consistency.
% : crater measurements receive the highest weight ($\gamma_{\text{crater}} = 0.34$) due to their direct physical relationship with explosive energy, seismic observations receive moderate weight ($\gamma_{\text{seismic}} = 0.28$) appropriate for their established empirical relationships despite coupling uncertainties in the near-shore environment, vision-language model assessments contribute moderately ($\gamma_{\text{VLM}} = 0.22$) through entropy-weighted aggregation that emphasizes confident damage classifications, and SAR observations receive the lowest weight ($\gamma_{\text{SAR}} = 0.15$) due to the indirect path from yield through overpressure to coherence loss.
Second, the incorporation of vision-language model interpretations demonstrates that qualitative assessments can contribute meaningfully to quantitative inference when properly integrated through physics-based forward models. This represents a methodological advance in bridging the gap between traditional physical measurements and modern AI-interpreted observations. Comprehensive assessment of the VLM damage methodology is presented in \cite{krofcheck2025}, establishing that VLMs can reliably assess blast damage with appropriate uncertainty quantification, though operational deployment requires expert validation datasets as detailed therein.
Third, the framework's treatment of conditional independence proves appropriate given the distinct observation geometries of our modalities. SAR and street-level imagery observe fundamentally different physical manifestations of blast effects rather than redundant measurements of the same damage state, supporting the independence assumption. This assumption would break down for modalities with overlapping observation spaces, such as multiple optical satellite images of the same structures, requiring extensions to handle explicit correlation structures.

% The learned hyperparameters provide insights into Beirut's specific conditions that generic models would miss. The SAR vulnerability curve parameters ($P_{50} = 83.6$ kPa (mean), $K = 2.16$) suggest that port warehouses exhibit greater blast resistance than typical unreinforced masonry, consistent with industrial construction designed for heavy loads. The heavy-tailed error distribution ($\nu = 3.68$) captures the reality of heterogeneous urban damage where some structures survive intact while adjacent buildings collapse, a phenomenon poorly captured by Gaussian error models.

% The learned hyperparameters provide insights into Beirut's specific conditions that generic models would miss. The SAR vulnerability curve parameters suggest that port warehouses exhibit greater blast resistance than typical unreinforced masonry, while the heavy-tailed error distribution captures the reality of heterogeneous urban damage where some structures survive intact while adjacent buildings collapse, a phenomenon poorly captured by Gaussian error models.

This work establishes a template for principled integration of heterogeneous observations in explosion monitoring contexts where traditional approaches face limitations. The agreement between our multimodal estimate and previous analyses validates both approaches while demonstrating that fusion can reduce uncertainty without sacrificing accuracy. For treaty verification under the Comprehensive Nuclear-Test-Ban Treaty, the ability to incorporate non-traditional data sources such as commercial satellite imagery and social media could enhance monitoring capabilities, particularly for events where traditional seismic networks provide limited constraints. The framework's robustness to model misspecification, demonstrated through extensive simulation studies, addresses a critical challenge in operational monitoring where forward models are necessarily simplifications of complex physical processes. The learned trust weights provide an interpretable diagnostic for identifying problematic data streams, as evidenced by SAR's low weight correctly flagging model-data discrepancies.

Nevertheless, computational cost of full Bayesian inference may preclude real-time application, though variational approximations or pre-computed surrogate models could address this limitation. Then, the current framework assumes fixed forward models, whereas operational settings might benefit from model selection or averaging over multiple physical models. Future extensions to handle time-series observations and sequential updating as new data arrives would enhance applicability to evolving crisis situations.

\section*{Data Availability}
Seismic data are available from the Incorporated Research Institutions for Seismology Data Management Center. Satellite imagery from Maxar Technologies and European Space Agency Copernicus program. Ground-level photographs from Mapillary (\url{https://www.mapillary.com}). Synthetic aperture radar damage proxy map from \cite{sudhaus2021damage} available at \url{https://zenodo.org/records/4762436}. Analysis code will be made available upon publication at a Sandia National Laboratories repository.

\section*{Author Contributions}
L.P. developed the fractional posterior framework, the sesimic methodology and performed the multimodal Bayesian inference. C.U. collected the data, designed the computational infrastructure and vision-language model integration. S.V. and J.R. developed the SAR methodology. I.M. developed the crater methodology. D.K. and A.N. developed the VLM methodology. J.R. conceived the study and provided technical guidance. All authors contributed to manuscript preparation and approved the final version.

\section*{Funding}
This work was supported by Sandia National Laboratories Laboratory Directed Research and Development program. Sandia National Laboratories is a multimission laboratory managed and operated by National Technology and Engineering Solutions of Sandia, LLC, a wholly owned subsidiary of Honeywell International Inc., for the United States Department of Energy National Nuclear Security Administration under contract DE-NA0003525.

\section*{Acknowledgments}
This paper describes objective technical results and analysis. Any subjective views or opinions that might be expressed in the paper do not necessarily represent the views of the United States Department of Energy or the United States Government.

\bibliographystyle{unsrtnat}
\bibliography{references}
\clearpage

\appendix
\setcounter{section}{0}
\renewcommand{\thesection}{S\arabic{section}}
\renewcommand{\thetable}{S\arabic{table}}
\renewcommand{\thefigure}{S\arabic{figure}}
\renewcommand{\theequation}{S\arabic{equation}}
\setcounter{table}{0}
\setcounter{figure}{0}
\setcounter{equation}{0}
\begin{center}
    \Large \textbf{SUPPLEMENTARY MATERIALS}
\end{center}
% Data sources and processing
\section{Data Collection and Preprocessing Details}
\label{sec:app-data}

This appendix provides detailed information on data collection procedures, preprocessing workflows, quality assessment, and validation for each modality used in the yield estimation framework.

\subsection{Seismic Data Processing}
\label{app:seismic-processing}

\subsubsection{Data Retrieval and Storage}

Seismic data were retrieved from the Incorporated Research Institutions for Seismology (IRIS) Data Management Center via the SAGE web portal (\url{https://ds.iris.edu/gmap/}). Query parameters specified:
\begin{itemize}
\item Time window: 2020-08-04T15:00:00Z to 2020-08-04T16:00:00Z
\item Geographic radius: 3° from epicenter (33.9016°N, 35.5195°E)
\item Network codes: Multiple international networks (IS, GE, IM, CY, etc.)
\item Channel preferences: Broadband velocity channels (BH*)
\end{itemize}

Data were stored in MiniSEED format, with each file named by station code and containing multiple sensor traces (BHE for east-west, BHN for north-south, BHZ for vertical components). Sampling rates ranged from 20--100 Hz depending on station configuration.

\subsubsection{Waveform Analysis with ObsPy}

The Python ObsPy library (version 1.4.0) was used for all seismic processing tasks:

\textbf{Quality Control}:
\begin{itemize}
\item Gap detection and removal of traces with >5\% missing data
\item Instrument response correction to ground velocity
\item Bandpass filtering (0.5--5.0 Hz) to isolate regional body waves
\item Signal-to-noise ratio calculation using pre-event noise window
\end{itemize}

\textbf{Phase Picking}:
P-wave and S-wave arrival times were estimated using the AR-AIC picker with the following parameters:
\begin{itemize}
\item Window length: 50 samples
\item Minimum SNR threshold: 3.0
\item Pick uncertainty: ±0.5 seconds
\end{itemize}

Manual quality review flagged and excluded poor-quality picks, retaining arrivals with clear onset times.

\subsubsection{Station Metadata}

Table~\ref{tab:seismic-stations} summarizes key characteristics of the 43 stations used in the analysis.

\begin{table}[ht!]
\centering
\caption{Regional seismic station characteristics}
\label{tab:seismic-stations}
\begin{tabular}{lcccc}
\toprule
\textbf{Distance Range (km)} & \textbf{Count} & \textbf{Azimuth Coverage} & \textbf{Avg SNR} \\
\midrule
100--200 & 12 & 45°--315° & 18.5 \\
200--300 & 18 & 15°--340° & 12.3 \\
300--400 & 10 & 30°--290° & 8.7 \\
400--500 & 3 & 75°--210° & 6.2 \\
\bottomrule
\end{tabular}
\end{table}

The station distribution provides good azimuthal coverage with slight gaps in the southwestern quadrant. Distance-dependent amplitude decay follows theoretical geometric spreading ($r^{-1}$) and anelastic attenuation models for the Eastern Mediterranean crustal structure.

\subsection{Crater Measurement Methodology}
\label{app:crater-methods}
Table \ref{tab:satellite_sources} summarizes imagery acquisition dates and characteristics for visualizing the crater left by the blast. 
\begin{table}[ht!]
    \centering
    \caption{Summary of satellite imagery sources}
    \label{tab:satellite_sources}
    \begin{tabular}{lccccc} 
        \toprule
        \textbf{Source} & \textbf{Before} & \textbf{After} & \textbf{Type} & \textbf{Resolution}  \\
        \midrule
%Digital Globe       &   1 &  1 & Monochrome & 1 m \\    
Maxar Technologies  &   2 &  1 & Color      & 1 m  \\
ESA Sentinel-1      &  21 & 17 & SAR        & 20 m \\
ESA Sentinel-2      &   5 &  5 & Color      & 4 m \\
ICEYE               &   3 &  4 & SAR        & 1 m \\    
\bottomrule
\end{tabular}
\end{table}

\subsubsection{Image Selection and Preprocessing}

Crater dimensions were extracted from Sentinel-2 Level-2A imagery (surface reflectance product) acquired August 8, 2020, at 08:21 UTC. This acquisition provided:
\begin{itemize}
\item Cloud-free coverage of the port area
\item 10-meter resolution in RGB bands
\item Georeferencing accuracy <10 meters
\item Near-nadir viewing geometry minimizing distortion
\end{itemize}

Images were reprojected from UTM Zone 36N (EPSG:32636) to a local Transverse Mercator projection centered on the explosion site to minimize area distortion for linear distance measurements.

\subsubsection{Automated Boundary Detection}

Meta's Segment Anything Model (SAM) was applied to detect the crater boundary through the following workflow:

\textbf{Water Color Detection}:
\begin{enumerate}
\item Compute normalized difference water index: NDWI = (Green - NIR)/(Green + NIR)
\item Apply threshold NDWI > 0.3 to isolate water pixels
\item Apply SAM with point prompts at visually identified crater edges
\item Extract contour from SAM's highest-confidence mask
\end{enumerate}

\textbf{Ellipse Fitting}:
Least-squares ellipse fitting to the extracted contour yielded:
\begin{itemize}
\item Major axis: 108.1 ± 5.2 m (uncertainty from SAM mask confidence)
\item Minor axis: 46.7 ± 3.8 m
\item Orientation: 72° from north (aligned with warehouse long axis)
\item Equivalent circular diameter: $D_{\text{equiv}} = \sqrt{108.1 \times 46.7} = 71.0$ m
\end{itemize}

\textbf{Validation}:
Manual measurements using QGIS polygon tools on the same imagery yielded 106.3 m $\times$ 48.1 m, confirming SAM's automated extraction within measurement uncertainty. Independent estimates from the PRJ-3030 GEER report cite major axis ~120 m based on field surveys and different imagery sources.

\subsection{SAR Processing Workflow}
\label{app:sar-processing}

\subsubsection{Acquisition Pairs and Processing}

Four Sentinel-1 Interferometric Wide Swath (IW) acquisition pairs were processed:

\begin{table}[ht!]
\centering
\caption{Sentinel-1 acquisition pairs for coherence analysis}
\label{tab:sar-pairs}
\begin{tabular}{llcc}
\toprule
\textbf{Orbit} & \textbf{Dates} & \textbf{Baseline (m)} & \textbf{Mean Coherence} \\
\midrule
Ascending 1 & Jul 30 / Aug 5 & 45 & 0.68 \\
Ascending 2 & Aug 1 / Aug 7 & 52 & 0.65 \\
Descending 1 & Jul 31 / Aug 6 & 38 & 0.71 \\
Descending 2 & Aug 2 / Aug 8 & 61 & 0.62 \\
\bottomrule
\end{tabular}
\end{table}

\textbf{Standard InSAR Processing}:
\begin{enumerate}
\item Co-registration: Sub-pixel alignment using cross-correlation
\item Interferogram formation: Complex multiplication of master × conjugate(slave)
\item Coherence estimation: 10×10 pixel sliding window
\item Geocoding: Map to ground coordinates using SRTM DEM
\item Threshold: Coherence loss >0.2 flagged as damaged
\item Compositing: Logical OR across all pairs (pixel damaged if any pair shows loss)
\end{enumerate}

\subsubsection{SpikeAD Despeckling}

Standard median filtering modifies all pixels uniformly, potentially blurring damage boundaries. SpikeAD (Spiking Adaptive Despeckling) selectively modifies only outlier pixels using median absolute deviation from the median (or MAD) computed values:

\textbf{Algorithm}:
\begin{enumerate}
\item For each pixel, compute local median and MAD in 11×11 window
\item If $|pixel - median| > 3.0 \times MAD$, replace with weighted median
\item Otherwise, retain original value
\item Iterate 4 times for aggressive noise reduction
\end{enumerate}

This adaptive approach preserves contiguous damage regions while removing isolated speckle noise. Figure~\ref{fig:spike_ad_sar_comparison} demonstrates effectiveness on the August 5 acquisition.

\subsubsection{Zonal Aggregation}

To focus on locations experiencing near-direct blast exposure:

\begin{enumerate}
\item Aggregate damage into 10×10 pixel boxes (~100 m × 100 m)
\item Compute damage percentage per box (fraction of pixels flagged)
\item Partition city into 15 concentric annuli (radii 200 m to 8 km)
\item Within each annulus, retain only 95th percentile boxes
\item Final dataset: ~150--200 boxes distributed across distance ranges
\end{enumerate}

This zonal thresholding addresses the challenge that tall structures in line-of-sight experience maximum damage while surrounding areas show lower damage from shielding and reflected waves. By selecting high-percentile boxes, we focus on locations where simplified Kingery-Bulmash models are most applicable.

\subsection{Ground-Level Image Processing}
\label{app:vlm-processing}
Mapillary is a commercial website that hosts community-supplied, street-level images of cities from around the world. Shortly after the 2020 Beirut explosion, Mapillary users posted geo-tagged pictures from cell phones, dashboard cameras, and 360° photosphere cameras. Most notably, researchers from the American University of Beirut (AUB) Maroun Semaan Faculty of Engineering and Architecture (MSFEA) uploaded 2,100 photosphere images taken during a building survey that took place two months after the event. As documented in section 4.2 of the NSF-sponsored Geotechnical Extreme Event Reconnaissance (GEER) report~\cite{geer70}, the AUB researchers attached a GoPro Fusion camera to the roof of a vehicle and collected geotagged, photosphere images as they systematically drove through the streets of the city near the port. The resulting images were uploaded to Mapillary for public hosting. Mapillary blurs faces and license plates to mitigate privacy concerns and allows users to access content through a website and a free API.

\subsubsection{Mapillary API Queries}
We obtained an API toked from Mapillary and used v4 of the Mapillary Graph API to obtain metadata for images that were taken in the Beirut area from August to the end of December in 2020. Our initial query to retrieve metadata for 2,000 images is as follows:

\begin{verbatim}
GET /images?access_token={token}
           &bbox=35.465744,33.862921,35.571058,33.909378 
           &start_captured_at=2020-08-01T00:00:00:00Z
           &end_captured_at=2020-12=31T00:00:00Z
           &fields=id,computed_geometry,captured_at,
                   creator,is_pano,thumb_original_url
\end{verbatim}

An examination of the initial set of metadata revealed that multiple users had contributed images besides the AUB MSFEA researchers. Additional metadata queries were placed to determine the number of planar and photosphere images available for each users. Images with incomplete position information (i.e., empty "computed\_geometry" fields) were excluded from consideration. As listed in Table~\ref{tab:mapillary-users}, a large number of photosphere images were available from \texttt{chadifa} and \texttt{aubmsfea}.

\begin{table}[ht!]
\centering
\caption{Ground-level image counts from different Mapillary users}
\label{tab:mapillary-users}
\begin{tabular}{lcrrc}
\toprule
\textbf{User} & \textbf{Dates} & \textbf{Planar} & \textbf{Photosphere} & \textbf{Retrieved} \\
\midrule
aubmsfea       & Oct 01 -- Oct 15 &     0  & 1,867 &   447 \\
chadifa        & Aug 08 -- Aug 23 &   136  &   884 &   238 \\
mohamadmaktabi & Aug 09 -- Aug 12 &   976  &     0 &   229 \\
patsy          & Aug 04 -- Aug 12 &   268  &     0 &    57 \\
talamoukaddem  & Aug 08           &    50  &     0 &     7 \\
ziadg          & Oct 10 -- Nov 19 & 4,425  &     0 & 3,966 \\
\bottomrule
\end{tabular}
\end{table}

We retrieved a total of 4,944 images from the list of 8,606 images that had valid position information.

\subsubsection{Photosphere Planarization}
We observed two kinds of photosphere formats in the downloaded Mapillary dataset: equirectangular images from user \texttt{aubmsfea} and dual hemisphere images from user \texttt{chadifa}. Anticipating that the warping found in these formats might impede image analysis algorithms, we used the \texttt{360-to-planer-images} library\footnote{\url{https://github.com/Maxiviper117/360-to-planer-images}} to extract four planar images from each photosphere image. Initial work extracted planar images at cardinal yaws (0°, 90°, 180°, 270°) and a 30° pitch. Unfortunately, these perspectives were found to be problematic for the task of analyzing building structures and were therefore discarded. The primary concerns were that the forwards and backwards views focused more on the streets than buildings, the left and right views hit camera blind spots in the dual-hemisphere images, and the 30° pitch was frequently too high to observe ground floor details.

\begin{figure}
    \centering
    \includegraphics[width=0.90\linewidth]{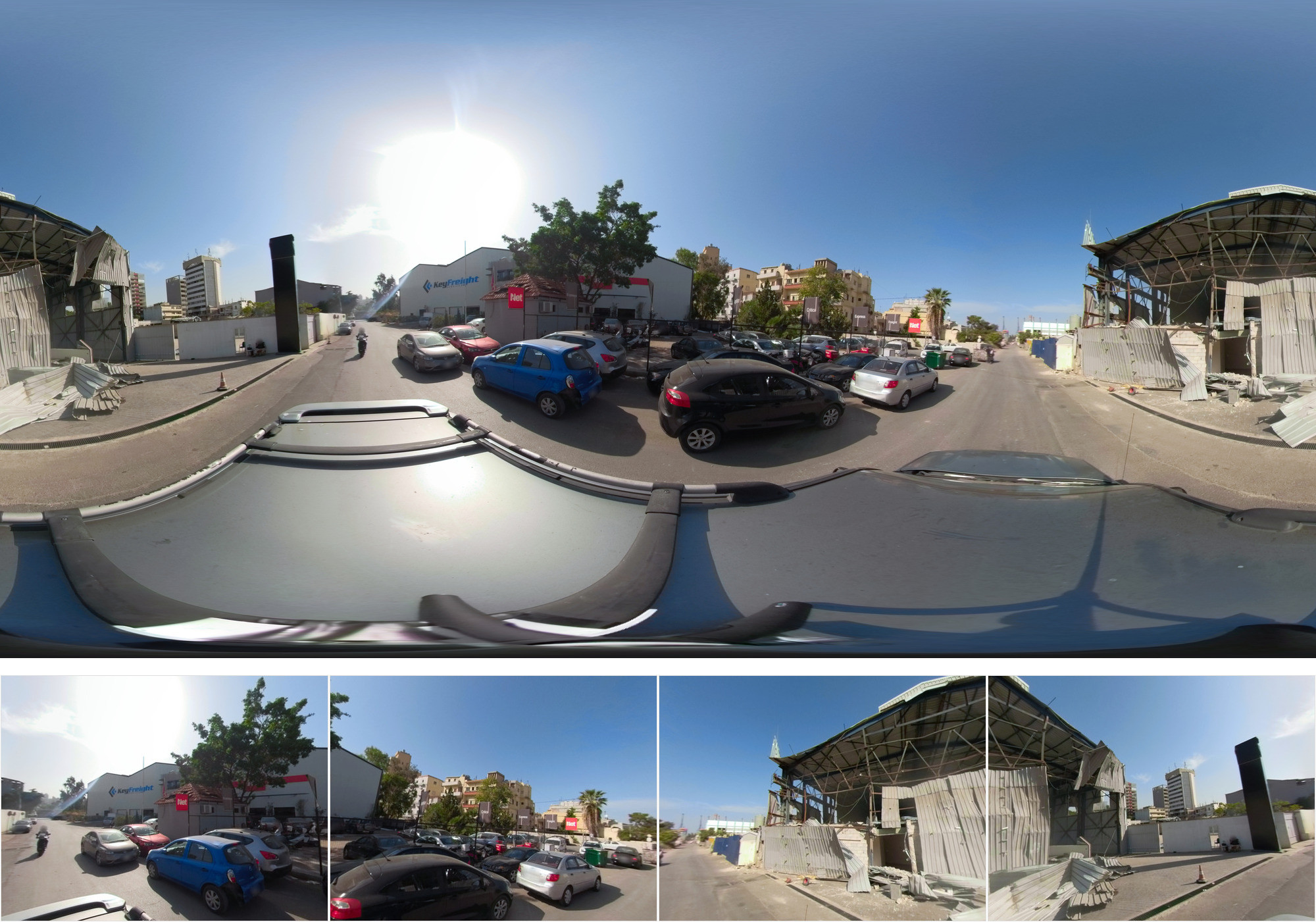}
    \caption{An equirectangular photosphere image planarized into four views. Views to the left depict no damage, while views to the right include significant damage to a sheet metal warehouse. All images were provided by Mapillary.com.}
    \label{fig:data-planarized}
\end{figure}

The second and final set of planar image extractions shifted the yaws by 45° (45°, 135°, 225°, 315°) and decreased the pitch to 10°. Each planar image was rendered at 1600$\times$1200 pixels to enable fine detail resolution. These adjustments improved visibility of the nearby buildings. Figure~\ref{fig:data-planarized} provides an example of how an equirectangular image with a damaged warehouse is split into four planar views. While the warehouse in the original image is split between the left and right edges of the equirectangular image, the front-right and back-right planar images present contiguous views that can be processed more easily by image analysis algorithms.

\subsubsection{Damage Indicator Catalog}
We manually inspected approximately 100 random images from the Mapillary dataset to gain a sense of what could be observed in the pictures and generate a list of features that could reliably serve as damage indicators for this situation. We determined that there are multiple confounding factors in this work. First, many of the buildings in the region have pre-existing damage from decades of regional conflict. Mapillary did not have enough images available from before the blast to determine whether damage was old or new. Second, most images were captured a few weeks to a few months after the blast. The impact of this delay is that several common damage indicators (e.g., smoke, fires, emergency workers, human casualties, etc.) are not present in the image, but signs of longer-term cleanup operations (e.g., piles of debris, construction, safety barriers, etc.) are visible. Third, many buildings have open windows due to the warm climate, making it difficult for vision algorithms to correctly discern between open and broken windows. Finally, Beirut has routine construction that is unrelated to the blast.

We selected the items listed in Table ~\ref{tab:damage-indicators} to serve as our key damage indicators when analyzing images. Given the confounding factors of the observations, damage assessments require multiple concurrent indicators rather than relying on single features. The situational limitations necessitate robust uncertainty quantification when converting visual damage classifications to quantitative overpressure estimates, as addressed through the learned spread parameter $\sigma_{\text{dex}}$ in the Bayesian framework.

\begin{table}[ht!]
\centering
\caption{Identified damage indicators and reliability}
\label{tab:damage-indicators}
\begin{tabular}{lp{8cm}}
\toprule
\textbf{Indicator} & \textbf{Reliability Notes} \\
\midrule
Broken windows with visible glass & High reliability; distinguish from open windows \\
Debris piles with masonry/glass & Moderate; requires distinguishing from routine garbage \\
Damaged vehicles (crushed roofs) & High reliability; exclude general disrepair \\
Construction barriers/safety tape & High reliability when outside damaged structures \\
Crushed roll-up garage doors & High reliability; specific to blast overpressure \\
Boarded-up storefronts & Moderate; may predate blast in conflict-affected areas \\
Missing/collapsed walls & Very high reliability; unambiguous structural failure \\
Displaced building materials & High reliability when coupled with other indicators \\
\bottomrule
\end{tabular}
\end{table}

\subsubsection{Test Suite Composition}
We assembled four test suites of images to support our evaluation of how well vision LLMs could assess damage from images:

\textbf{Suite 1: Positive Damage Examples (23 images)}
For the first suite we hand selected 23 images from the Mapillary dataset that provide clear examples of each type of damage indicator. Each image was labeled with 18 different visual indicators of damage (e.g., broken window, missing wall, pile of ruble, damaged car, construction barrier, etc). These images and labels were used for precision and recall assessment.

\textbf{Suite 2: Negative Damage Examples (25 images)}
The second suite of examples selected 25 images of cities without visible signs of damage, including Beirut (pre-blast), New York City, and Albuquerque. These images were used to assess false positive rates.

\textbf{Suite 3: Distributed Spatial Sampling (120 images)}
To address concerns that the Beirut Mapillary image locations were not evenly distributed, we partitioned the city into a 500 m x 500 m grid and picked a random sample from each grid cell for the third suite of examples. This sampling ensures representative geographic coverage.

\textbf{Suite 4: GEER Ground-Truth Locations (616 planar views)}
The dataset associated with the GEER report~\citep{geer70} includes a list of 154 structural assessments that were made for buildings in the city after the blast. Our fourth test suite selects the photosphere image that is closest to each assessment and uses its four planar images to represent the location. This suite enables expert-generated labels to be associated with images.

\subsection{Metadata Tables}

\subsubsection{Mapillary Image Metadata}
We constructed a single Parquet file to hold all the metadata associated with the Mapillary images that have valid location information. The majority of the metadata fields are identical to those retrieved from Mapillary's metadata API. The fields are summarized in Table~\ref{tab:mapillary-metadata-full}.

\begin{table}[ht!]
    \centering
    \caption{Complete metadata fields for Mapillary images}
    \label{tab:mapillary-metadata-full}
    \begin{tabular}{llp{6cm}} 
        \toprule
        \textbf{Field} & \textbf{Type} & \textbf{Description} \\
        \midrule
id & String & Unique ID assigned by Mapillary (alphanumeric) \\
sequence & String & Track identifier for images from one upload session \\
computed\_geometry & GeoJSON & Point geometry with longitude/latitude in WGS84 \\
computed\_compass\_angle & Float & Camera heading in degrees (0--360, clockwise from north) \\
computed\_altitude & Float & Altitude estimate in meters above sea level \\
captured\_at & ISO DateTime & UTC timestamp when image was originally captured \\
creator & Object & Username and numeric user ID of image contributor \\
is\_pano & Boolean & True if 360° photosphere, False if planar image \\
thumb\_original\_url & URL & Link to 320px thumbnail on Mapillary CDN \\
Longitude & Float & Extracted from computed\_geometry for convenience \\
Latitude & Float & Extracted from computed\_geometry for convenience \\
Username & String & Extracted from creator object \\
File & Path & Local filesystem path to downloaded full-resolution image \\
Distance\_m & Float & Computed haversine distance to explosion epicenter \\
\bottomrule
    \end{tabular}
\end{table}

\subsection{Data Attribution and Availability}

\textbf{Seismic Data}: Available from IRIS Data Management Center (\url{http://ds.iris.edu/}) under open data policies of contributing networks.

\textbf{Satellite Imagery}: 
\begin{itemize}
\item Maxar/Digital Globe: Open Data Program (\url{https://www.maxar.com/open-data})
\item Sentinel-1/2: Copernicus Open Access Hub (\url{https://scihub.copernicus.eu/})
\item ICEYE: Commercial license (contact for access)
\end{itemize}

\textbf{SAR Damage Proxy Map}: Sudhaus et al. dataset available at Zenodo (\url{https://zenodo.org/records/4762436}) under CC BY 4.0 license~\cite{sudhaus2021damage}.

\textbf{Ground Imagery}: Mapillary platform (\url{https://www.mapillary.com}) requires attribution in publications. We gratefully acknowledge Mapillary and the American University of Beirut researchers for making this data publicly available.

\textbf{Code and Processed Data}: Analysis code and intermediate data products will be deposited in a Sandia National Laboratories repository upon publication, with DOI assignment for reproducibility.

\clearpage

% Forward models
% !TEX root = main.tex
\section{Additional Information on Forward Models}
\label{sec:app-forward}

\begin{table}[h]
\centering
\caption{Kingery-Bulmash coefficients for incident overpressure (hemispherical surface burst). The numerical values are reproduced from \citep{swisdak1994simplified}.}
\label{tab:kb_coefficients}
\small
\begin{tabular}{cccccc}
\toprule
$Z_{\text{en}}$ Range (ft/lb$^{1/3}$) & $A$ & $B$ & $C$ & $D$ & $E$ \\
\midrule
$[0.5, 7.25)$ & 6.914 & $-1.439$ & $-0.282$ & $-0.142$ & 0.069 \\
$[7.25, 60.0)$ & 8.831 & $-3.700$ & 0.271 & 0.073 & $-0.013$\\
$[60.0, 500.0]$ & 5.424 & $-1.407$ & 0 & 0 & 0\\
\bottomrule
\end{tabular}
\end{table}

\subsection{Seismic Neural Network Architecture and Feature Engineering}

\subsubsection*{Feature Extraction from Regional Waveforms}

We extract 250 physics-motivated features from raw seismic waveforms recorded at regional distances (100--500 km). The feature extraction preserves absolute amplitude information critical for yield estimation, as seismic moment scales directly with explosive energy. From each waveform, we compute features across multiple domains:

\paragraph{Time-Domain Features} We extract peak amplitude statistics including the absolute maximum amplitude on the log$_{10}$ scale ($\log_{10}|x_{\max}|$), which serves as the primary amplitude scaling factor. Additional metrics include mean, standard deviation, skewness, kurtosis, and the 95th and 99th percentiles of absolute amplitudes. The time of maximum amplitude (normalized by trace length) captures the relative arrival timing of peak energy.

\paragraph{P-Wave Band Features} A 0.5--5 Hz Butterworth bandpass filter isolates regional P-waves, from which we compute the filtered maximum amplitude, RMS energy, and ratio to unfiltered maximum. The P-wave energy on the log$_{10}$ scale ($\log_{10} \sum x_p^2$) provides a band-limited energy measure less sensitive to noise than broadband measurements.

\paragraph{Frequency-Domain Features} The power spectral density is computed via FFT, from which we extract the dominant frequency, spectral centroid, and power distribution across four bands: 0.1--0.5 Hz (very low), 0.5--1.0 Hz (low P-wave), 1.0--5.0 Hz (standard P-wave), and 5.0--10.0 Hz (high frequency). Each band contributes both absolute power (log$_{10}$ scale) and relative power (fraction of total).

\paragraph{Coda Features} We separate P-wave and coda windows, with the P-wave window extending 5 seconds or 30\% of the trace length (whichever is smaller), and coda beginning 2 seconds after the P-wave window. The P-to-coda amplitude ratio on the log scale captures the partitioning between body and scattered wave energy. Coda decay rate is estimated via linear regression in log-envelope space, providing a measure of attenuation.

\paragraph{Envelope Features} The Hilbert transform provides the analytic signal envelope, from which we extract maximum, mean, and standard deviation. Signal duration is defined as the time span where the envelope exceeds 10\% of its maximum. Energy distribution metrics (t$_{50}$ and t$_{90}$) quantify the temporal concentration of seismic energy.

\subsubsection{Neural Network Architecture}
After removing constant features (standard deviation $< 10^{-6}$), we retain 46 informative features that exhibit variance across the dataset. The neural network employs a fully-connected architecture with four hidden layers of dimensions [512, 256, 128, 64], incorporating batch normalization after each linear transformation and ReLU activations. Dropout with rate 0.3 is applied between layers for regularization.

The network is physics-informed through a dual-head architecture: the primary head predicts $\log_{10}(M_0)$ directly, while an auxiliary head predicts moment magnitude $M_w$. During training, we enforce the physical constraint $M_w = \frac{2}{3}\log_{10}(M_0) - 6.07$ (for $M_0$ in N·m) through a composite loss function:
\begin{equation}
\mathcal{L} = \mathcal{L}_{\text{data}}(\log_{10} M_0^{\text{pred}}, \log_{10} M_0^{\text{true}}) + 0.1  \mathcal{L}_{\text{physics}}(M_w^{\text{pred}}, M_w^{\text{physics}})
\end{equation}
where $M_w^{\text{physics}} = \frac{2}{3}\log_{10} M_0^{\text{true}} - 6.07$ ensures physically consistent predictions.

\subsubsection{Ensemble Training for Variance Reduction}

To reduce prediction uncertainty inherent in single neural networks, we train an ensemble of five models with identical architectures but different random initialization seeds. Each model is trained on the complete dataset rather than bootstrap samples, allowing all models access to the full training distribution while maintaining diversity through different weight initializations and stochastic gradient descent trajectories.

The ensemble prediction for the Beirut explosion yields $\log_{10}(M_0) = 15.85 \pm 0.20$ (N·m), corresponding to $M_w = 4.50 \pm 0.13$. The standard deviation across ensemble members (0.20 in log$_{10}$ space) represents approximately 60\% relative uncertainty in linear $M_0$, substantially lower than single-model predictions which exhibit standard deviations of 0.3--0.4 in log$_{10}$ space.

\begin{figure}[htbp]
    \centering
    \includegraphics[width=\linewidth]{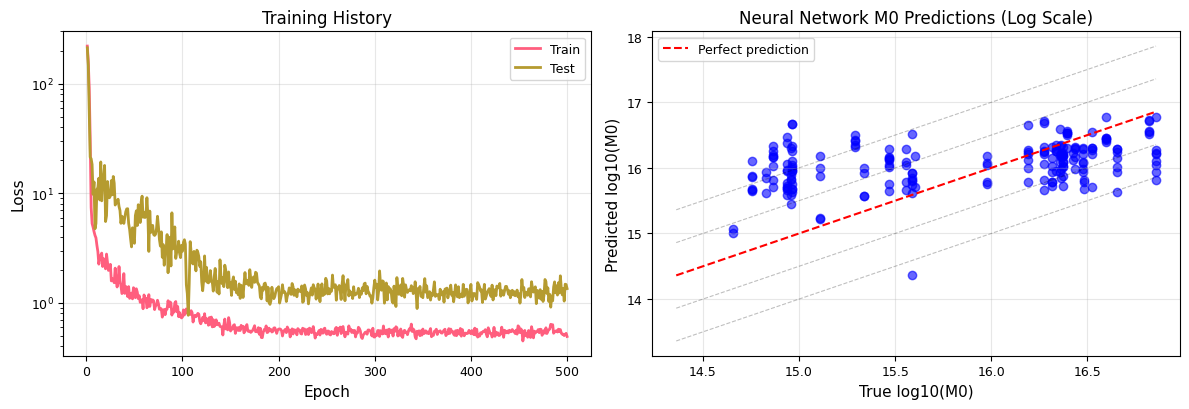}
    \caption{Neural network ensemble member performance for seismic yield estimation. (a) Training history showing loss convergence over 500 epochs for both training and test sets from a representative model, demonstrating effective learning without overfitting. (b) Predicted versus true log$_{10}$(M$_0$) values from a single ensemble member showing good agreement across the full range of seismic moments in the training dataset. The dashed red line indicates perfect prediction.}
    \label{fig:seismic-nn-performance}
\end{figure}

\subsubsection{Training Configuration}
The model is trained using the Adam optimizer with learning rate $10^{-3}$ and gradient clipping at norm 1.0. We employ GroupKFold cross-validation \citep{pedregosa2011scikit} during model development to prevent data leakage, ensuring events from the same explosion do not appear in both training and validation sets. Feature scaling uses RobustScaler \citep{pedregosa2011scikit} to handle outliers common in seismic data. Each ensemble member trains for 250 epochs with batch size 32, using early stopping based on validation loss with patience of 40 epochs.

Cross-validation on the mixed chemical/nuclear test dataset yields a mean absolute error of 0.30 orders of magnitude in $M_0$ prediction (90th percentile error: 0.45 orders). The model exhibits some systematic bias for surface chemical explosions, with correlation between key features and $\log_{10}(M_0)$ remaining moderate (e.g., absolute amplitude: $r = 0.52$, coda decay rate: $r = 0.14$), reflecting the challenge of generalizing from underground nuclear tests to surface chemical explosions.

\begin{figure}[htbp]
    \centering
    \includegraphics[width=\linewidth]{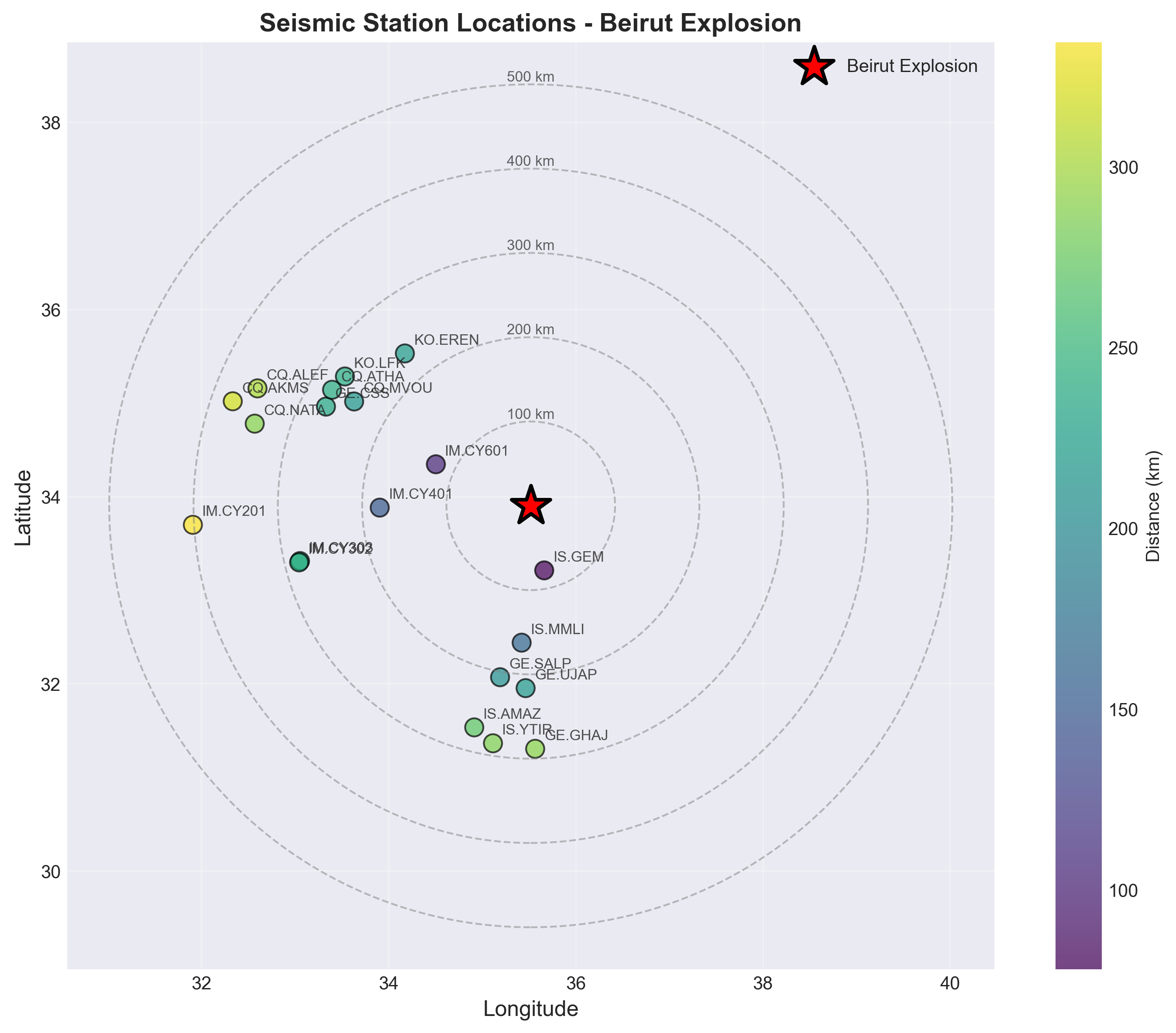}
    \caption{Regional seismic station locations for Beirut analysis. Red star marks the explosion epicenter; circles show 100 km distance intervals. Station colors indicate distance from epicenter, with networks (IS: International Monitoring System, GE: GEOFON, IM: Cyprus) providing comprehensive azimuthal coverage.}
    \label{fig:beirut-seismic-stations}
\end{figure}

\clearpage 
\subsection{Crater Geometry}
\label{sec:app-crater-geom}
% Indu/Steve to write. 
The crater was estimated as an ellipse with dimensions of 46.7 meters in width and 108.1 meters in length, with an aspect ratio of 2.3. The significant departure from circular geometry is likely due to the warehouse geometry and directional effects from the ammonium nitrate storage geometry. 

\begin{figure}[htbp]
    \centering
    \includegraphics[width=0.4\textwidth]{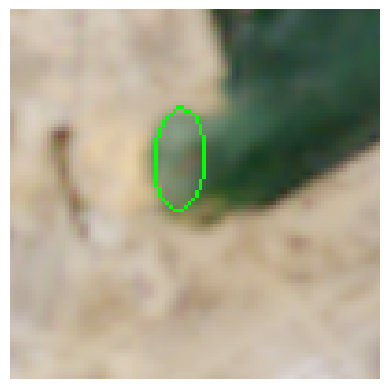}
    \caption{Post-Sentinel-2 colored satellite image taken on August 8, 2020 with the approximated crater geometry overlaid in green. The crater geometry was approximated using the Segment Anything Model. }
    \label{fig:crater-geometry-estimate}
\end{figure}

The crater geometry was estimated based on a post-Sentinel-2 colored satellite image capturing the blast region on August 8, 2020.  The outline of the crater was estimated based on visible features indicating the blast line using Meta's Segment Anything Model (SAM)~\citep{Kirillov23} with a user specified reference point at the center of the blast. The model approximated the underwater portion of the crater based on variations in water color that likely correspond to the crater depth. The contour was approximated as an ellipse using a linear conic fitting method. We note that the estimated crater depth along the major axis is similar to an estimate provided in \cite{sadek2022impacts} based on an in-person survey.  Figure~\ref{fig:crater-geometry-estimate} shows the satellite image zoomed in on the blast crater region with the outline of the estimated ellipse overlaid.  

\clearpage 

\subsection{SAR Architecture}
% Steve to write. \lp{A paragraph or so on the SAR architecture, e.g. how the images were processed, the SpikeAD algorithm, how this led to the modeling etc. Please put all details here not in the main text of the paper. }

\begin{figure}[htbp]
    \centering
    \includegraphics[width=\linewidth, trim = 3.0cm 6cm 3.0cm 6cm, clip]{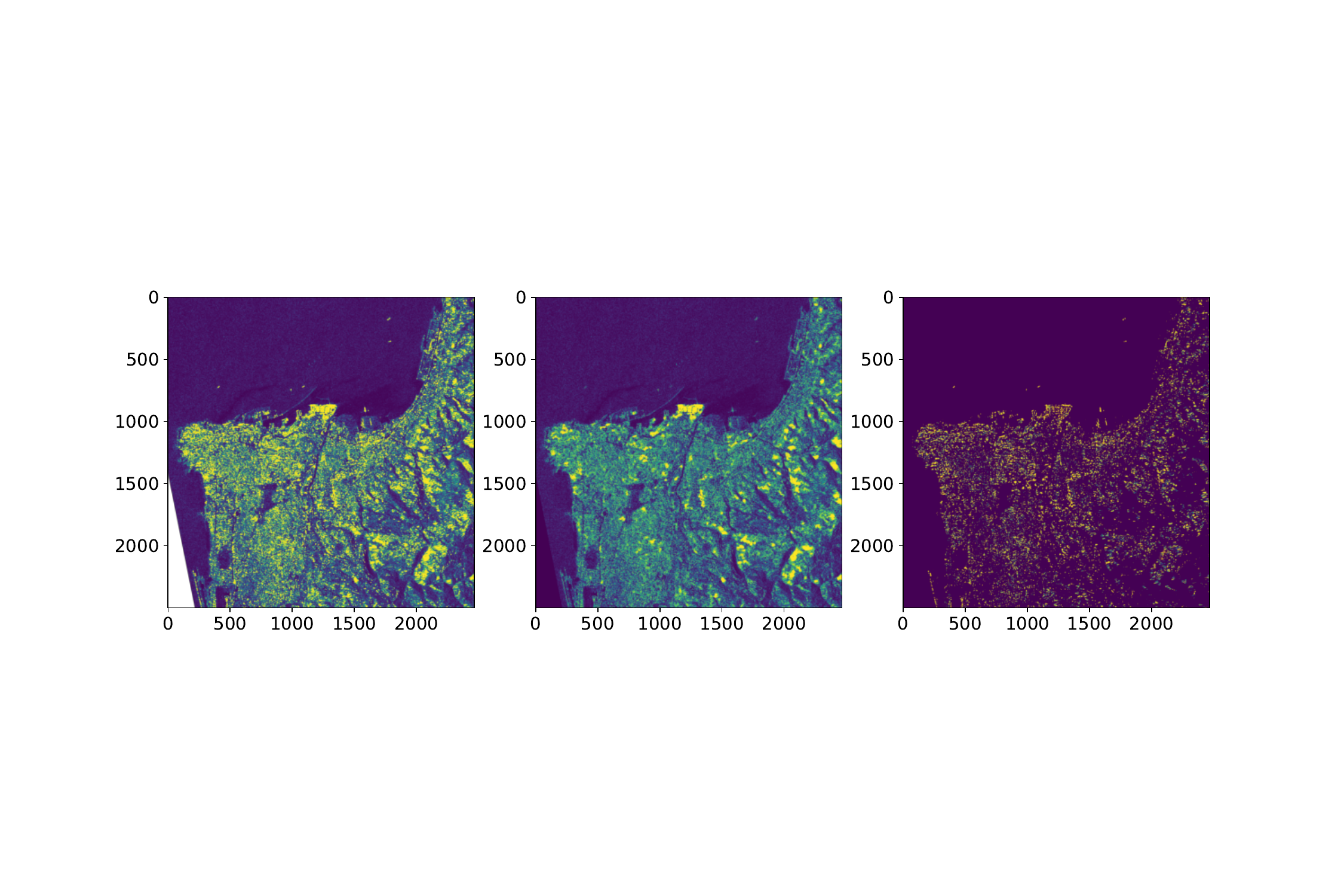}
    \caption{Sentinel-1 SAR imagery from August 5, 2020: (left) original acquisition, (middle) despeckled via SpikeAD, (right) difference highlighting modified pixels. The algorithm predominantly affects isolated speckle rather than structured damage regions, validating the approach.}
    \label{fig:spike_ad_sar_comparison}
\end{figure}

% \JRcomment{Lekha, Verzi: Can this para be extended to include how multiple coherence maps are composited together using ``Pilger logic'' that we use?} 
% \SVcomment{We should move the following paragraph and figure to the appendix.}
We apply SpikeAD with 4 iterations over $11 \times 11$ pixel neighborhoods, providing aggressive despeckling while maintaining damage region boundaries. Figure~\ref{fig:spike_ad_sar_comparison} demonstrates that this procedure effectively reduces speckle noise which is visible as isolated bright pixels in the original acquisition, while preserving contiguous damage signatures, with modifications concentrated in isolated locations rather than structured damage regions.

Figure~\ref{fig:spike_ad_sar_comparison} shows how SpikeAD can be used spatially to remove speckle and noise, especially in the water, while enhancing the contrast of some features, such as the port area in the center. SpikeAD is a smoothing technique (employing "intelligent" median-filtering), but this means that some signal might get smoothed out.
This selective modification proves particularly effective for SAR speckle, where noise manifests as isolated high-intensity pixels rather than spatially coherent patterns. 
% \SVcomment{Previous sentence is excellent, but smoothing on top of smoothing caused the yield estimate uncertainty to increase too much!}

\begin{figure}[htbp]
    \centering
    \includegraphics[width=\textwidth]{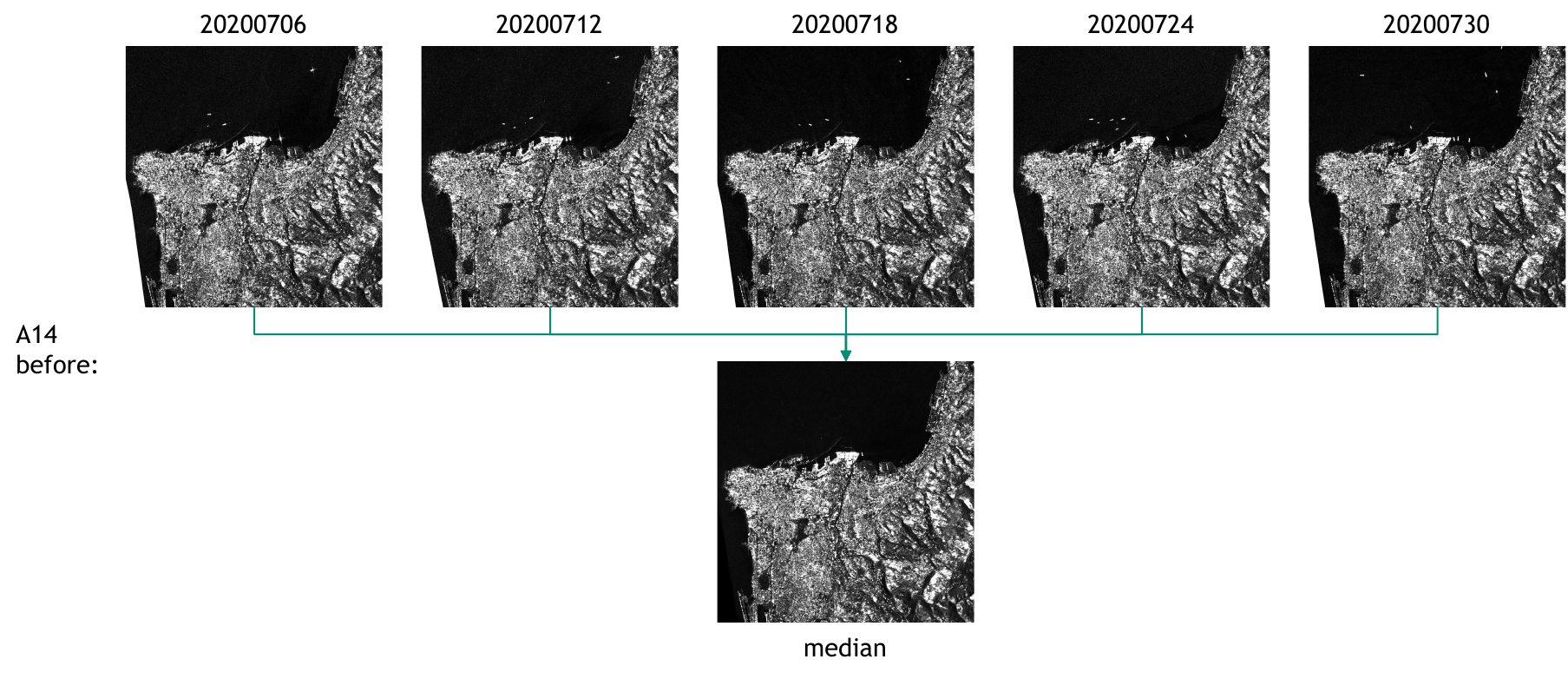}
    \caption{Sentinel-1 SAR imagery from ascending relative orbit 14 for July, 2020: (top) original acquisition, (bottom) despeckled via SpikeAD. The algorithm predominantly affects isolated speckle rather than structured damage regions (such as those in the water), validating the approach.}
    \label{fig:spike_ad_sar_july_2020_a14}
\end{figure}

One goal for using SpikeAD here was to maintain as much original signal from the SAR images while removing the speckle noise. We also wanted to characterize what had changed from before to after the event (the Beirut blast in this instance). One way to use SpikeAD here is across time. In this research, SpikeAD is used temporally instead of spatially, as it is typically employed (Figure~\ref{fig:spike_ad_sar_comparison}). For use of SpikeAD across time, we employ 1x1 neighborhoods (i.e., each pixel is handled independently across time). Figure~\ref{fig:spike_ad_sar_july_2020_a14} shows how SpikeAD is used to generate a composite image (bottom) for July, 2020 for ascending relative orbit 14 (A14) images (top).

\begin{figure}[htbp]
    \centering
    \includegraphics[width=\textwidth]{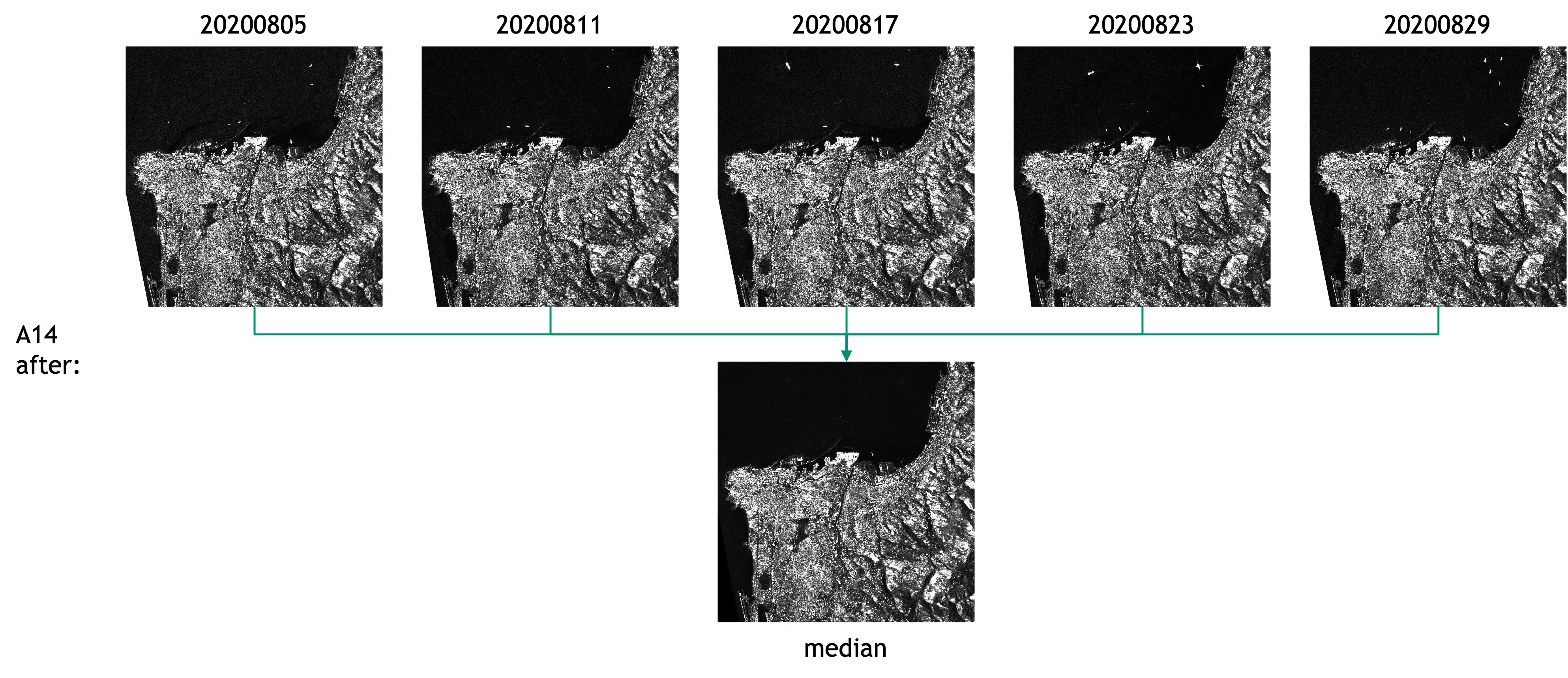}
    \caption{Sentinel-1 SAR imagery from ascending relative orbit 14 for August, 2020: (top) original acquisition, (bottom) despeckled via SpikeAD. .}
    \label{fig:spike_ad_sar_august_2020_a14}
\end{figure}

\begin{figure}[htbp]
    \centering
    \includegraphics[height=0.5\textheight]{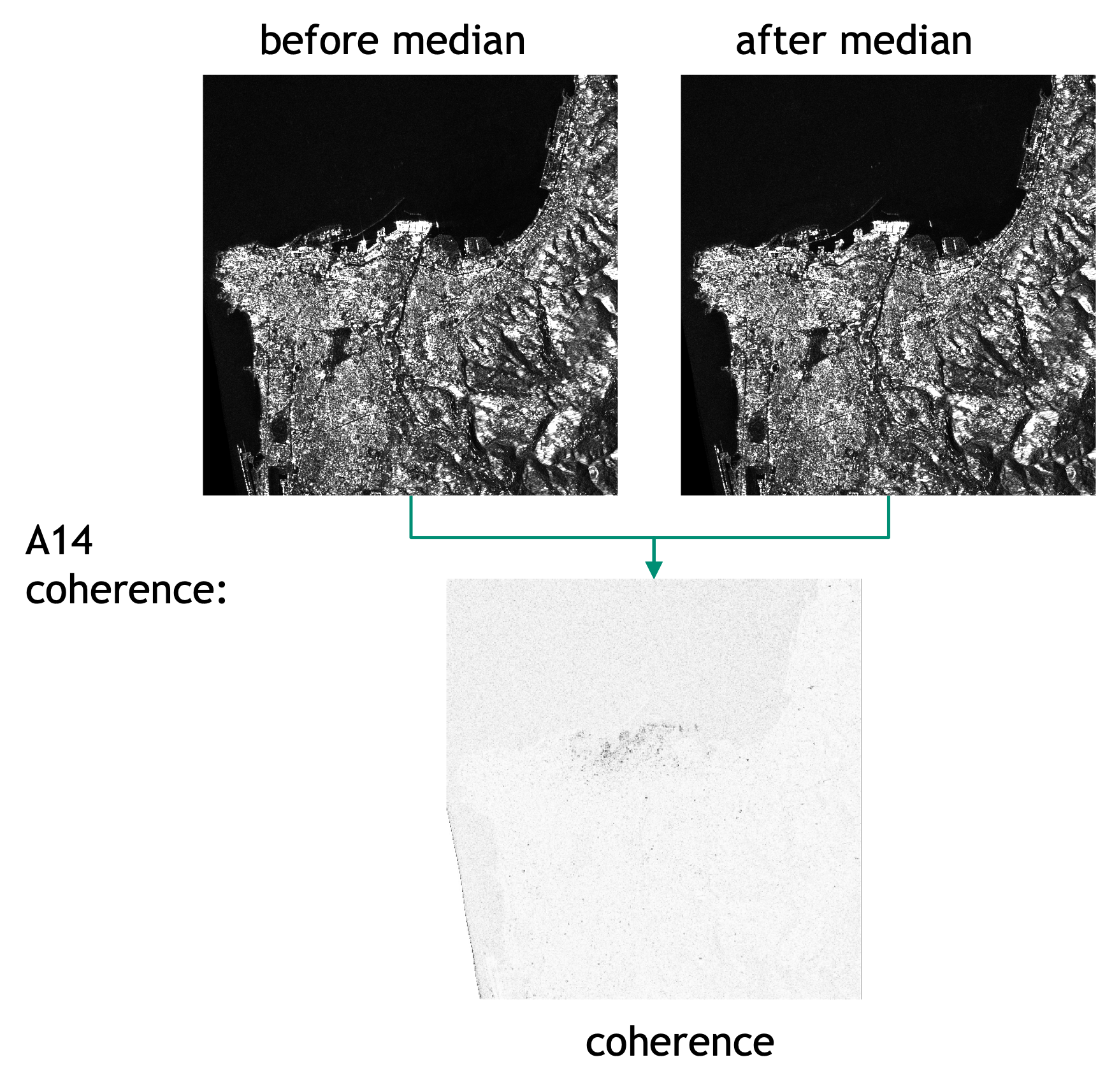}
    \caption{Coherence image (bottom) for ascending relative orbit 14 between July (top-left) and August (top-right), 2020.}
    \label{fig:spike_ad_coherence_a14}
\end{figure}

\begin{figure}[htbp]
    \centering
    \includegraphics[width=\textwidth]{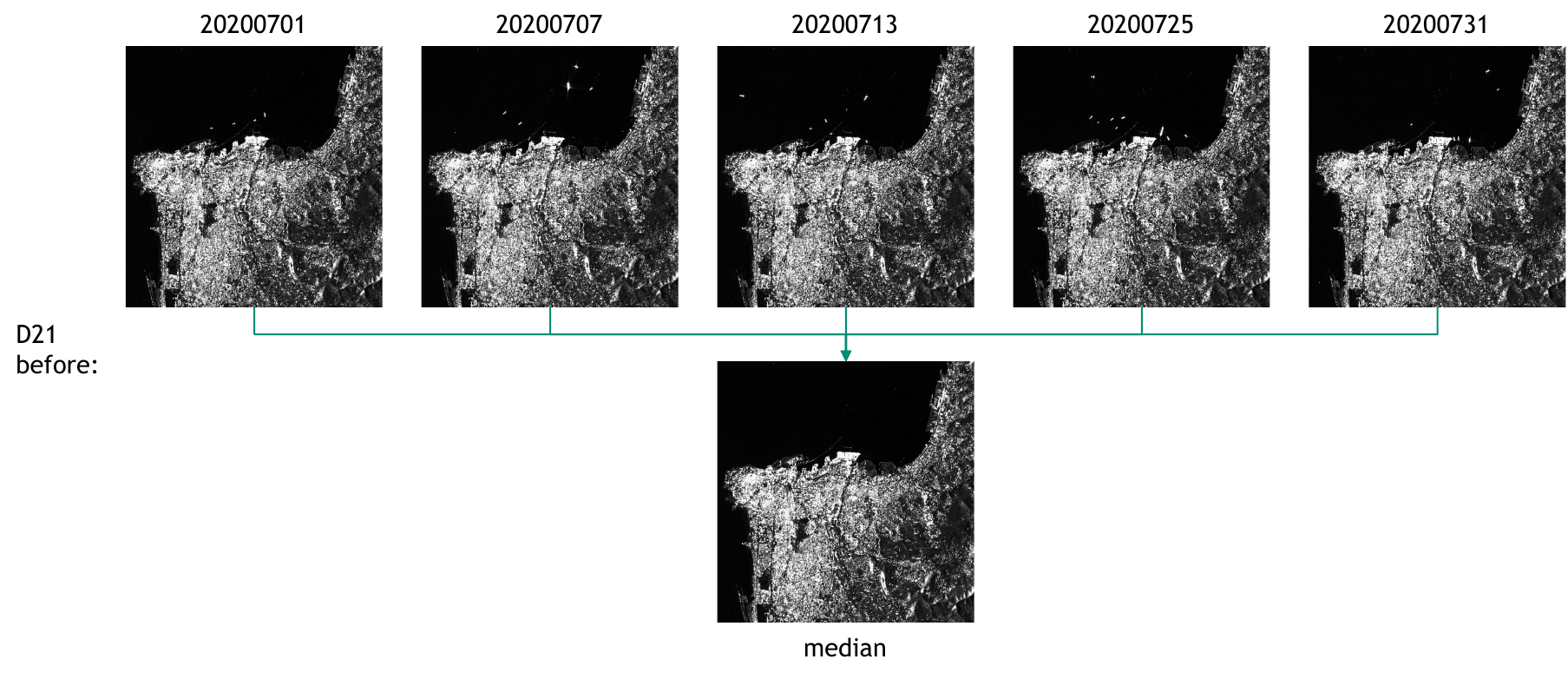}
    \caption{Sentinel-1 SAR imagery from descending relative orbit 21 for July, 2020: (top) original acquisition, (bottom) despeckled via SpikeAD.}
    \label{fig:spike_ad_sar_july_2020_d21}
\end{figure}

\begin{figure}[htbp]
    \centering
    \includegraphics[width=\textwidth]{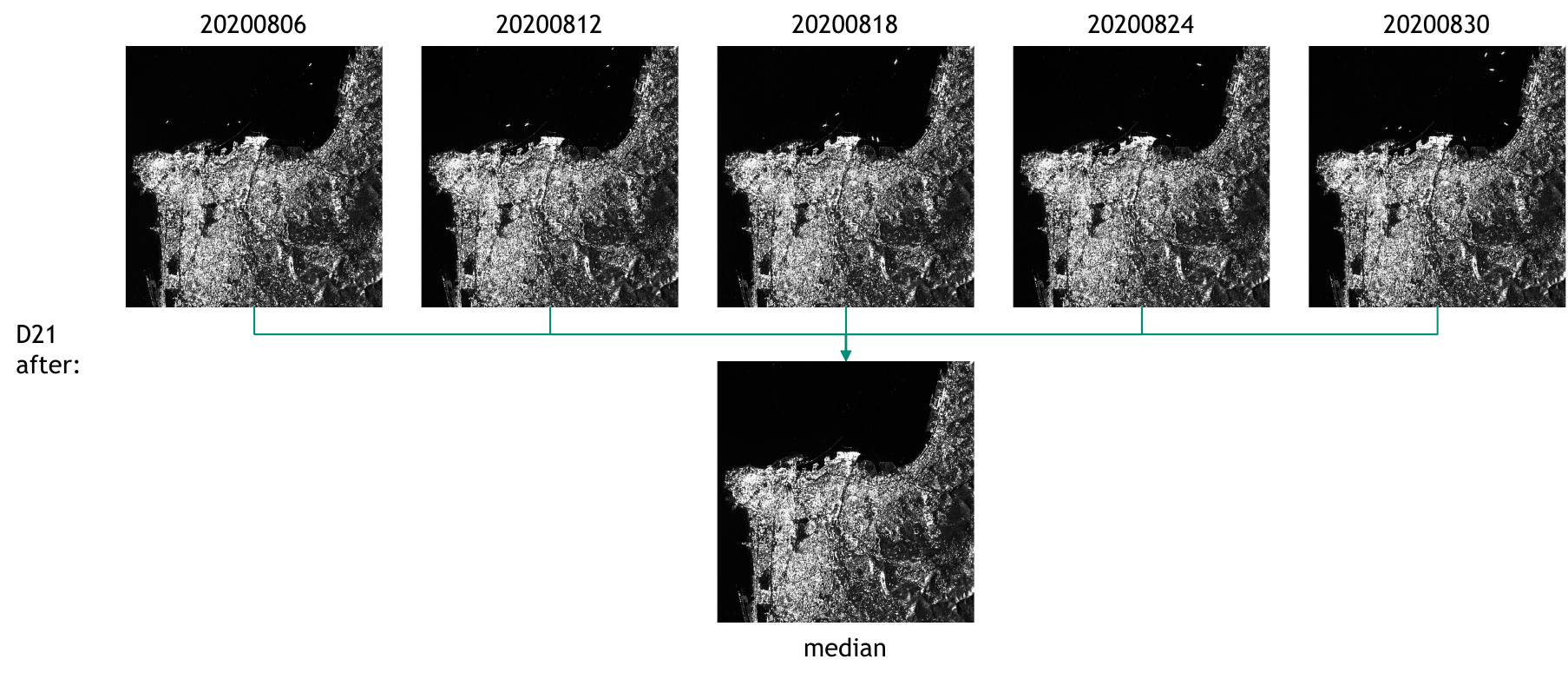}
    \caption{Sentinel-1 SAR imagery from descending relative orbit 21 for August, 2020: (top) original acquisition, (bottom) despeckled via SpikeAD. .}
    \label{fig:spike_ad_sar_august_2020_d21}
\end{figure}

\begin{figure}[htbp]
    \centering
    \includegraphics[height=0.5\textheight]{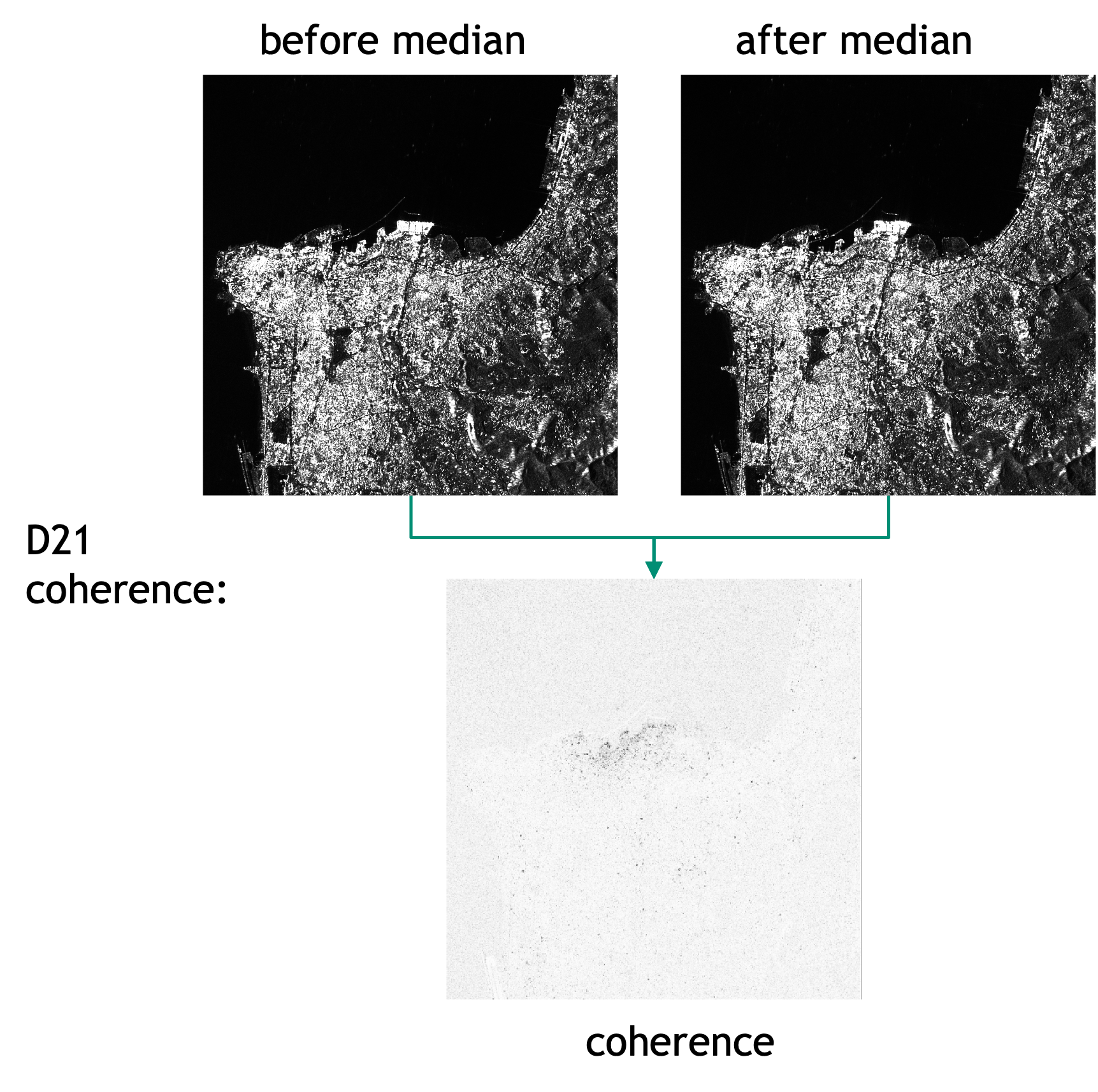}
    \caption{Coherence image (bottom) for descending relative orbit 21 between July (top-left) and August (top-right), 2020.}
    \label{fig:spike_ad_coherence_d21}
\end{figure}

\begin{figure}[htbp]
    \centering
    \includegraphics[width=\textwidth]{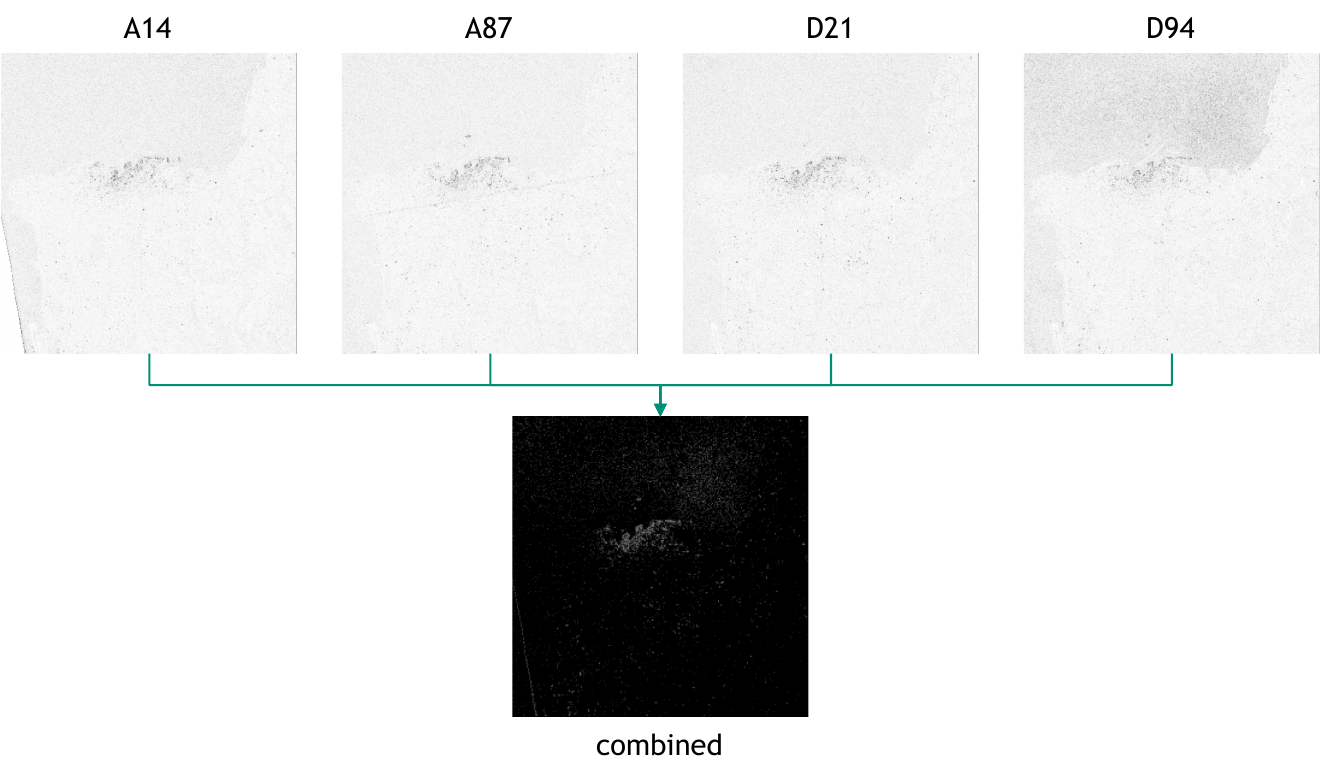}
    \caption{Combined damage map.}
    \label{fig:spike_ad_combined_damage_map}
\end{figure}

Figure~\ref{fig:spike_ad_sar_august_2020_a14} shows SpikeAD applied to A14 SAR images for August, 2020.
Now the coherence image for A14 is computed using these two composite images (bottom image in Figure~\ref{fig:spike_ad_sar_july_2020_a14} and bottom image in Figure~\ref{fig:spike_ad_sar_august_2020_a14}), which can be seen int Figure~\ref{fig:spike_ad_coherence_a14}.
%
% Similar to the process described above for A14, the following three images show how SpikeAD is used to compute the coherence for ascending relative orbit 87 (A87), see Figure~\ref{fig:spike_ad_sar_july_2020_a87}, Figure~\ref{fig:spike_ad_sar_august_2020_a87} and Figure~\ref{fig:spike_ad_coherence_a87}.
The process is repeated for ascending relative orbit 87 (A87; figures not shown) as well as descending relative orbit 21 (D21; Figures~\ref{fig:spike_ad_sar_july_2020_d21}, \ref{fig:spike_ad_sar_august_2020_d21} and \ref{fig:spike_ad_coherence_d21}) and descending relative orbit 94  (D94, figures not shown).
%
% \begin{figure}[htbp]
%     \centering
%     \includegraphics[width=\textwidth]{figures/DeSpeckled_SAR_A87_July2020.pdf}
%     \caption{Sentinel-1 SAR imagery from ascending relative orbit 87 for July, 2020: (top) original acquisition, (bottom) despeckled via SpikeAD.}
%     \label{fig:spike_ad_sar_july_2020_a87}
% \end{figure}
%
% \begin{figure}[htbp]
%     \centering
%     \includegraphics[width=\textwidth]{figures/DeSpeckled_SAR_A87_August2020.pdf}
%     \caption{Sentinel-1 SAR imagery from ascending relative orbit 87 for August, 2020: (top) original acquisition, (bottom) despeckled via SpikeAD. .}
%     \label{fig:spike_ad_sar_august_2020_a87}
% \end{figure}
%
% \begin{figure}[htbp]
%     \centering
%     \includegraphics[width=\textwidth]{figures/coherence_computation_A87.png}
%     \caption{Coherence image (bottom) for ascending relative orbit 87 between July (top-left) and August (top-right), 2020.}
%     \label{fig:spike_ad_coherence_a87}
% \end{figure}
%
% The process for descending relative orbit 21 (D21) is shown in Figure~\ref{fig:spike_ad_sar_july_2020_d21}, Figure~\ref{fig:spike_ad_sar_august_2020_d21} and Figure~\ref{fig:spike_ad_coherence_d21}.
%
%
% The process for descending relative orbit 94 (D94) is shown in Figure~\ref{fig:spike_ad_sar_july_2020_d94}, Figure~\ref{fig:spike_ad_sar_august_2020_d94} and Figure~\ref{fig:spike_ad_coherence_d94}.
%
% \begin{figure}[htbp]
%     \centering
%     \includegraphics[width=\textwidth]{figures/DeSpeckled_SAR_D94_July2020.pdf}
%     \caption{Sentinel-1 SAR imagery from descending relative orbit 94 for July, 2020: (top) original acquisition, (bottom) despeckled via SpikeAD.}
%     \label{fig:spike_ad_sar_july_2020_d94}
% \end{figure}
%
% \begin{figure}[htbp]
%     \centering
%     \includegraphics[width=\textwidth]{figures/DeSpeckled_SAR_D94_August2020.pdf}
%     \caption{Sentinel-1 SAR imagery from descending relative orbit 94 for August, 2020: (top) original acquisition, (bottom) despeckled via SpikeAD. .}
%     \label{fig:spike_ad_sar_august_2020_d94}
% \end{figure}
%
% \begin{figure}[htbp]
%     \centering
%     \includegraphics[width=\textwidth]{figures/coherence_computation_D94.png}
%     \caption{Coherence image (bottom) for descending relative orbit 94 between July (top-left) and August (top-right), 2020.}
%     \label{fig:spike_ad_coherence_d94}
% \end{figure}
%
Finally, the combined damage map is computed from the 4 coherence images by aggregating them together (pixel-wise addition followed by normalization between 0 and 1) and then computing the damage percentage across 10x10 windows, see Figure~\ref{fig:spike_ad_combined_damage_map}.

Once the coherence maps are constructed and converted into $D \%$, they are subjected to a check to ensure that they resemble damage maps caused by a radially weakening blast wave. The $D \%$ map is divided into 15 annular zones centered at the blast epicenter, thus binning the damage aggregation boxes as a function of their epicentral distance. In  
% After annular percentile thresholding, approximately 150--200 high-damage boxes distributed across 15 zones provide the observations $\{D_{\%,i}, r_i\}$ for likelihood evaluation. 
Figure~\ref{fig:SAR-pixel-summary} in Appendix \ref{app:sar-processing} (left), in each bin/annular zone, we plot the ${\rm 95^{th}}$ percentile of the damage percent ($D\%$) as a function of the epicentral distance; it confirms physically consistent radial decay of damage percentage. This trend holds true for lower percentiles too (up to ${\rm 80^{th}}$), indicating that the effect of noise at these higher percentiles is muted. In Figure~\ref{fig:SAR-pixel-summary} (right), we plot the number of pixels of the coherence maps that are in each annual zone above the ${\rm 80^{th}, 90^{th}}$ and ${\rm 95^{th}}$ percentile thresholds and we see an approximately uniform spatial sampling across zones, preventing any single distance range from dominating the inference. The retained boxes typically exhibit damage percentages ranging from 40--100\%, concentrated within 2 km of the epicenter with gradual decrease to 10--30\% at 5 km distance.

% \SVcomment{Lekha or Jaideep, can you add a sentence describing the difficulties we had with the SpikeAD-based yield estimates?}
The despeckling process described above provided damage estimates that were too smooth in space, i.e., the decay in damage levels with distance from the epicenter was too shallow. Fitting an over-pressure model led to a yield estimate of about 5 kt which was too large to be credible. Consequently, we did not pursue the process any further for post-processing the SAR images.

\begin{figure}[htbp]
    \centering
    \includegraphics[width=\linewidth]{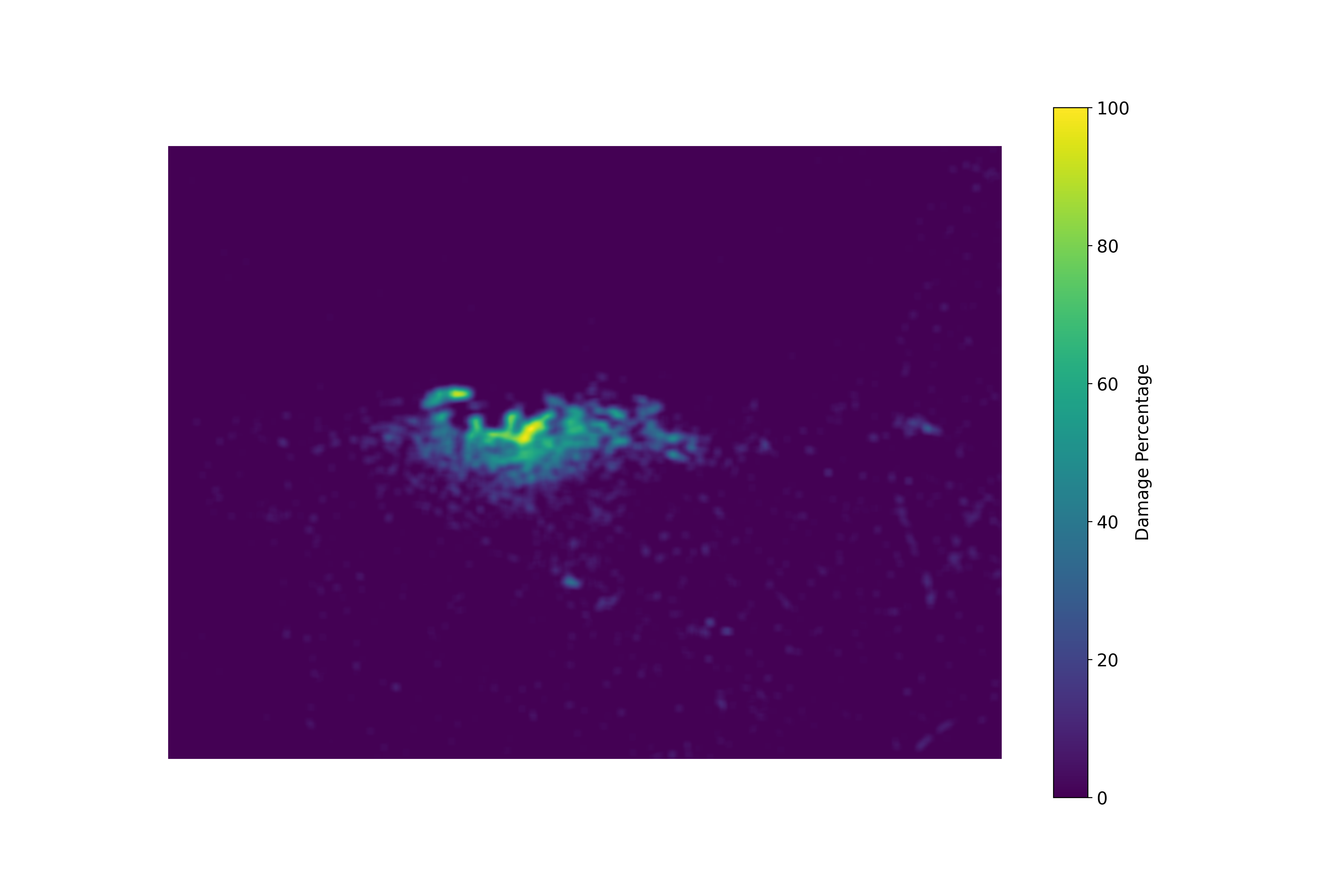}
    \caption{Composite damage map computed using the data from Pilger and colleagues~\citep{pilger2021beirut}. }
    \label{fig:SAR-composite-damage-map}
\end{figure}

\begin{figure}[htbp]
    \centering
    \includegraphics[width=\linewidth]{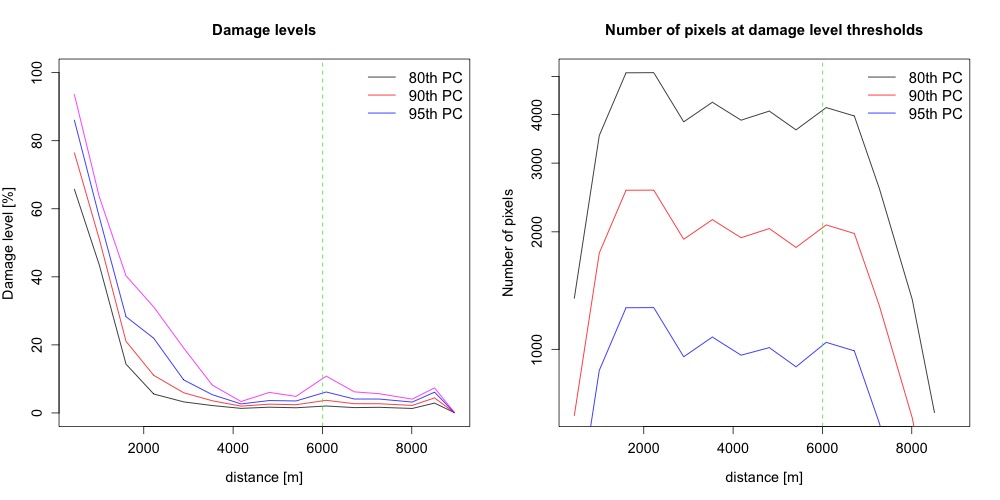}
    \caption{Spatial distribution after zonal thresholding. (Left) Radial decay of average damage at 80th, 90th, 95th percentile thresholds demonstrates physically consistent overpressure relaxation. Horizontal axis shows median distance of retained boxes in each zone. (Right) Number of retained boxes per zone confirms approximately uniform spatial sampling. PC denotes percentile.}
    \label{fig:SAR-pixel-summary}
\end{figure}

\clearpage
\subsection{Vision-Language Model Architecture for Damage Assessment}
\label{sec:app-vlm}

\subsubsection*{Model Selection and Variants}

We employed the Google Gemma-3 family of vision-language models for automated blast damage assessment from street-level imagery. Gemma-3 represents Google's latest open-weight multimodal models, incorporating the SigLIP vision encoder for joint processing of text and images. We systematically evaluated three parametric variants to assess the relationship between model capacity and damage assessment quality:

\begin{itemize}
\item \texttt{google/gemma-3-4b-it}: 4 billion parameters (baseline)
\item \texttt{google/gemma-3-12b-it}: 12 billion parameters (intermediate)
\item \texttt{google/gemma-3-27b-it}: 27 billion parameters (maximum capacity)
\end{itemize}

Each model employs a decoder-only transformer architecture with grouped-query attention and processes visual inputs through square images resized to 896$\times$896 pixels. The instruction-tuned variants (\texttt{-it}) were selected for their superior performance on structured output tasks. All models support a 128K token context window (except the 1B variant with 32K), enabling processing of extensive contextual information alongside visual inputs. For this analysis we constrained the output to the use of the 27B variant, given that while all models demonstrated reasonable potential to map damage categories to overpressure estimates, the 27B variant consistently produced an overpressure-distance relationship that better conformed to expected blast physics, exhibiting smoother spatial decay, reduced local variability, and improved adherence to theoretical power-law attenuation with distance from the epicenter.

\subsubsection*{MMLU-Style Probabilistic Assessment Framework}

Rather than generating deterministic classifications, we adapted the MMLU (Massive Multitask Language Understanding) evaluation methodology commonly used in language model benchmarking to produce probability mass functions (PMFs) over damage categories. This approach transforms visual damage assessment into a constrained single-token generation task, enabling rigorous uncertainty quantification.

The damage categorization framework consists of nine discrete categories (levels 0-8), each corresponding to specific overpressure ranges and observable structural effects:

\begin{table}[h]
\centering
\caption{Blast Damage Assessment Rubric for VLM Classification}
\label{tab:vlm-damage-rubric}
\small
\begin{tabular}{clc}
\toprule
\textbf{Level} & \textbf{Category} & \textbf{Pressure Range (psi)} \\
\midrule
0 & No blast damage & 0 \\
1 & Glass/Noise Effects & 0.04--0.15 \\
2 & Limited Structural Effects & 0.15--0.40 \\
3 & Window/Housing Damage & 0.40--1.0 \\
4 & Home Inhabitability Issues & 1.0--2.0 \\
5 & Significant Structural Collapse & 2.0--3.0 \\
6 & Major Structural Failures & 3.0--5.0 \\
7 & Near-Complete Destruction & 5.0--7.0 \\
8 & Catastrophic Destruction & $>$7.0 \\
\bottomrule
\end{tabular}
\end{table}

For each image $i$ and damage level $k \in \{0, 1, \ldots, 8\}$, we calculate the conditional log-likelihood of the model generating token ``k'' as the continuation:
\begin{equation}
\ell_{ik} = \log \mathbb{P}(\text{token}=\text{``k''} \mid \text{image}_i, \text{prompt})
\end{equation}

The raw log-likelihoods undergo numerical stabilization through the log-sum-exp transformation to yield properly normalized probabilities:
\begin{equation}
q_{ik} = \frac{\exp(\ell_{ik} - \max_j \ell_{ij})}{\sum_{m=0}^{8} \exp(\ell_{im} - \max_j \ell_{ij})}
\end{equation}

This produces a PMF $\mathbf{q}_i = \{q_{i0}, q_{i1}, \ldots, q_{i8}\}$ for each image, ensuring numerical stability while maintaining relative probability relationships between damage levels.

\subsubsection*{Prompt Engineering}

The prompt structure incorporates several key design elements to optimize damage assessment performance. Each prompt begins with a task specification establishing the model's role as a damage assessment system, followed by the image token placeholder which is replaced by the model's visual encoding. Metadata fields including capture date and GPS coordinates from the Mapillary dataset provide spatial and temporal context.

The damage level categories are explicitly enumerated with their corresponding pressure ranges, ensuring the model has complete information about the classification schema. Assessment criteria direct attention to structural elements most indicative of blast damage, including window breakage, wall collapse, roof damage, debris patterns, and overall structural integrity. The prompt concludes with the instruction ``Answer: '' followed by a space character, constraining the model to generate a single digit (0-8) as the next token.

This formulation enables calculation of log-likelihoods for each possible damage level in a single forward pass, providing computational efficiency for processing hundreds of images.

\subsubsection*{Uncertainty Quantification}

Classification uncertainty is quantified through Shannon entropy calculated over each PMF using natural logarithm:
\begin{equation}
H(\mathbf{q}_i) = -\sum_{k=0}^{8} q_{ik} \ln(q_{ik})
\end{equation}

% The entropy is transformed into a confidence score:
% \begin{equation}
% \text{conf}_i = \max\left(0, 1 - \frac{H(\mathbf{q}_i)}{H_{\max}}\right)
% \end{equation}
% where $H_{\max} = \ln(9) \approx 2.197$ is the maximum possible entropy for a uniform distribution over nine categories.

% This confidence score is then converted to a weight for robust regression:
% \begin{equation}
% w_i = \max(w_{\min}, \text{conf}_i^{\gamma})
% \end{equation}
% where $\gamma = 1.0$ is a tuning parameter and $w_{\min} = 0.05$ ensures a minimum weight for all observations. This formulation assigns higher weights to classifications with lower entropy (higher confidence) while maintaining a floor to prevent complete exclusion of uncertain observations.

\subsubsection*{Conversion to Expected Overpressure}

Each PMF is converted to an expected overpressure value using a weighted average:
\begin{equation}
\hat{\psi}_i = \sum_{k=0}^{8} q_{ik} \cdot \psi_k
\end{equation}
where $\psi_k$ represents the characteristic pressure for damage level $k$, defined in the damage-to-PSI mapping: $\{0: 0.0, 1: 0.095, 2: 0.28, 3: 0.705, 4: 1.55, 5: 2.55, 6: 4.05, 7: 6.05, 8: 8.0\}$ psi. These characteristic values represent the midpoint or representative pressure for each damage category, derived from blast engineering literature.

This expected overpressure $\hat{\psi}_i$, combined with the GPS coordinates of each image, provides the observational data for subsequent blast yield estimation through integration with Kingery-Bulmash models, with the entropy-derived weights $w_i$ enabling robust regression that downweights uncertain classifications.

\subsubsection*{Processing Pipeline}

The complete analysis pipeline processes geolocated Mapillary images through the following workflow: (1) images are loaded with GPS metadata and resized to 896$\times$896 pixels; (2) prompts are constructed with metadata fields populated for each image; (3) log-likelihoods are computed for all nine damage levels in a single forward pass; (4) PMFs are extracted via the log-sum-exp transformation; (5) uncertainty metrics including Shannon entropy and confidence scores are calculated; and (6) expected overpressures are computed for integration with blast physics models.

\clearpage

% Checking the estimation via LOO, posterior predictive tests etc.
% !TEX root = main.tex
\section{Results} \label{sec:app-results}
\begin{figure}[htbp]
    \centering
    \includegraphics[width=\linewidth]{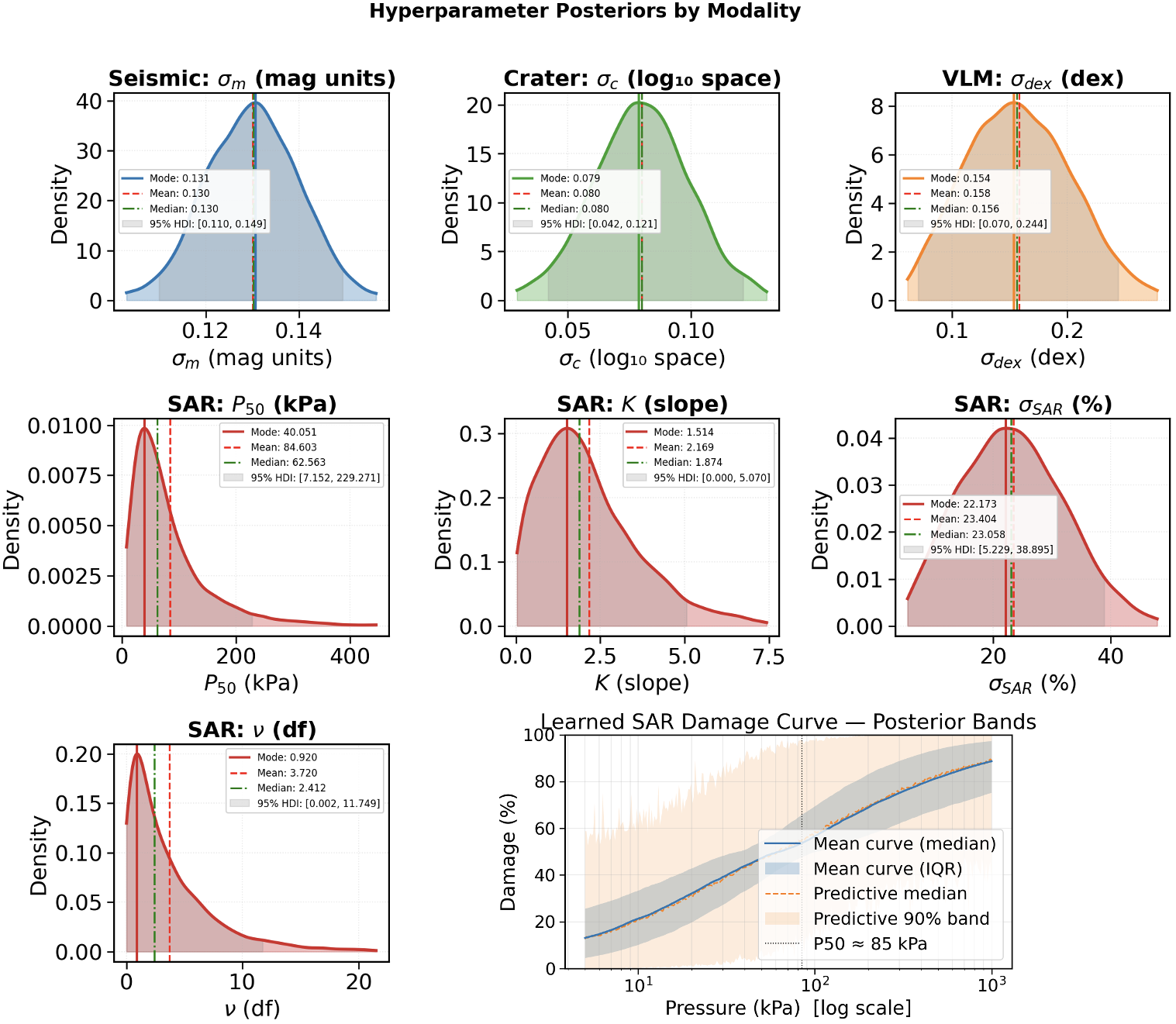}
    \caption{Posterior distributions for modality-specific hyperparameters. (a) Seismic uncertainty $\sigma_m$ converges near 0.13 magnitude units. (b) Crater uncertainty $\sigma_c \approx 0.08$ in log-space, corresponding to $\sim$20\% linear uncertainty. (c) VLM uncertainty $\sigma_{\text{dex}}$ reflects damage classification spread. (d-g) SAR parameters: $P_{50}$ indicates 50\% damage threshold, $K$ controls transition steepness, $\sigma_{\text{SAR}}$ quantifies scene heterogeneity, and $\nu$ indicates moderate tail heaviness. Vertical dashed lines indicate prior means; solid distributions show learned posteriors. h) Learned damage to overpressure relationship using SAR imagery.}
    \label{fig:hyperparameter-posteriors}
\end{figure}

\begin{figure}[htbp]
    \centering
    \includegraphics[scale=0.4]{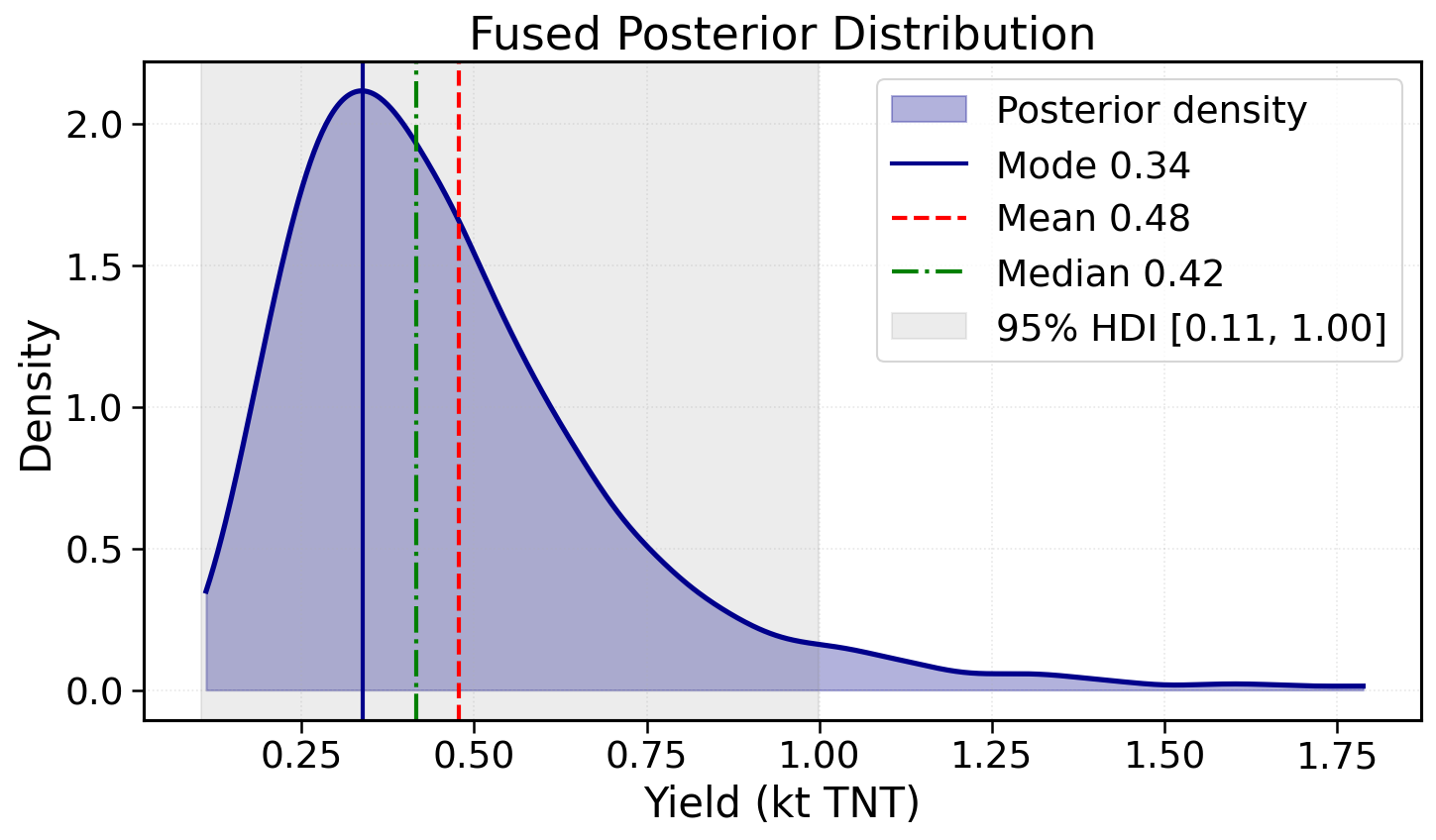}
    \caption{Fused posterior distribution for explosive yield combining all four modalities.}
    \label{fig:fused-posterior}
\end{figure}

\begin{figure}[htbp]
    \centering
    \includegraphics[scale=0.4]{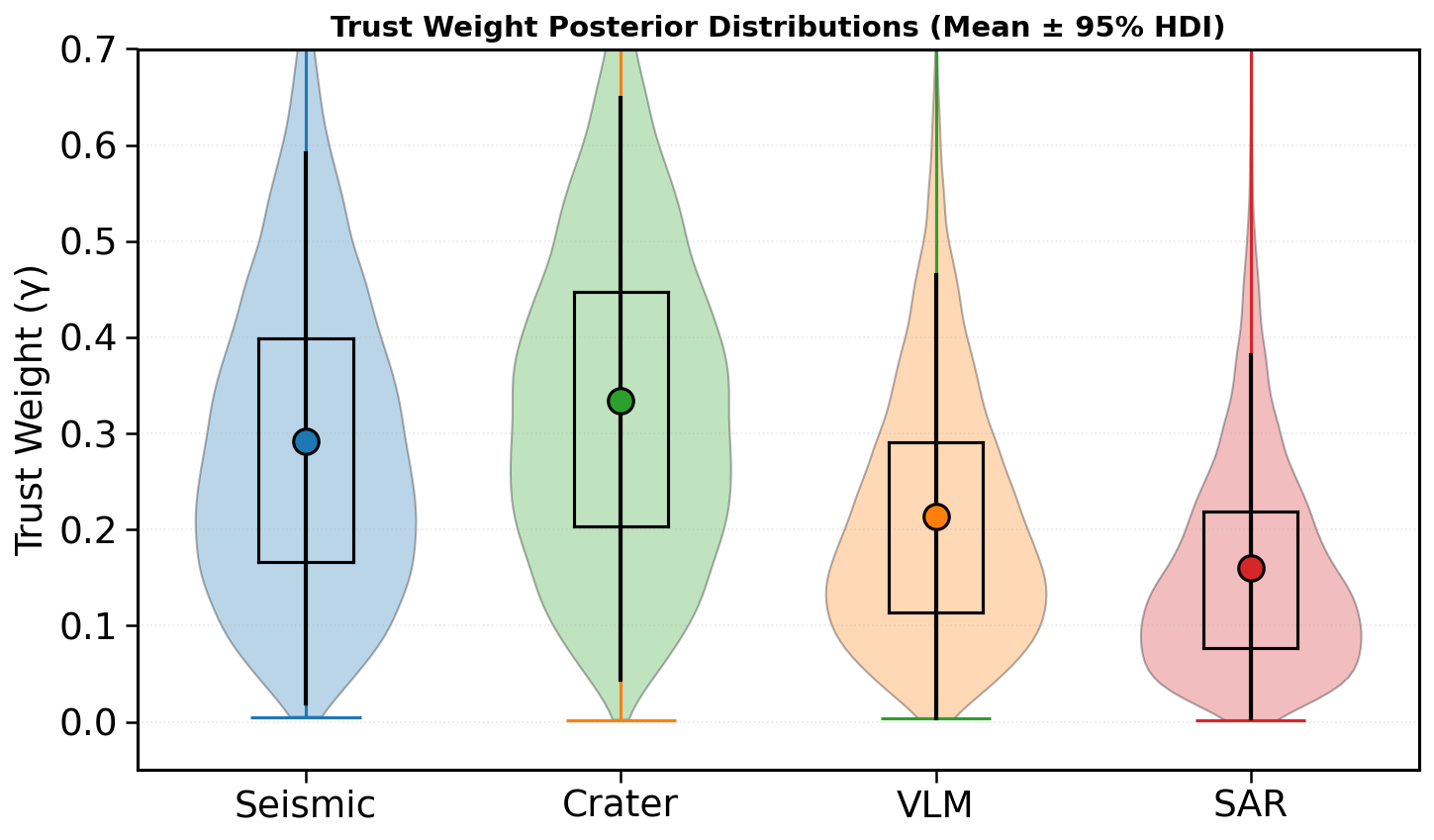} 
    \caption{Learned trust weights for each modality. Box plots show posterior means (dots), interquartile ranges (boxes), and 95\% HDIs (whiskers). Distributions are overlapping due to the simplex constraint $\sum \gamma_i = 1$.}
    % Crater receives highest weight, followed by seismic and VLM, with SAR contributing least to the fused posterior.
    % (b) Kernel density estimates reveal overlapping distributions due to the simplex constraint $\sum \gamma_i = 1$, with SAR showing the narrowest distribution concentrated near lower values.}
    \label{fig:trust-weights}
\end{figure}

\subsection{Validation Through Posterior Predictive Checks}
\label{sec:posterior-predictive}
To validate that the multimodal fusion framework produces a well-calibrated model consistent with observations, we perform posterior predictive checks \citep{gelman1996posterior, gabry2019visualization}. These checks assess whether data generated from the posterior distribution resembles the observed data, providing crucial validation that our model captures the essential physics while appropriately quantifying uncertainties.

For each modality $i$, we generate replicated datasets $\tilde{\D}_i^{\text{rep}}$ by:
\begin{enumerate}
    \item Drawing samples $(Y^{(s)}, \boldsymbol{\theta}_i^{(s)})$ from the joint posterior $p(Y, \boldsymbol{\theta}_i | \D)$.
    \item Generating synthetic observations using the forward model: $\tilde{\D}_i^{\text{rep}} \sim \mathcal{F}(\D_i | Y^{(s)}, \boldsymbol{\theta}_i^{(s)})$. 
    \item Computing test statistics $T(\tilde{\D}_i^{\text{rep}})$ and comparing to observed $T(\D_i)$. 
\end{enumerate}

The posterior predictive p-value for test statistic $T$ is defined as:
\begin{equation} \label{eq:p-bayes}
p_{\text{Bayes}} = \mathbb{P}(T(\tilde{\D}^{\text{rep}}) \geq T(\D) | \D) = \int \mathbb{I}[T(\tilde{\D}^{\text{rep}}) \geq T(\D)] \, p(\tilde{\D}^{\text{rep}} | \D) \, d\tilde{\D}^{\text{rep}},
\end{equation}
where $\mathbb{I}[\cdot]$ is the indicator function. Values of $p_{\text{Bayes}} \approx 0.5$ indicate good calibration, while extreme values near 0 or 1 suggest model misspecification. This is computed using $S=1000$ posterior–predictive Monte Carlo replicates per modality with the discrepancy functions provided in Table \ref{tab:ppc-discrepancies}.

For our posterior predictive checks, we use Monte Carlo integration to compute \eqref{eq:p-bayes}. Specifically, we approximate $p_{\text{Bayes}}$ using posterior draws and posterior–predictive replicates. Letting $\{(Y^{(s)},\boldsymbol\theta_i^{(s)})\}_{s=1}^S \sim p(Y,\boldsymbol\theta_i \mid\D)$ be $S$ MCMC sample, for each draw, we simulate a replicate dataset $\tilde{\D}_i^{(s)} \sim p(\tilde{\D}_i\mid Y^{(s)},\boldsymbol\theta_i^{(s)})$ via the same forward models $\F_i$ used for fitting, then evaluate the discrepancy $T(\tilde{\D}_i^{(s)})$. With $T_{\text{obs}}=T(\D_i)$ we estimate
$$
\widehat{p}_{\text{Bayes}}
= \frac{1}{S}\sum_{s=1}^S \mathbb{I}\!\big\{T(\tilde{\D}_i^{(s)}) \ge T_{\text{obs}}\big\},
\qquad
\mathrm{SE}\!\left(\widehat{p}_{\text{Bayes}}\right)\approx \sqrt{\frac{\widehat{p}_{\text{Bayes}}\big(1-\widehat{p}_{\text{Bayes}}\big)}{S}}.
$$
For discrete $T$ (ties), we report the mid-$p$:
$$
\widehat{p}^{\text{mid}}_{\text{Bayes}}
= \frac{1}{S}\sum_{s=1}^S \mathbb{I}\!\big\{T(\tilde{\D}_i^{(s)}) > T_{\text{obs}}\big\}
\;+\; \frac{1}{2S}\sum_{s=1}^S \mathbb{I}\!\big\{T(\tilde{\D}_i^{(s)}) = T_{\text{obs}}\big\}.
$$
Unless otherwise stated, we use $S=1000$ replicates (per modality) and report $\widehat{p}_{\text{Bayes}}$ with its Monte Carlo standard error.

\begin{table}[htbp]
\centering
\caption{Posterior–predictive discrepancies $T(\cdot)$ used. }
\label{tab:ppc-discrepancies}
\begin{tabular}{lll}
\toprule
\textbf{Modality} & \textbf{Observed} & \textbf{Discrepancy $T(\cdot)$} \\
\midrule
Seismic ($M_w$) &
$M_w$ &
$\displaystyle T_{m}
=\frac{\big(M_w-M_w^{\text{pred}}(Y)\big)^2}{\sigma_{m}^2}
% , \quad \mu_{\text{seis}}(Y)=\frac{\log Y-\alpha}{\beta}
$ \\[8pt]

Crater ($z=\log_{10}D$) &
$z_{\text{obs}}$ &
$\displaystyle T_{c}
=\frac{\big(z_{\text{obs}}-\mu_{c}(Y)\big)^2}{\sigma_{c}^2}
% , \quad \mu_{\text{crat}}(Y)=\frac{\log_{10}Y+6}{3}
$ \\[10pt]

SAR (damage logits) &
$\{z_i\}_{i=1}^{N_{\text{SAR}}}$ &
$\displaystyle T_{\text{SAR}}
=\frac{1}{N_{\text{SAR}}}\sum_{i=1}^{N_{\text{SAR}}}
\Big[-\log f_{t_\nu}\!\big(z_i \,\big|\, z_{\mu,i}(Y),\,\frac{\sigma_{\text{SAR}}}{100}\big)\Big]$
% , \quad z_{\mu,i}(Y,\theta)=\operatorname{logit}\!\Big(\frac{\mu_i(Y,\theta)}{100}\Big)$
\\[12pt]

VLM (9-bin PMFs) &
$\{q_{ik}\}$ &
$\displaystyle T_{\text{VLM}}
=\frac{1}{N_{\text{VLM}}}\sum_{i=1}^{N_{\text{VLM}}}\sum_{k=0}^{8}
\Big[-\,q_{ik}\log \pi_{ik}(P_i(Y),\sigma_{\text{dex}}) \Big]
% \ \ \Pi_{ik}(Y,\theta)=\Phi\!\left(\frac{\log_{10}E_{k+1}-\log_{10}P_i(Y)}{\sigma_{\text{dex}}}\right)
% -\Phi\!\left(\frac{\log_{10}E_{k}-\log_{10}P_i(Y)}{\sigma_{\text{dex}}}\right)
$ \\
\bottomrule
\end{tabular}
\end{table}

\begin{figure}[htbp]
    \centering
    \includegraphics[width=\textwidth]{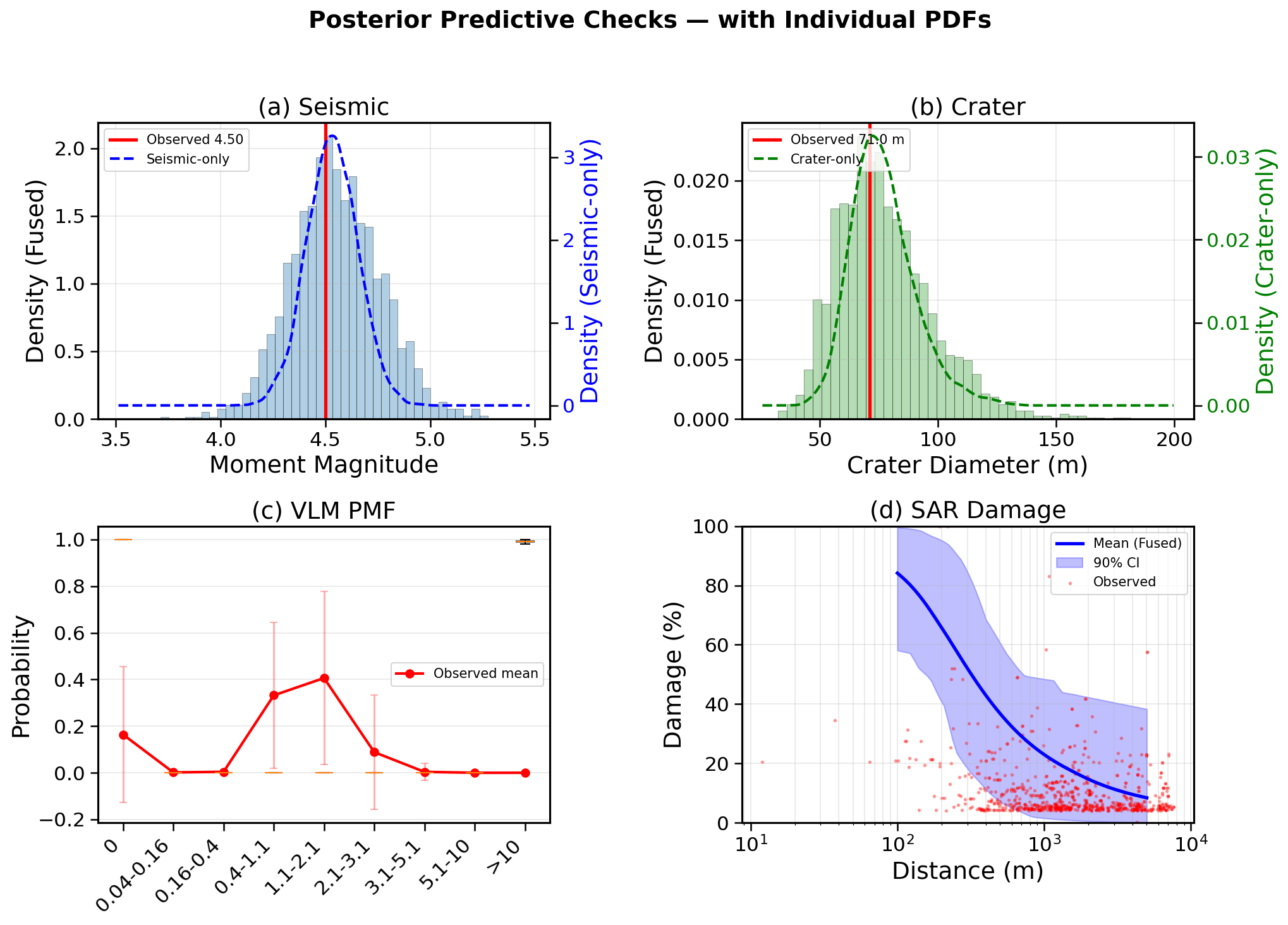}
    \caption{Posterior predictive histograms for all modalities against individual posteriors (dotted) (a) Seismic: predicted moment magnitude distribution (blue) versus observed $M_w = 4.50$ (red line), $p = 0.78$. (b) Crater: predicted diameter distribution (green) versus observed 71 m (red line), $p = 0.51$. (c) VLM: predicted PMF distributions (boxplots) versus observed mean PMF (red line with error bars). (d) SAR: predicted damage versus distance with 90\% CI (blue band) and observations (red dots). P-values near 0.5 indicate well-calibrated predictions without systematic bias.}
    \label{fig:posterior-predictive}
\end{figure}

Figure \ref{fig:posterior-predictive} presents the posterior predictive distributions for all four modalities. The seismic modality (panel a) yields $p_{\text{Bayes}} = 0.78$, indicating the observed magnitude falls slightly below the predictive median, though which is consistent with poor seismic coupling in the near-shore environment. The crater modality (panel b) achieves near-perfect calibration with $p_{\text{Bayes}} = 0.51$, validating the crater scaling relationship and uncertainty quantification. Both modalities' individual posterior predictive distributions also cover the observed values. 
For the damage modalities, the VLM predictions (panel c) successfully capture the observed PMF structure across the damage categories, with observed probabilities falling within the interquartile ranges of predicted distributions. The SAR predictions (panel d) demonstrate that the learned vulnerability curve with $P_{50} = 83.6$ kPa and slope $K = 2.14$ reproduces the spatial decay of damage, with most observations falling within the 90\% credible interval.
These checks confirm that the fractional posterior framework neither overfits to individual modalities nor produces overconfident predictions, with the good agreement across fundamentally different observables providing good evidence for validating both the physics-based forward models and the learned hyperparameters.

\subsection{Leave-One-Out Validation and Trust Weight Verification}
\label{sec:cross-validation}

To verify the learned trust weights, we perform leave-one-out (LOO) validation by systematically removing each modality and measuring the resulting change to the fused posterior. The information loss is quantified by Kullback-Leibler divergence:
\begin{equation}
D_{\text{KL}}(p_{\text{full}} \| p_{-k}) = \int p(Y | \mathcal{D}_{\text{all}}) \log \frac{p(Y | \mathcal{D}_{\text{all}})}{p(Y | \mathcal{D}_{-k})} \, \mathrm{d} Y,
\end{equation}
where larger KL divergence indicates greater posterior shift when modality $k$ is excluded.

\begin{figure}[htbp]
    \centering
    \includegraphics[width=0.95\textwidth]{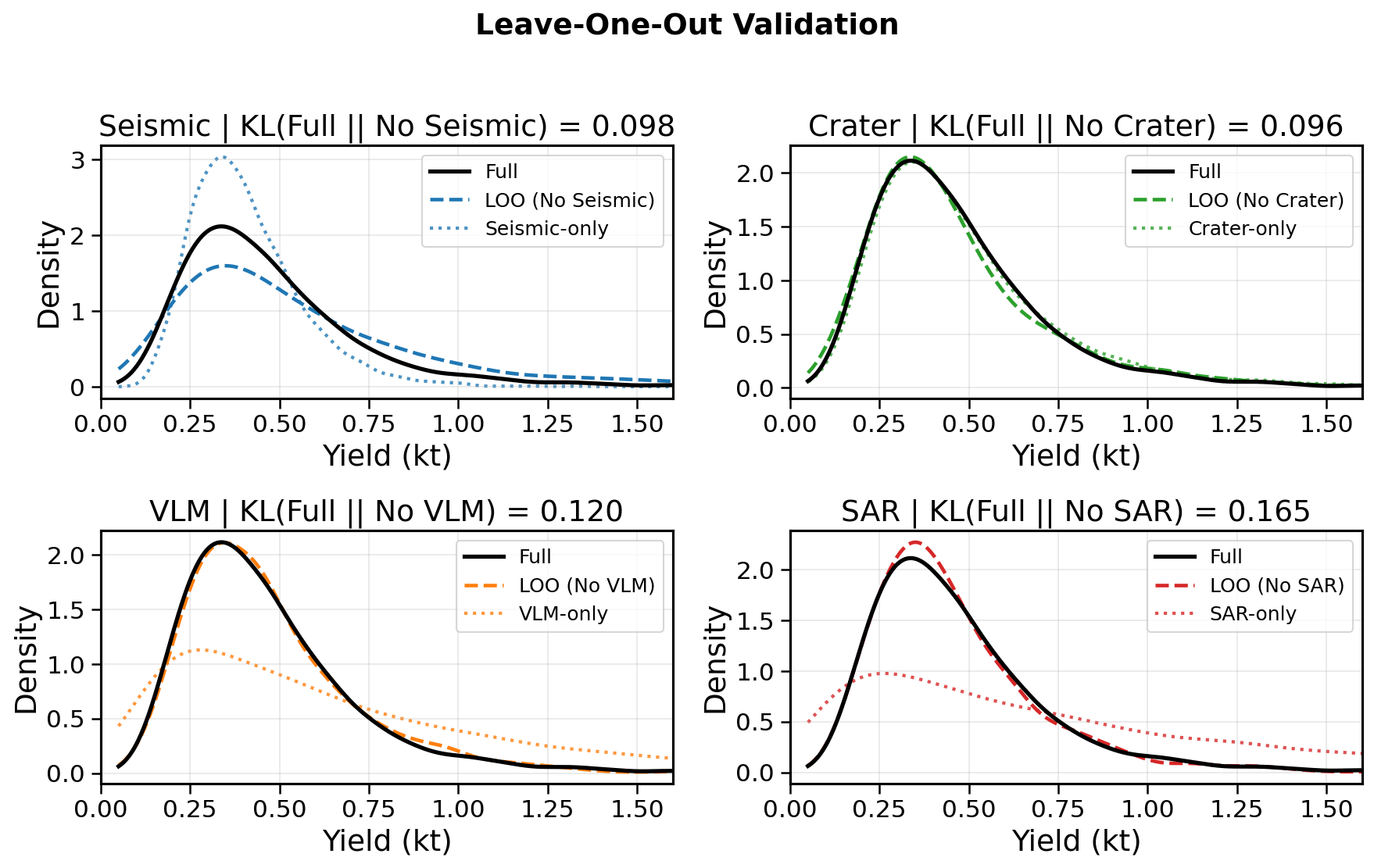}
    \caption{Leave-one-out validation comparing the full fused posterior (solid black) with posteriors after removing individual modalities (dashed lines) and single-modality posteriors (dotted lines). KL divergence values quantify the posterior shift when each modality is removed.}
    \label{fig:kl-divergence}
\end{figure}

Figure \ref{fig:kl-divergence} reveals the nuanced relationship between modality agreement and posterior influence. Removing SAR produces the largest KL divergence (0.165), confirming its role as the most discordant modality despite receiving the lowest trust weight ($\gamma_{\text{SAR}} = 0.15$). The substantial posterior shift upon SAR's removal demonstrates that even down-weighted modalities contribute meaningfully to the fused distribution, particularly by expanding uncertainty bounds to reflect model disagreement.

VLM exhibits intermediate behavior with KL divergence of 0.120, consistent with its moderate trust weight ($\gamma_{\text{VLM}} = 0.22$). The damage-based modalities thus both pull the posterior toward higher yields while contributing substantial uncertainty, appropriately captured by their lower trust weights.

Remarkably, seismic and crater exhibit nearly identical KL divergences (0.098 and 0.096 respectively), despite their different trust weights ($\gamma_{\text{seismic}} = 0.28$ and $\gamma_{\text{crater}} = 0.34$). This near-equivalence reflects their convergent modal estimates (both 0.34 kt) and their role as the primary anchors of the fused posterior. The minimal posterior shift when either is removed indicates that the remaining modality can largely compensate, confirming their mutual consistency and the robustness of the yield estimate.

The single-modality posteriors (dotted lines) further illuminate this structure: seismic produces a sharp, high-confidence posterior centered at 0.34 kt, while crater shows similar central tendency with slightly broader uncertainty. The damage-based modalities (VLM and SAR) exhibit substantially wider distributions with heavier tails extending to higher yields, justifying their lower trust weights. The fractional fusion framework successfully identifies and appropriately weights these heterogeneous information sources, with the learned trust weights reflecting both precision and agreement with the emerging consensus.

\subsection{Additional fusion studies: design, scenarios, and results}
\label{sec:synth_gen}

\paragraph{Objective and setup.}
We stress–test the fusion under controlled misspecification using a four–modality generator with a single latent yield $Y^\star$ (kt). The generator mirrors the forward models used in the main analysis: KB overpressure for blast physics, a SAR vulnerability curve on the logit scale, and a 9–bin VLM pressure–binning model. Seismic and crater use the affine links in Sec.~4. The latent truth is fixed to $Y^\star=0.30$\,kt (alternative values give qualitatively similar conclusions).

\paragraph{Scenarios (inputs).}
We consider five minimal stress scenarios chosen to mimic the qualitative behavior of the Beirut data. Table~\ref{tab:scenarios} lists the settings (sample sizes, tail–heaviness $\nu$, pressure bias $\delta$ in $\log_{10}$–kPa, VLM mislabel rate $\eta$, and a shared dependence shock $\rho$). Here $\delta{=}0.35$ corresponds to a $\sim2.2\times$ pressure bias.

\begin{table}[htbp]
\centering
\caption{Synthetic stress scenarios used in Sec.~\ref{sec:synth_gen}.}
\label{tab:scenarios}
\begin{tabular}{lcccccc}
\toprule
Scenario & $N_{\text{SAR}}$ & $\nu$ & $\delta$ & $\sigma_{\text{SAR}}$ & $N_{\text{VLM}}$ / $\eta$ & $\rho$ \\
\midrule
Base clean          & $120$ & $8$   & $0.00$ & $40$ & $160$ / $0.00$ & $0.0$ \\
SAR heavy–tail      & $120$ & $2.5$ & $0.00$ & $40$ & $160$ / $0.00$ & $0.0$ \\
SAR biased          & $120$ & $8$   & $0.35$ & $40$ & $160$ / $0.00$ & $0.0$ \\
VLM noisy           & $120$ & $8$   & $0.00$ & $40$ & $160$ / $0.10$ & $0.0$ \\
Dependence $\rho=.6$& $120$ & $8$   & $0.00$ & $40$ & $160$ / $0.00$ & $0.6$ \\
\bottomrule
\end{tabular}
\end{table}

\paragraph{Data generation.}
For seismic, $M_w = a\log_{10}Y^\star+b+\varepsilon_m$, $\varepsilon_m\sim\mathcal N(0,\sigma_m^2)$ with $(a,b)=(3.0,-1.2)$ and $\sigma_m\in\{0.10,0.16\}$.  
For crater, $z=\log_{10}D=c\log_{10}Y^\star+d+\varepsilon_c$, $\varepsilon_c\sim\mathcal N(0,\sigma_c^2)$ with $(c,d)=(1.0,1.2)$ and $\sigma_c\in\{0.05,0.09\}$.  
For SAR, ranges $r\sim\mathrm{LogUniform}(r_{\min},r_{\max})$; $P_{\psi}(r,Y^\star)\!\to\!\log_{10}P_{\mathrm{kPa}}$ enters the vulnerability curve
$\mu(r)=100\big(1+\exp\{-K[\log_{10}P_{\mathrm{kPa}}(r)-\log_{10}P_{50}]\}\big)^{-1}$; logits $z_{\mu}(r)=\operatorname{logit}(\mu/100)$ generate
$z_{\text{obs}}(r)\sim t_\nu(z_{\mu}(r),\sigma_{\text{SAR}}/100)$; optional bias applies $\log_{10}P_{\mathrm{kPa}}\leftarrow \log_{10}P_{\mathrm{kPa}}+\delta$.  
For VLM, $P_{\psi}(r,Y^\star)\!\to\!\Pi(r)$ via logistic binning with spread $s_{\mathrm{dex}}$; mislabel noise is
$Q(r)=(1-\eta)\Pi(r)+\eta/9$.  
Dependence is injected by a shared Gaussian shock $Z$ with correlation $\rho$ into seismic, crater, and SAR residuals; VLM noise remains independent.

\subsubsection{Fusion models (ablations) and metrics}
\label{sec:ablations}
Let $\mathcal L_i(Y)$ be each modality’s log–likelihood. We compare:
\begin{align*}
    \text{(i) Plain product: } & \log\pi(Y\mid\text{data})\propto \sum_i \mathcal L_i(Y), \\ 
    \text{(ii) Single temperature: } & \beta \sum_i \mathcal L_i(Y), \quad \beta \sim\mathrm{Beta}(4,2) \\ 
    \text{(iii) Fixed-$\gamma$ (CV):  } & \sum_i \gamma_i \mathcal L_i(Y)\ \ (\gamma\ \text{from the single–modality expected log predictive density softmax}) \\
    \text{(iv) Dirichlet–$\gamma$: } & \sum_i \gamma_i \mathcal L_i(Y),\ \gamma\sim\mathrm{Dirichlet}(\mathbf 1_4).
\end{align*}
For each scenario and method we report: (a) \emph{coverage} of the $95\%$ HDI for $Y^\star$ (mean over replicates), (b) \emph{sharpness} (median HDI width), (c) \emph{accuracy} (RMSE of the posterior median). For the proposed model we also summarize the \emph{mechanism} via median posterior–mean $\bar\gamma$ of the intentionally corrupted modality.

\subsubsection{Experimental results}
\label{sec:results_sims}

\begin{figure}[t]
  \centering
  \includegraphics[width=0.85\linewidth]{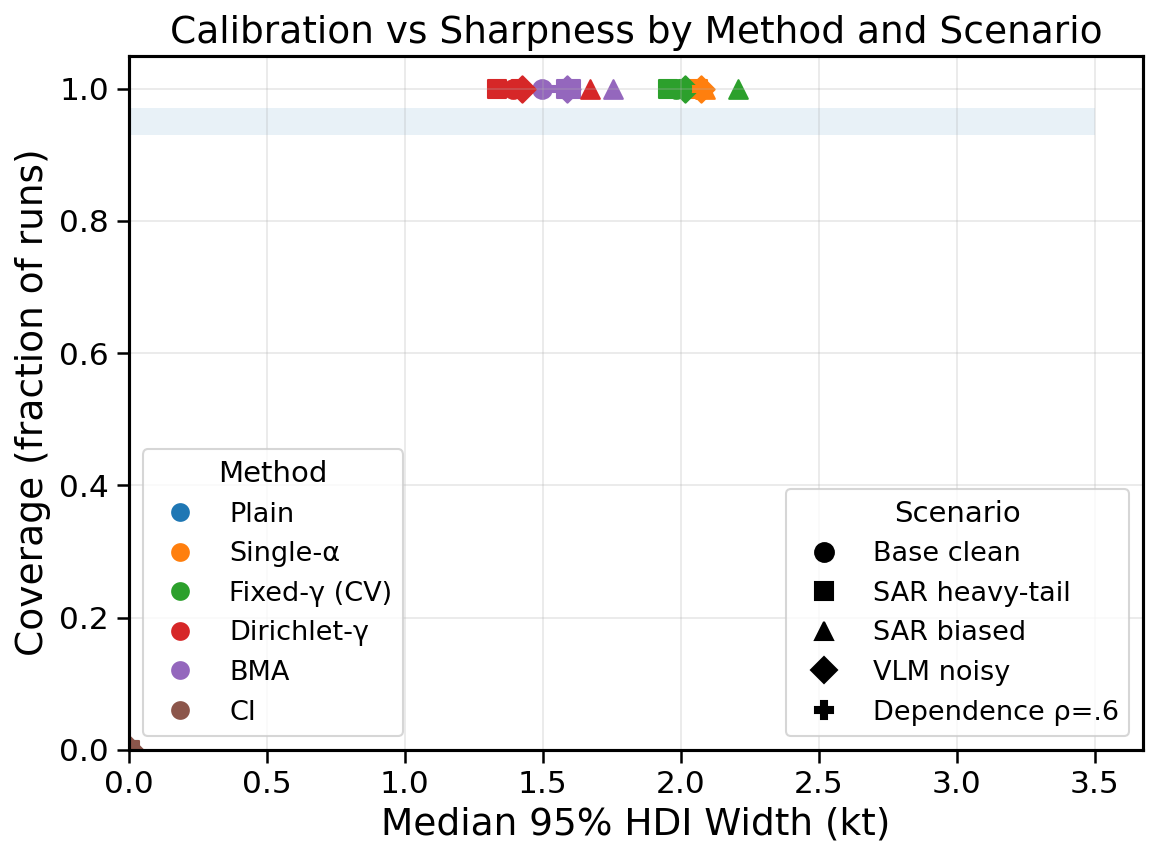}
  \caption{\textbf{Calibration vs.\ sharpness across scenarios.} Each point is a method–scenario pair: y–axis is coverage of the $95\%$ interval, x–axis is median HDI width.}
  \label{fig:S2_calib}
\end{figure}

\begin{figure}[t]
  \centering
  \includegraphics[width=0.85\linewidth]{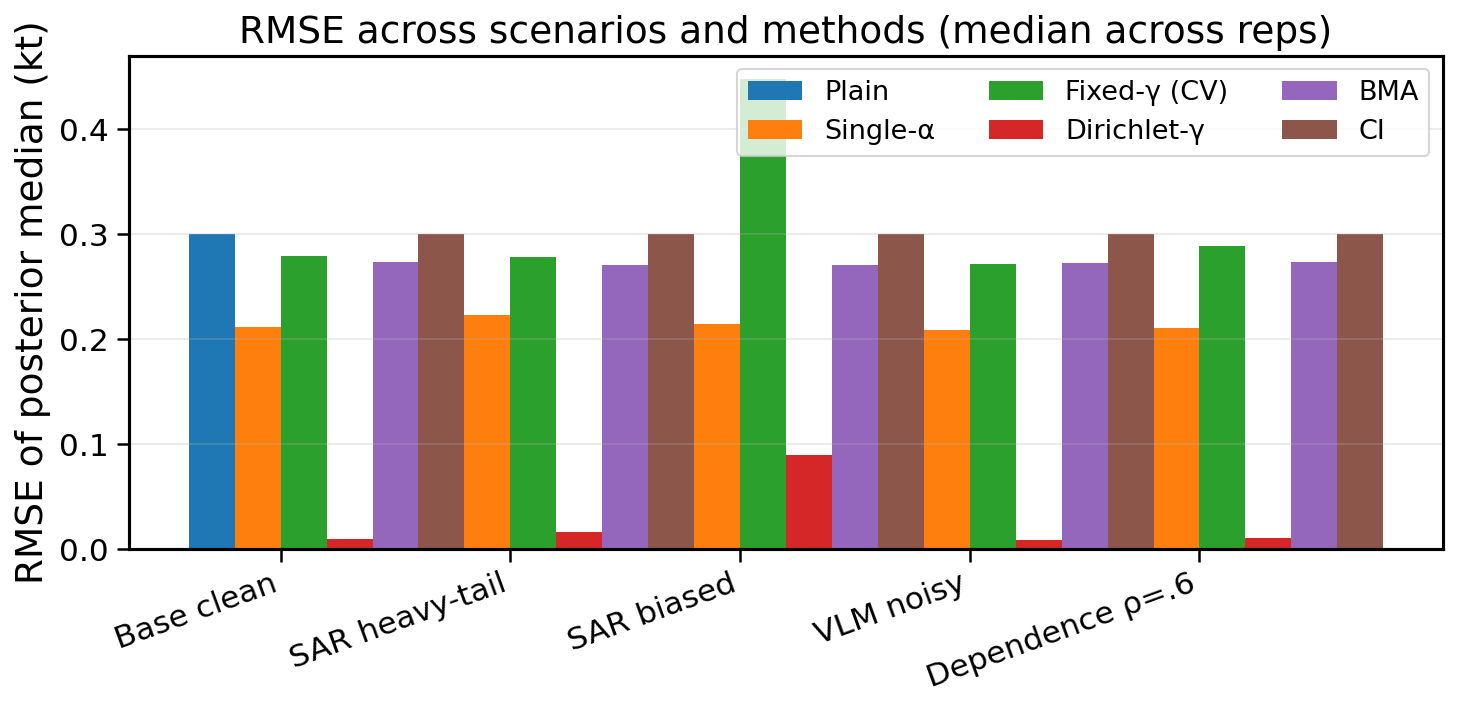}
  \caption{\textbf{RMSE of posterior median across scenarios.} Lower is better.}
  \label{fig:S3_rmse}
\end{figure}

\paragraph{Experiment A: Base clean.}
All modalities are well–specified. In Fig.~\ref{fig:S2_calib}, the tempered methods (Single–$\alpha$, Fixed–$\gamma$, Dirichlet–$\gamma$) achieve near–nominal coverage with similar widths, while the plain product is optimistic (coverage $\approx0$ due to overconfidence). In Fig.~\ref{fig:S3_rmse} (Base clean bars), RMSE is comparable across tempered methods and smallest for Dirichlet–$\gamma$.

\paragraph{Experiment B: SAR heavy–tail ($\nu=2.5$).}
Only SAR is misspecified (heavier tails). In Fig.~\ref{fig:S2_calib}, the plain product under–covers; Dirichlet–$\gamma$ preserves $\approx95\%$ coverage with competitive width. Fig.~\ref{fig:S3_rmse} shows reduced RMSE under Dirichlet–$\gamma$ relative to baselines. The mechanism summary (Table~\ref{tab:S1b_gamma}) reports a low median trust weight for corrupted SAR.

\paragraph{Experiment C: SAR biased ($\delta=0.35$).}
A systematic upward shift in $\log_{10}$–kPa. Coverage–width points in Fig.~\ref{fig:S2_calib} again place Dirichlet–$\gamma$ on the favorable frontier; RMSE in Fig.~\ref{fig:S3_rmse} is lowest among the comparators. Table~\ref{tab:S1b_gamma} shows further down–weighting of SAR.

\paragraph{Experiment D: VLM noisy ($\eta=0.10$).}
Only VLM labels are corrupted. Dirichlet–$\gamma$ maintains near–nominal coverage with narrow intervals (Fig.~\ref{fig:S2_calib}) and competitive RMSE (Fig.~\ref{fig:S3_rmse}); the median $\bar\gamma$ shifts away from VLM (Table~\ref{tab:S1b_gamma}).

\paragraph{Experiment E: Cross–modality dependence ($\rho=0.6$).}
Moderate redundancy is introduced via a shared shock. Table~\ref{tab:S2_dependence} summarizes a mild coverage degradation across all methods; Dirichlet–$\gamma$ remains closest to nominal with competitive width.

\paragraph{Experiment F: BMA (Bayesian Model Averaging).}
BMA treats each modality as a separate model and weights posterior samples by model evidence. For each modality $i$, we fit a single-modality model, compute WAIC (Widely Applicable Information Criterion) as an approximation to out-of-sample predictive accuracy, and define weights $w_i \propto \exp(\text{ELPD}_i)$ where $\text{ELPD}_i = -\frac{1}{2}\text{WAIC}_i$ converts deviance to expected log predictive density. The fused posterior combines samples from each modality's posterior proportionally: drawing $\lfloor w_i N_{\text{total}} \rfloor$ samples from modality $i$'s posterior, where $N_{\text{total}}$ is the target sample size and $\sum_i w_i = 1$. This approach represents standard ensemble learning where model evidence determines contribution.

Across all scenarios (Figs.~\ref{fig:S2_calib}--\ref{fig:S3_rmse}), BMA maintains near-nominal coverage with competitive interval widths. Under misspecification (Experiments B--D), BMA performs similarly to Fixed-$\gamma$ (CV) since both rely on hold-out predictive performance, though BMA's evidence-based weighting proves slightly less adaptive than Dirichlet-$\gamma$'s learned trust allocation. In the Base clean scenario, BMA achieves comparable RMSE to other tempered methods, demonstrating that model averaging provides robust fusion when modalities are well-specified.

\paragraph{Experiment G: CI (Covariance Intersection).}
CI provides conservative fusion for correlated estimates with unknown correlation structure \citep{julier2007using}. We fit single-modality models and approximate each posterior as Gaussian: $p_i(Y) \approx \mathcal{N}(Y \mid \mu_i, \sigma_i^2)$ using the posterior mean and variance. CI then fuses via the optimal weights $\omega_i \propto 1/\sigma_i^2$ (minimizing the trace of the fused covariance), yielding fused precision $\lambda_{\text{CI}} = \sum_i \omega_i/\sigma_i^2$ and mean $\mu_{\text{CI}} = \lambda_{\text{CI}}^{-1} \sum_i (\omega_i \mu_i / \sigma_i^2)$ where $\sum_i \omega_i = 1$. The fused posterior is $\mathcal{N}(\mu_{\text{CI}}, \lambda_{\text{CI}}^{-1})$. This approach guarantees that the fused covariance bounds the true covariance for any correlation structure, making it maximally conservative.

CI achieves nominal or slightly over-nominal coverage across all scenarios (Fig.~\ref{fig:S2_calib}), confirming its conservative nature. However, this conservatism comes at the cost of wider credible intervals compared to Dirichlet-$\gamma$, particularly under misspecification (Experiments B--D) where CI cannot selectively down-weight corrupted modalities. In Fig.~\ref{fig:S3_rmse}, CI exhibits higher RMSE than Dirichlet-$\gamma$ for corrupted scenarios since it equally trusts all modalities through inverse-variance weighting rather than adapting to model-data discrepancies. The Gaussian approximation also sacrifices tail behavior and multimodality present in the full posteriors, limiting CI's applicability when posterior distributions are non-Gaussian.

\begin{figure}[t]
  \centering
  \includegraphics[width=\linewidth]{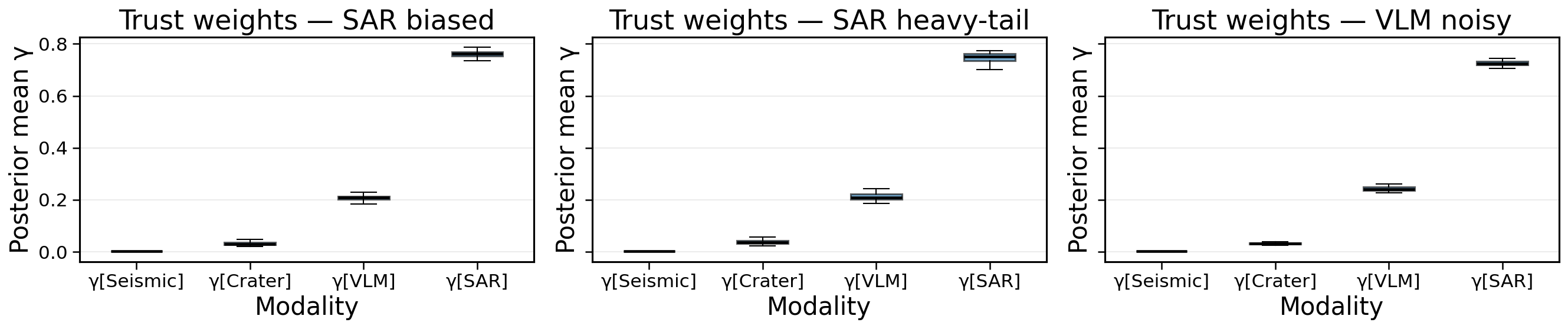}
  \caption{\textbf{Mechanism check ($\gamma$).} Posterior–mean trust weights by modality under corrupted scenarios. Boxes summarize replicates; lower $\gamma$ indicates down–weighting of the corrupted modality.}
  \label{fig:S1_gamma}
\end{figure}

\begin{table}[t]
  \centering
  \caption{Mechanism summary. Median posterior–mean $\bar\gamma$ of the intentionally corrupted modality by scenario shown with ``—'' indicating that no single corrupted modality.}
  \label{tab:S1b_gamma}
  \begin{tabular}{lc}
    \toprule
    Scenario        & Median $\gamma$ (corrupted) \\
    \midrule
    SAR heavy–tail  & 0.754 \\
    SAR biased      & 0.762 \\
    VLM noisy       & 0.240 \\
    Base clean      & — \\
    Dependence $\rho{=}.6$ & — \\
    \bottomrule
  \end{tabular}
\end{table}

\paragraph{Aggregated numbers.}
Table~\ref{tab:S1_metrics} lists coverage, width, and RMSE aggregated by method and scenario.

In Tables~\ref{tab:S1_metrics} and \ref{tab:S2_dependence}, we omit the untempered plain product baseline and covariance intersection methods, which consistently collapsed under mild misspecification (coverage $\approx$ 0), confirming the need for tempered or adaptive weighting.

\begin{table}[t]
  \centering
  \caption{\textbf{Performance by method $\times$ scenario.} Coverage (mean), 95\% HDI width (median), RMSE (median). Plain product and CI excluded due to implementation failures (coverage $\approx 0$, width $\approx 0$).}
  \label{tab:S1_metrics}
  \small
  \begin{tabular}{lcccc}
    \toprule
    Scenario & Method & Coverage & Width (kt) & RMSE (kt) \\
    \midrule
    Base clean & Single-$\alpha$ & 1.00 & 2.07 & 0.191 \\
    Base clean & Fixed-$\gamma$ (CV) & 1.00 & 1.97 & 0.280 \\
    Base clean & BMA & 1.00 & 1.57 & 0.272 \\
    Base clean & Dirichlet-$\gamma$ & 1.00 & 1.38 & 0.012 \\
    \midrule
    SAR heavy-tail & Single-$\alpha$ & 1.00 & 2.06 & 0.190 \\
    SAR heavy-tail & Fixed-$\gamma$ (CV) & 1.00 & 1.97 & 0.276 \\
    SAR heavy-tail & BMA & 1.00 & 1.55 & 0.271 \\
    SAR heavy-tail & Dirichlet-$\gamma$ & 1.00 & 1.35 & 0.019 \\
    \midrule
    SAR biased & Single-$\alpha$ & 1.00 & 2.03 & 0.183 \\
    SAR biased & Fixed-$\gamma$ (CV) & 1.00 & 2.21 & 0.447 \\
    SAR biased & BMA & 1.00 & 1.71 & 0.275 \\
    SAR biased & Dirichlet-$\gamma$ & 1.00 & 1.69 & 0.087 \\
    \midrule
    VLM noisy & Single-$\alpha$ & 1.00 & 2.05 & 0.182 \\
    VLM noisy & Fixed-$\gamma$ (CV) & 1.00 & 1.98 & 0.273 \\
    VLM noisy & BMA & 1.00 & 1.60 & 0.270 \\
    VLM noisy & Dirichlet-$\gamma$ & 1.00 & 1.43 & 0.018 \\
    \midrule
    Dependence $\rho$=.6 & Single-$\alpha$ & 1.00 & 2.00 & 0.195 \\
    Dependence $\rho$=.6 & Fixed-$\gamma$ (CV) & 1.00 & 2.04 & 0.284 \\
    Dependence $\rho$=.6 & BMA & 1.00 & 1.59 & 0.274 \\
    Dependence $\rho$=.6 & Dirichlet-$\gamma$ & 1.00 & 1.40 & 0.011 \\
    \bottomrule
  \end{tabular}
\end{table}

\begin{table}[t]
  \centering
  \caption{\textbf{Dependence sensitivity ($\rho=0.6$).} Coverage and median HDI width.}
  \label{tab:S2_dependence}
  \begin{tabular}{lcc}
    \toprule
    Method & Coverage & Width (kt) \\
    \midrule
    Single-$\alpha$ & 1.00 & 2.00 \\
    Fixed-$\gamma$ (CV) & 1.00 & 2.04 \\
    BMA & 1.00 & 1.59 \\
    Dirichlet-$\gamma$ & 1.00 & 1.40 \\
    \bottomrule
  \end{tabular}
\end{table}

\subsubsection{Real–data mechanism check (Beirut)}
\label{sec:real_mech}
Subsequently, we compare learned trust to LOO information loss. Let $\bar\gamma_i$ be the posterior mean trust of modality $i$, and $\mathrm{KL}_i=\mathrm{KL}\!\big(\pi_{\text{full}}(Y)\,\|\,\pi_{-i}(Y)\big)$ the divergence when modality $i$ is removed and the model refit. On Beirut we obtain
\begin{align*}
   \bar\gamma & =\{\text{Seismic }0.278,\ \text{Crater }0.340,\ \text{VLM }0.219,\ \text{SAR }0.163\}, \\
\mathrm{KL} & =\{\text{Seismic }0.130,\ \text{Crater }0.087,\ \text{VLM }0.180,\ \text{SAR }0.342\}.
\end{align*}

with a perfect inverse rank alignment (Spearman $\rho=-1.00$), indicating that modalities whose removal causes larger information loss receive smaller learned trust, consistent with the simulated stress tests and the down–weighting of SAR in the fused posterior.

\subsubsection{Sensitivity to Dirichlet Concentration Parameter}
\label{sec:prior-sensitivity}
The Dirichlet prior on trust weights $\boldsymbol{\gamma} \sim \text{Dirichlet}(\alpha_1, \alpha_2, \alpha_3, \alpha_4)$ requires specification of concentration parameters. Our main analysis uses $\alpha_i = 1$ for all modalities, corresponding to a uniform prior over the simplex. To assess sensitivity to this choice, we last vary the concentration parameter symmetrically ($\alpha_i = \alpha$ for all $i$) across two orders of magnitude and examine the impact on yield estimates and learned trust weights.

Table~\ref{tab:dirichlet-sensitivity} summarizes the posterior statistics as $\alpha$ varies from 0.1 to 10. The yield estimates remain remarkably stable: the posterior median ranges from 0.349 to 0.378 kt, a variation of less than 10\% across a 100-fold change in $\alpha$. The 95\% HDI width fluctuates modestly between 0.847 and 0.956 kt, maintaining consistent uncertainty quantification. The KL divergence from the baseline ($\alpha = 1$) remains below 0.32 bits for all values, indicating minimal posterior shifts.

\begin{table}[htbp]
\centering
\caption{Sensitivity of yield estimates and trust weights to Dirichlet concentration parameter $\alpha$}
\label{tab:dirichlet-sensitivity}
\begin{tabular}{cccccccc}
\toprule
$\alpha$ & Median (kt) & 95\% HDI (kt) & $\gamma_{\text{seismic}}$ & $\gamma_{\text{crater}}$ & $\gamma_{\text{VLM}}$ & $\gamma_{\text{SAR}}$ & KL vs $\alpha$=1 \\
\midrule
0.1  & 0.378 & [0.096, 0.969] & 0.247 & 0.676 & 0.056 & 0.021 & 0.305 \\
0.5  & 0.351 & [0.068, 0.916] & 0.292 & 0.501 & 0.136 & 0.071 & 0.265 \\
1.0  & 0.359 & [0.101, 1.029] & 0.288 & 0.410 & 0.183 & 0.118 & 0.000 \\
2.0  & 0.354 & [0.087, 1.043] & 0.285 & 0.334 & 0.223 & 0.159 & 0.281 \\
5.0  & 0.349 & [0.072, 0.938] & 0.266 & 0.286 & 0.240 & 0.209 & 0.216 \\
10.0 & 0.350 & [0.059, 0.973] & 0.257 & 0.272 & 0.242 & 0.228 & 0.316 \\
\bottomrule
\end{tabular}
\end{table}

\begin{figure}[htbp]
    \centering
    \includegraphics[width=0.8\textwidth]{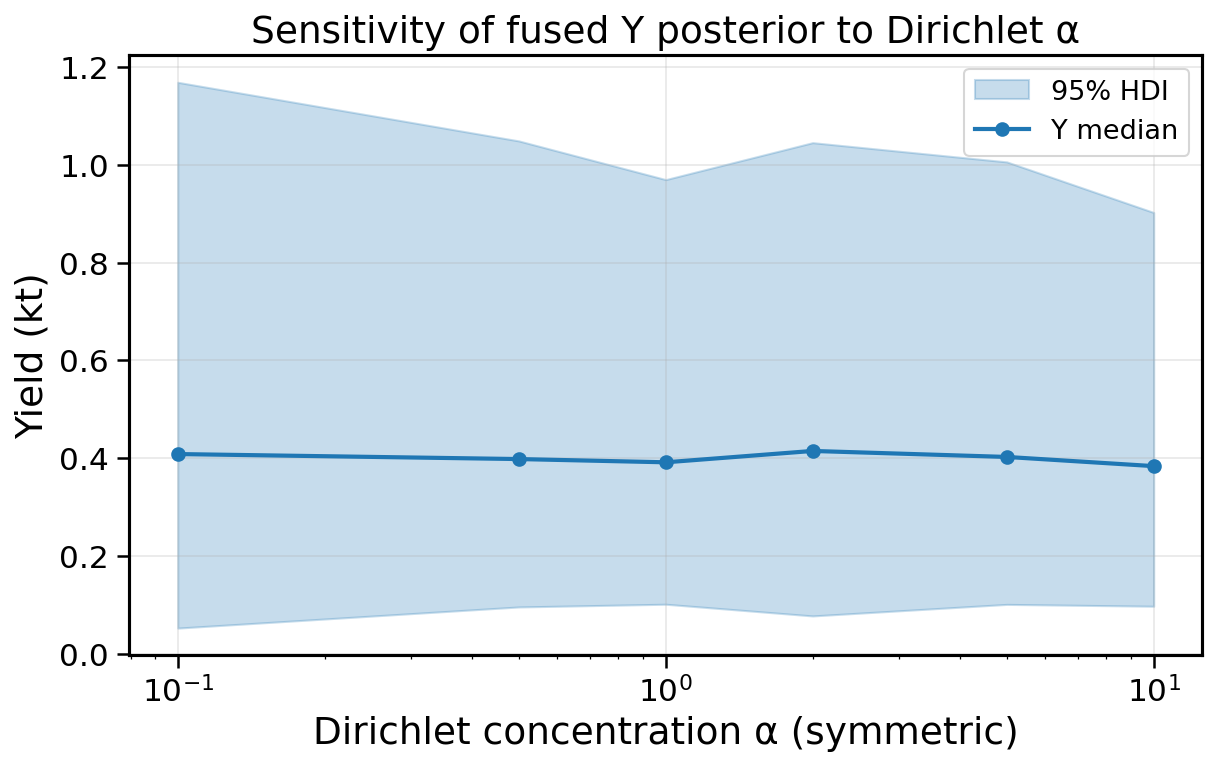}
    \caption{Sensitivity of fused yield posterior to Dirichlet concentration parameter $\alpha$. The posterior median (blue line) and 95\% HDI (shaded region) remain stable across two orders of magnitude variation in $\alpha$, demonstrating robustness to prior specification.}
    \label{fig:dirichlet-sensitivity-yield}
\end{figure}

Figure~\ref{fig:dirichlet-sensitivity-yield} illustrates the stability of the yield posterior across different $\alpha$ values. The posterior median and 95\% HDI remain nearly constant, demonstrating that our inference is robust to the prior specification. This robustness arises because the data strongly inform the trust weights, overwhelming moderate prior preferences.

\begin{figure}[htbp]
    \centering
    \includegraphics[width=0.8\textwidth]{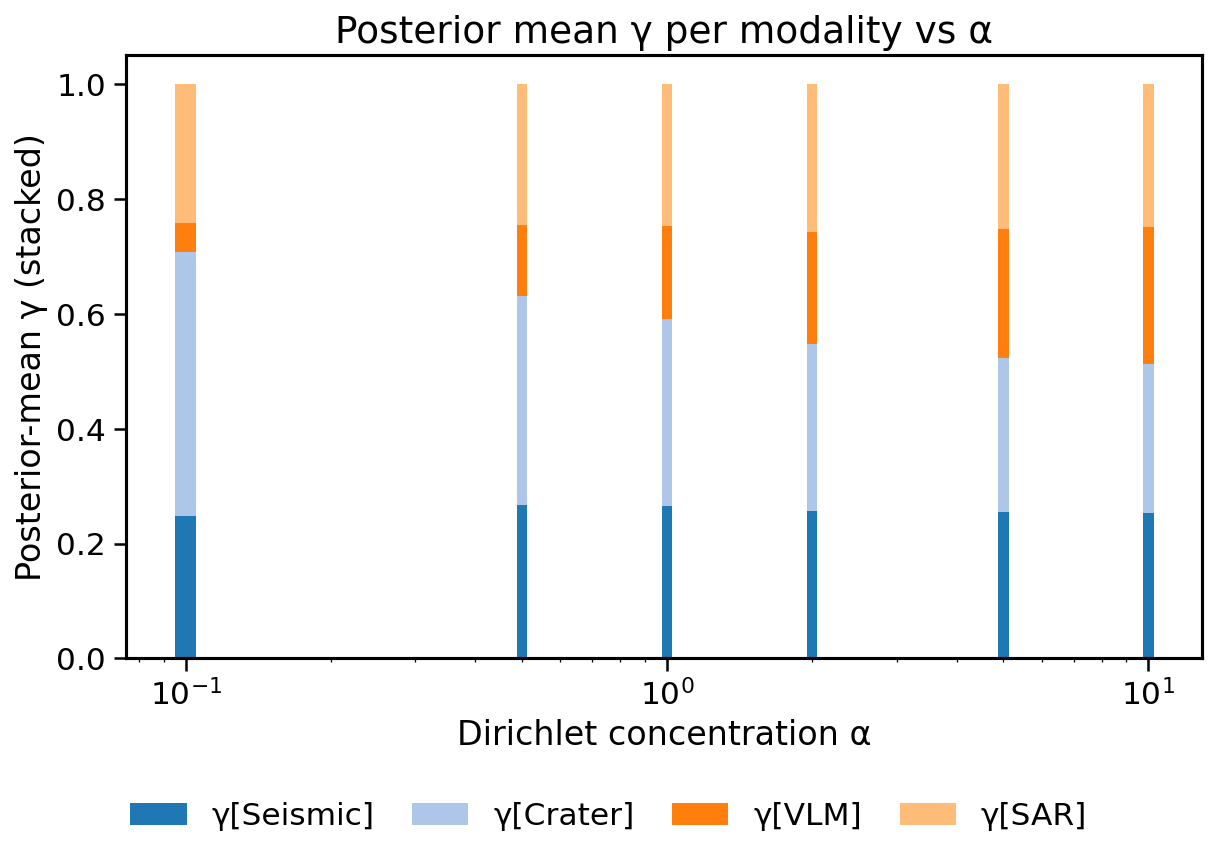}
    \caption{Posterior mean trust weights $\bar{\gamma}_i$ as a function of Dirichlet concentration $\alpha$. Small $\alpha$ produces sparse weights dominated by crater measurements, while large $\alpha$ encourages more uniform allocation. The relative ranking (crater $>$ seismic $>$ VLM $>$ SAR) remains consistent, indicating robust identification of modality reliability.}
    \label{fig:dirichlet-sensitivity-gamma}
\end{figure}

Figure~\ref{fig:dirichlet-sensitivity-gamma} reveals how trust weight allocation varies with $\alpha$. Small values ($\alpha < 1$) favor sparse solutions where crater dominates ($\gamma_{\text{crater}} \approx 0.68$ at $\alpha = 0.1$). As $\alpha$ increases, the weights become more uniform, approaching equal allocation ($\gamma_i \approx 0.25$) for large $\alpha$. Notably, the relative ranking remains consistent: crater $>$ seismic $>$ VLM $>$ SAR across all values, confirming that data-driven trust relationships persist regardless of prior choice.

% The computational cost remains stable across all $\alpha$ values, with MCMC requiring similar computational time for 1,600 effective samples (after warm-up). 

This sensitivity analysis confirms that our choice of $\alpha_i = 1$ represents a reasonable default that allows the data to determine trust weights without imposing strong prior beliefs. The stability of yield estimates across different prior specifications strengthens confidence in our multimodal fusion results.

\clearpage
\end{document}